\documentclass{jfm}
\usepackage{graphicx, amsmath}
\usepackage{natbib}
\DeclareGraphicsExtensions{.pdf,.png,.jpg}
\usepackage{epsfig,colortbl,array}
\newcolumntype{M}{>{\centering\arraybackslash}m{\dimexpr.5\linewidth-2\tabcolsep}}

\graphicspath{{jpg_figures/}}

\newcommand\del{\bb{\nabla}}
\newcommand\bcdot{\bb{\cdot}}

\newcommand\vv{\bb{v}}
\newcommand\uu{\bb{u}}
\newcommand{\BV}{Brunt-V\"ais\"al\"a\ }

\newcommand\op {\omega}
\newcommand\opt {\widetilde\omega}
\newcommand\msun {$M_\odot$}

\newcommand\ylm{ Y_\ell^m}            
\newcommand\ulm{ u^\ell_m}            
\newcommand\vlm{ v^\ell_m}            
\newcommand\wlm{ w^\ell_m}            
\newcommand\rlm{ \bb{R}_\ell^m}            
\newcommand\slm{ \bb{S}_\ell^m}            
\newcommand\tlm{ \bb{T}_\ell^m}            
\newcommand\thlm{ t^\ell_m}

\newcommand\bb[1] {   \mbox{\boldmath{$#1$}}  }

\newcommand{\beq}{\begin{equation}}
\newcommand{\eeq}{\end{equation}}

\usepackage[usenames,dvipsnames]{xcolor}

\newcommand{\prt}{\partial}
\newcommand{\dnr}[1]{\frac{d  #1}{dr}}
\newcommand{\ddnr}[1]{\frac{d^2  #1}{dr^2}}

\ifCUPmtlplainloaded \else
  \checkfont{eurm10}
  \iffontfound
  \IfFileExists{upmath.sty}
  {\typeout{^^JFound AMS Euler Roman fonts on the system, using the 'upmath' package.^^J}%
\usepackage{upmath}}
{\typeout{^^JFound AMS Euler Roman fonts on the system, but you dont seem to have the}%
 \typeout{'upmath' package installed. JFM.cls can take advantage
    of these fonts,^^Jif you use 'upmath' package.^^J}%
}
\else
\fi
\fi

\ifCUPmtlplainloaded \else
  \checkfont{msam10}
    \iffontfound
       \IfFileExists{amssymb.sty}
{\typeout{^^JFound AMS Symbol fonts on the system, using the
'amssymb' package.^^J}%
\usepackage{amssymb}%
\let\le=\leqslant  \let\leq=\leqslant
\let\ge=\geqslant  \let\geq=\geqslant
}{}
\fi
\fi

\ifCUPmtlplainloaded \else
  \IfFileExists{amsbsy.sty}
    {\typeout{^^JFound the 'amsbsy' package on the system, using it.^^J}%
    \usepackage{amsbsy}}
    {}
\fi

\title[Gravito-inertial waves in a differentially rotating spherical shell]{Gravito-inertial waves in a differentially rotating spherical shell}

\author[G. M. Mirouh, C. Baruteau, M. Rieutord and J. Ballot]%
{G. M. Mirouh$^{1,2}$%
\thanks{Email address for correspondence: giovanni.mirouh@irap.omp.eu},\ns C. Baruteau$^{1,2}$, \ns M. Rieutord$^{1,2}$ and J. Ballot$^{1,2}$}

  \affiliation{$^1$IRAP; CNRS; 14, avenue Edouard Belin, 31400 Toulouse, France\\[\affilskip]
    $^2$Universit\'e de Toulouse; UPS-OMP; IRAP; Toulouse, France}
  \pubyear{2014}
  \volume{42}
  \pagerange{42--42}

  \date{?; revised \today; accepted ?. - To be entered by editorial office}

\begin{document}
  \maketitle

\begin{abstract}
The gravito-inertial waves propagating over a shellular baroclinic
flow inside a rotating spherical shell are analysed using the Boussinesq
approximation. The wave properties are examined by computing paths
of characteristics in the non-dissipative limit, and by solving the full
dissipative eigenvalue problem using a high-resolution spectral method.
Gravito-inertial waves are found to obey a mixed-type second-order
operator and to be often focused around short-period attractors of
characteristics or trapped in a wedge formed by turning surfaces and
boundaries. We also find eigenmodes that show a weak dependence with
respect to viscosity and heat diffusion just like truly regular modes.
Some axisymmetric modes are found unstable and likely destabilized by
baroclinic instabilities. Similarly, some non-axisymmetric modes that
meet a critical layer (or corotation resonance) can turn unstable at
sufficiently low diffusivities. In all cases, the instability is driven
by the differential rotation. For many modes
of the spectrum, neat power laws are found for the dependence of the damping
rates with diffusion coefficients, but the theoretical explanation
for the exponent values remains elusive in general. The eigenvalue spectrum 
turns out to be very rich and complex, which lets us suppose
an even richer and more complex spectrum for rotating stars or planets that own
a differential rotation driven by baroclinicity.
\end{abstract}



\section{Introduction}
Stars of mass above 2.5 solar masses (2.5\msun), also known as
early-type stars, are basically made of a central convective core (where
nuclear reactions take place) and a radiative envelope. When the star is
rapidly rotating, as it is often the case for this type of stars,
the radiative envelope is differentially rotating as a result of the baroclinic
torque that comes from the stable stratification in the
envelope \cite[][]{zahn92,R06}. Besides, the envelope is the seat of
oscillations driven by the so-called $\kappa$-mechanism \cite[][]{Z63,unno_etal89}. 
This mechanism is a
consequence of the variations of heat conductivity with temperature that
make, in some places, hotter material less conductive\footnote[2]{Recall
  that heat conduction in stars is due to the diffusion of photons and
  is therefore controlled by the opacity of the medium. Opacity varies
  rapidly with temperature in regions where some abundant element changes
  its ionization degree. Most frequently, regions of the transition
  between ionization states of hydrogen or helium are the drivers of the
  $\kappa$-mechanism.} \cite[see][]{GD08}. 
These $\kappa$-driven oscillations have been observed
in many stars. Astrophysicists have classified these stars in various
families, like $\delta$ Scuti stars, Slowly Pulsating B-stars, $\beta$ Cephei
stars etc. This latter family are main sequence (in-core hydrogen burning) stars
of mass around 9~\msun. They usually rotate quickly but some of them are
slow rotators. 
The analysis of the eigenmodes is much easier for the slowly rotating 
$\beta$ Cephei stars than for the fast rotating ones, and astrophysicists have
been able to identify the modes of oscillation \cite[][]{dupret_etal04},
which turn out to be gravity modes (where buoyancy is the restoring
force). However, the vast majority of $\beta$ Cephei stars rotate rapidly and
their oscillation frequencies remain hardly interpreted because
of a lack of tools to identify the corresponding modes.

The foregoing problem comes from the fact that gravity modes are strongly
perturbed by the Coriolis acceleration. In fact they combine with inertial
modes, which are restored by the Coriolis force, and form a vast set of
low-frequency modes usually called gravito-inertial modes, the properties
of which are still poorly known. Gravito-inertial modes deserve studies
because they can inform us on the physical conditions at the boundary
between the convective core and the radiative envelope of the stars, a
location where mixing is important. Indeed, convective cores mix material
of the envelope through convective overshooting  \cite[e.g.][]{maeder09},
a phenomenon that crucially impacts the lifetime of this type of stars.
Hence, gravito-inertial modes offer a precious window on the core of
massive stars.

Beside the possibility of observing stellar interiors \cite[e.g.][]{mathis_etal14}, 
gravito-inertial modes are also thought to play an important role in the dissipation
and momentum transfer processes that are associated with the
tidal interaction between stars \cite[][]{WS99a}, between stars and planets
\cite[][]{O14}, or between planets and their satellites \cite[for the example of Saturn, 
see][]{lainey_cassini15}. For instance, the fate of Jupiter-like
planets on close-in orbits around fast rotating early-type stars is a main
concern for our understanding of the evolution of planetary systems. The
tidal interaction between the star and the planet is strongly influenced
by the excitation of gravito-inertial modes.  This interaction determines
the long-term evolution of a planet's orbital elements \cite[][]{O14}, 
this evolution and the associated time scales are still poorly known.
Determining the oscillations' properties may also help constrain
models of the interior of the planet itself \cite[][]{F14}, by determining
the presence and the location of stably stratified fluid layers, for instance.
Gravito-inertial modes are also thought to play a role in the generation
and evolution of vortices in stably stratified protoplanetary disks \cite[][]{marcus13, marcus15}.

As alluded above, the propagation and dissipation properties of
gravito-inertial modes are still poorly understood. As shown by early work
\cite[][]{Friedl82,Friedl87}, these waves are controlled by a mixed-type
operator: oscillatory solutions (i.e. modes) may be supported in only
part of the shell.  \cite{DRV99} have shown that  modes may be singular
with an amplitude focused around an attractor of characteristics. Indeed,
\cite{DRV99} have shown that characteristics of the hyperbolic region,
where waves propagate, often tend to be focused, either along a periodic
orbit or in a wedge formed by the boundaries and by the turning surfaces,
which separate the hyperbolic and elliptic domains.  These so-called
attractors \cite[][]{ML95} appear as shear layers when viscosity is
taken into account. They systematically shape pure inertial modes
\cite[][]{RGV01,BR13}. The role of these singularities in shaping the
spectrum of a stellar envelope is not clear. However, in the studies
that have focused on gravito-inertial modes, the background rotation has
always been taken as solid body. In stellar envelopes the situation is
not so simple since, as mentioned above, stellar envelopes are pervaded
by baroclinic flows (the so-called thermal wind in geophysics), that
impose a differential rotation. As shown by \cite{BR13}, differential
rotation may profoundly change the nature of oscillations at low
frequencies. Instabilities may arise, for instance, from critical layers
where the phase velocity of the waves equals that of the fluid, or from
the poorly known axisymmetric, baroclinic diffusive (ABCD) instability
\cite[][]{SK84}.

In the present work we investigate the properties of gravito-inertial
modes in the context of baroclinic stellar envelopes, but with a
simplified model. The radiative envelope of the star is modelled by
a stably stratified rotating incompressible fluid contained in a
spherical shell. Thus compressibility is ignored and the Boussinesq
approximation is used, as in \cite{R06}.  In this set-up baroclinic
flows are axisymmetric and are the superposition of a differential rotation 
and a weak meridional circulation. Despite these simplifications the study of the
properties of disturbances propagating over such a flow remains quite 
tricky, even if we can use the equatorial symmetry of the set-up. 
Fortunately, we can further simplify the background by using a result 
of \cite{HR14} who have shown that the differential rotation loses its 
latitude dependence and becomes ``shellular" (i.e. $\Omega\equiv\Omega(r)$ 
with $r$ being the radial spherical coordinate), when no-slip boundary 
conditions are used, and the inviscid limit is taken. 
Such boundary conditions are of course not realistic, 
but are worth being used: they reduce the complexity of the eigenvalue 
problem, while they allow us to keep the connection between the strength 
of the stratification and the strength of the differential rotation.  
However, as far as the disturbances are concerned, we impose that they 
match stress-free boundary conditions. This is more realistic, less 
demanding numerically and less dissipative, thus easing the detection of unstable modes.

Hence, our model, though quite simplified compared to a real stellar
envelope, retains the essential features that affect gravito-inertial
waves: stratification, rotation, spherical geometry and the coupling
between stratification and differential rotation. In the next section we
shall give the mathematical formulation of this problem and then focus
on the properties of the governing operator and associated waves in the
non-dissipative limit (section~\ref{sec:nondiss}).  The role of viscosity
and heat diffusion will be investigated in section~\ref{sec:diss}.
Conclusions and outlook on the astrophysical questions end the paper.

\section{Formulation}
\label{sec:math}
\subsection{Physical model}

We consider a thermally stratified, differentially rotating viscous
fluid inside a spherical shell.  The shell is located between radii
$\eta R$ and $R$, with $0\leq\eta<1$.  The flows are described using
the Boussinesq approximation, and the fluid is of constant kinematic
viscosity $\nu$ and thermal diffusivity $\kappa$.

The Navier-Stokes equation in an inertial frame reads
\begin{equation}
  \rho\left(\frac{\prt\vv}{\prt t} + \vv\bcdot\del\vv\right) = -\del P + \rho\nu\nabla^2\vv + \rho \bb{g},
\end{equation}
where $\vv$ is the fluid's velocity, $P$ the pressure, and $\rho$
the density.  We decompose all quantities as $x = x_0 + x_1$, where
$x_0$ is the unperturbed background quantity and $x_1$ the associated
disturbance such that $|x_1| \ll |x_0|$.

In spherical coordinates $(r, \theta, \phi)$, the background fluid's
velocity reads $\vv_0 = r \sin\theta\Omega_0(r) \bb{e}_\phi$, where
$\Omega_0(r)$ is the fluid's angular velocity. Incompressibility
implies that $\bb{g} = - g_0 \bb{r}/R$, where $g_0$ is the surface
gravity. From the Boussinesq approximation we get $\rho_1 \bb{g}_0
= \rho_0 \alpha g_0 T_1 \bb{r}/R$ where $\alpha$ is the dilation
coefficient and $T_1$ the temperature perturbation \cite[see~][]{chandra61}.

Keeping only the linear terms, the Navier-Stokes equation now reads
\begin{equation}
\label{eq:ns_dim}
   \frac{\prt\vv_1}{\prt t} + \Omega\frac{\prt \vv_1}{\prt\phi} + 2\bb{\Omega}\times\vv_1 + r\sin\theta \left( \vv_1\bcdot\del\Omega \right)\bb{e}_\phi
              = -\frac{1}{\rho_0}\del P_1 + \nu \nabla^2\vv_1 + \alpha g_0 T_1 \bb{r},
\end{equation}
and the continuity equation simply becomes 
\begin{equation}
\label{eq:cont_dim}
  \del\bcdot\vv_1 = 0.
\end{equation}
The heat equation reads
\begin{equation}
\label{eq:heat}
   \frac{\prt T}{\prt t} + \vv\bcdot\del T = \kappa \nabla^2 T +Q.
\end{equation}
We assume that heat sinks $Q$ are uniformly distributed throughout
the shell, so as to impose a stable temperature gradient $\del
T_0 = \beta\bb{r} / R$ where $\beta$ is a positive constant
\cite[][]{DRV99}. Equation (\ref{eq:heat}) becomes, once linearised,
\begin{equation}
\label{eq:heat_dim}
   \frac{\prt T_1}{\prt t} + \Omega_0 \frac{\prt T_1}{\prt \phi} + \frac{\beta}{R}\vv_1\bcdot\bb{r}  = \kappa \nabla^2 T_1.
\end{equation}

As anticipated in the introduction, and as will be detailed in
section~\ref{sec:bc}, our stratification model yields a radially
varying angular velocity $\Omega_0(r)$. It allows us to use $\Omega_s =
\Omega_0(R)$ as a frequency scale.  We further use $R$ to rescale lengths
and $\beta R$ to rescale temperatures, and define three dimensionless
parameters to rewrite the set of equations:
\begin{equation}
  \label{eq:parameters}
  N^2 = \frac{\alpha\beta g_0}{\Omega_s^2},\qquad \mathcal{P} = \frac{\nu}{\kappa},\qquad E = \frac{\nu}{\Omega_s R^2},
\end{equation}
which are respectively the dimensionless \BV frequency squared, the Prandtl number and the Ekman number.

We seek solutions to equations (\ref{eq:ns_dim}), (\ref{eq:cont_dim}) and
(\ref{eq:heat_dim}) proportional to $\exp(\lambda t+ im\phi)$.  We denote
by $\op$ the imaginary part of $\lambda$, it is the wave frequency
in the inertial frame. From now on, we make use of $p=P_1/\rho_0$,
the dimensionless reduced pressure perturbation. Dropping the 0 and 1
subscripts, our non-dimensional set of equations finally reads
\begin{eqnarray}
\label{eq:ns_a}
  (\lambda+im\Omega) \vv + 2 \bb{\Omega}\times\vv + r\sin\theta \left( \vv\bcdot\del\Omega \right)\bb{e}_\phi
       &=&  -\del p + E \nabla^2\vv + N^2 T \bb{r},\\
\label{eq:masscons_a}
  \del\bcdot\vv &=& 0,\\
\label{eq:heat_a}
  (\lambda+im\Omega) T + r v_r &=& \frac{E}{\mathcal{P}} \Delta T.
\end{eqnarray}

\subsection{Boundary conditions and stratification}
\label{sec:bc}
We study the problem of gravito-inertial modes over a
differentially-rotating background flow.  We impose a rotation
profile that depends on the radial coordinate $r$ only, i.e. a
shellular differential rotation.  Such a profile has been used
in all one-dimensional (spherically symmetric) stellar models
\cite[e.g. ][]{Morel97,paxton_etal11}, and we shall now explain its
origin.

\cite{R06} has shown that, in the Boussinesq approximation, for any
\BV frequency ${\mathcal N}(r)$, the steady flow resulting from the combined
effects of rotation and stratification, has the following differential
rotation profile:
\begin{equation}
  \delta\Omega = \int\limits^1_r{\frac{{\mathcal N}^2(r')}{r'} dr'} + F(s),
\end{equation}
where the scalar function $F(s)$ is determined by the viscous boundary
conditions.  Using no-slip boundary conditions on both the internal and
external boundaries of the shell, the $F(s)$-term vanishes when the
Ekman number vanishes. In this limit a purely radial
differential rotation is easily computed from the \BV frequency profile
in the shell \cite[][]{HR14}.

For the sake of simplicity, we therefore invoke no-slip boundary
conditions for our base flow, which allow us to relate the differential
rotation to the \BV frequency with the following equation
\begin{equation}
  \Omega(r) = 1 + \frac{N^2}{2} (1-r^2),
\label{eq:omega}
\end{equation}
where we assume that the non-dimensional \BV frequency grows linearly
with the radial distance ${\mathcal N}(r) = N\times r$ \cite[as a consequence of
the uniform distribution of heat sinks, see][]{DRV99}.

We note that no-slip boundary conditions may be used in stars to mimic
interfaces with turbulent layers, for instance near the core where
strong gradients of mean molecular weight may limit the wave propagation
\cite[][]{KS83}, or near the surface where a turbulent layer may appear
because of mass loss \cite[][]{RB14}.  Interfaces between convective and
radiative zones may also act as ``walls'' limiting the wave propagation
domain.  No-slip boundary conditions can also be used in some Jovian
and Saturnian moons, where it is thought that a liquid ocean is trapped
between a solid core and an outer ice layer \cite[][]{Carr98}.

\begin{table}
  \begin{center}
  \begin{tabular}{|c||c|c|}
    \hline
       & Stress-free conditions            & No-slip conditions\\
    \hline
    \hline
 $m=4, N^2=1.5$            &  $\tau=-2.623\times10^{-3},~\op=-2.67162 $ &  $\tau=-2.705\times10^{-3},~\op=-2.67155$      \\
 $m=0, N^2=0.45$           &   $\tau=-1.777\times10^{-3},~\op=1.75431 $      &  $\tau=-1.833\times10^{-3},~\op=1.75433$       \\
 $m=0, N^2=2.6$            &   $\tau=-1.572\times10^{-3},~\op=0.80413 $      &  $\tau=-2.226\times10^{-3},~\op=0.80434$       \\
 $m=0; N^2=2.3$            &   $\tau=-8.884\times10^{-4},~\op=0.30613 $      &  $\tau=-1.393\times10^{-3},~\op=0.30730$       \\
    \hline
  \end{tabular}
  \end{center}
  \caption{Eigenvalues computed for two different sets of boundary conditions for $P=10^{-2}$ and
  $E=10^{-9}$, except for the last one which uses $E=10^{-10}$.}
  \label{tab:NSvsFF}
\end{table}

On top of this base flow, oscillations should meet the same boundary
conditions.  However, \cite{FH98} have shown that the impact of boundary
conditions on inertial eigenmodes is small.  We confirm this result in
the case of gravito-inertial modes, as shown in Table \ref{tab:NSvsFF}.
As one would expect, stress-free conditions for the velocity are slightly
less dissipative, hence more permissive of possible instabilities.
As they are also less demanding numerically, we choose them to complete
equations (\ref{eq:ns_a})-(\ref{eq:heat_a}) along with fixed temperature
conditions, i.e. $T(\eta) = T(1) = 0$.

\subsection{Range of parameters}
\label{sec:param}
In rotating stars, viscous forces are small, so that the Ekman and
Prandtl numbers lie in the following range~\cite[][]{R08, REL13}.
\begin{equation}
  \mathcal{P} \sim 10^{-5}, \quad E \sim 10^{-15} - 10^{-9}.
\end{equation}
However, as it gets harder to resolve shear layers at low
diffusivities, the lowest values of the Ekman number are presently out of reach.
The fully-dissipative solutions of the equations have thus been computed
for values in the following range
\begin{equation}
  \mathcal{P} \sim 10^{-2} - 1,\quad E\sim 10^{-10} - 10^{-6}.
\end{equation}
We also choose $N^2<9$ so as to limit the shellular differential
rotation between the core and the surface to $\Omega_{\rm core} /
\Omega_{\rm surface} < 5$, as actually expected in main sequence
stars~\cite[][]{ELR13}.

\subsection{Stability of the flow}
\label{sec:abcd}

In a differentially-rotating stratified fluid, several instabilities
can set in.  The two relevant kinds in the astrophysical context are the
baroclinic instabilities and the shear instabilities.

\subsubsection{Baroclinic instabilities}
These instabilities emerge from the misalignment of the isobaric
and isothermal surfaces.  At a given location, these surfaces form a
wedge. If a fluid parcel is displaced inside the wedge, it ends up in
a colder environment, and buoyancy pushes the parcel farther away from
its initial position and amplifies the motion.
The interested reader is referred to the review by \cite{zahn93houches} for more details.

For axisymmetric perturbations, the possible baroclinic instabilities are 
the so-called Goldreich-Schubert-Fricke (GSF) instability \cite[][]{GS67} and
the axisymmetric baroclinic diffusive (ABCD) instability \cite[][]{KS83}.
Both of them yield the same local instability criterion
when the fluid is stably stratified, differentially rotating and
chemically homogeneous, namely
\begin{equation}
  - \frac{1}{4} A_s \left(A_s^2 + A_z^2\right) > \mathcal{P} N^2,
\label{eq:rayleigh}
\end{equation}
where $A_s$ and $A_z$ are related to the partial derivatives of the specific
angular momentum $s^2 \Omega$ through
\begin{equation}
  A_s = \frac{2\Omega}{s} \frac{\prt}{\prt s}\left(s^2\Omega\right),\quad A_z = \frac{2\Omega}{s} \frac{\prt}{\prt z}\left(s^2\Omega\right).
\end{equation}
In our case, equation (\ref{eq:rayleigh}) amounts to a generalized
Rayleigh criterion on the distribution of angular momentum.
To assess whether our base flow is stable or not with respect to these
instabilities, we consider the shellular rotation profile given by
equation (\ref{eq:omega}). We get
\begin{equation}
  A_s = 4\Omega^2 - 2\Omega N^2 s^2, \quad A_z = - 2\Omega N^2 sz.
  \label{eq:AsAz}
\end{equation}

For an inviscid fluid, the criterion (\ref{eq:rayleigh}) simply 
becomes the classical Rayleigh criterion for the centrifugal instability 
$A_s <0$ \cite[][]{DR81} which becomes 
$N^2 \ge \frac{2}{1-\eta^2}$ in our model.
Viscosity contributes to stabilizing the flow. 
To assess whether this baroclinic instability may destabilize the
flow, we identify the range of parameters $(N^2, \op)$ for which the
criterion~(\ref{eq:rayleigh}) is expected to be met in the
propagation region of the shell (see section~\ref{sec:caracs}).  Outside of 
this range, either there is no unstable domain in the shell, or the instability
domain is entirely inside the evanescent region, pointing to  an {\it a priori}
stable configuration.  Axisymmetric modes with a non-vanishing amplitude in the unstable
domain delineated by equation (\ref{eq:rayleigh}) may
therefore have a positive growth rate, as section~\ref{sec:diss}~will show
(e.g. figure~\ref{fig:wedgeinst}).

Non-axisymmetric baroclinic instabilities are well-studied in geophysics \cite[][]{zahn93houches}, 
and the local non-dissipative criterion yields
\begin{equation}
  \cos^2\theta \frac{r}{\rho} \frac{\prt}{\prt r} \left(\frac{\Omega^2}{N^2} \rho r \frac{\prt\Omega^2}{\prt r} \right) > \Omega^2,
\end{equation}
which simplifies, using the Boussinesq approximation and equation (\ref{eq:omega}), into
\begin{equation}
  \label{eq:zahn}
  z^2 > \frac{1}{6 N^2}.
\end{equation}
A non-axisymmetric baroclinic instability is therefore possible for $N^2>1/6$, 
and the modes can be destabilized in a polar region of the shell
delimited by a horizontal line. Keep in mind that this criterion
does not take into account viscosity or thermal dissipation, which are expected to 
stabilize the flow, as it is the case for the ABCD instability.

It is also important to notice that these instability criteria are local and do
not take boundary conditions into account.  Indeed, not only can boundary
conditions damp the instability, but also the predicted unstable zones may lie
in the evanescent region of the shell, and therefore have no impact on the mode
propagation.

\subsubsection{Shear instabilities}
\label{sec:shear}
Shear instabilities may create turbulence in stellar radiation zones. 
The buoyancy has a stabilizing effect, and the local instability criterion is 
\begin{equation}
  Ri = \frac{N^2}{(dv/dr)^2} < \frac{1}{4},
\end{equation}
where $Ri$ is the Richardson number which compares the buoyancy with the
shear of the background flow, and $v$ the mean flow velocity \cite[][]{DR81}.

In the presence of strong thermal dissipation, this criterion is generalized to 
\cite[][]{zahn74,LCM99}:
\begin{equation}
  Ri Pe = \frac{N^2}{(dv/dr)^2} \times \frac{v \ell}{\kappa} < \frac{1}{4},
\end{equation}
where $Pe$ is the P\'eclet number, and $\ell$ a characteristic scale of the flow.
As $v = r\sin\theta \Omega(r)$, the criterion for instability becomes
\begin{equation}
  \label{eq:richardson}
  \frac{\mathcal{P}}{E}  \frac{\Omega N^2r^3\ell}{\sin\theta \left(\Omega-N^2r^2\right)^2} < \frac{1}{4}.
\end{equation}

This criterion allows us to determine the typical size of the turbulent layers
that can emerge from shear instabilities. 
In a stellar radiative zone, the ratio $\mathcal{P}/E$ is of order $10^{4}$ to $10^{10}$
(see section~\ref{sec:param}), and the ratio ${\Omega N^2r^3}/{\left(\Omega-N^2r^2\right)^2}$
is of order unity. Thus, if $\ell =O(1)$, the criterion is never met and the whole shell is stable.  
The criterion is only met when $\ell = O\left(\frac{E}{\mathcal{P}}\right)$,
implying that only layers of thickness $\ell \sim 10^{-10} - 10^{-4}$ can be destabilized.
This is clearly a very small scale that may only lead to small-scale turbulence.
These remarks show that shear instabilities cannot explain positive growth
rates obtained for some of the modes we show in section~\ref{sec:diss}, but they
can impact large-scale modes propagating in actual stars through enhanced
(possibly anisotropic) diffusion coefficients.

\section{Non-dissipative problem}
\label{sec:nondiss}

To get a first insight into the full solutions of equations
(\ref{eq:ns_a})--(\ref{eq:heat_a}), we study the problem in the non-dissipative
limit, by setting  $\nu = \kappa =0$.  We use the dynamics of characteristics
as a proxy for the study of the propagation properties of gravito-inertial
waves, as it is known to play a major role in shaping the solutions of the full
dissipative problem \cite[e.g. ][]{DRV99}.  It allows us to perform a full
exploration of the parameter space.  

\subsection{Paths of characteristics}
\label{sec:caracs}

We rewrite equations (\ref{eq:ns_a} - \ref{eq:heat_a}) using the
cylindrical coordinates $(s,\phi,z)$.  We combine their components in
order to reduce the system to a partial differential equation for the
reduced pressure perturbation $p$. The detailed derivation is given in
Appendix A.  Focussing on the second-order terms in the pressure equation,
we get
\begin{equation}
  (\opt^2- N^2z^2) \frac{\prt^2 p}{\prt s^2} + (2N^2 sz + A_z) \frac{\prt^2 p}{\prt s \prt z} + (\opt^2 - A_s - N^2s^2) \frac{\prt^2 p}{\prt z^2} = 0,
  \label{eq:poincare}
\end{equation}
where $A_s$ and $A_z$ are defined by equation (\ref{eq:AsAz}), $\opt = \op
+m\Omega$ is the mode's Doppler-shifted frequency, which is the mode's
frequency in the frame corotating with the fluid's surface angular frequency.
$N$ is the \BV frequency at the surface of the shell.

Equation (\ref{eq:poincare}) is the generalization of the pressure
perturbation equation of gravito-inertial modes in solid-body
rotation \cite[][]{FS82b, DRV99}, and of inertial modes in a
differentially-rotating shell \cite[][]{BR13}.  In the case of solid-body
rotation ($A_s = 4\Omega^2, A_z=0$) and without stratification, this
equation reduces to the Poincar\'e equation.  We therefore call the
left-hand side part of (\ref{eq:poincare}) the generalized Poincar\'e
operator.

From equation (\ref{eq:poincare}) we obtain the following
ordinary differential equations for the paths of characteristics
\begin{eqnarray}
  \frac{dz}{ds} &=& \frac{N^2 sz + A_z/2 \pm \sqrt{\Delta}}{\opt^2-N^2z^2}, 
  \label{eq:dzds}\\
  \frac{ds}{dz} &=& \frac{N^2 sz + A_z/2 \mp \sqrt{\Delta}}{\opt^2-A_s-N^2s^2},
  \label{eq:dsdz}
\end{eqnarray}
with
\begin{equation}
  \Delta = A_z\left(A_z/4 + N^2sz\right) - A_s \left(N^2z^2-\opt^2 \right) - \opt^2 \left(\opt^2 - N^2 (s^2+z^2)\right),
  \label{eq:delta}
\end{equation}
where we recall that $\opt=\op+m\Omega$.  For symmetry reasons, these
equations are solved in only the Northern meridian plane of the shell,
delimited by the equator, the rotation axis, the inner and outer
boundaries of the shell.  We integrate equations (\ref{eq:dzds}) or
(\ref{eq:dsdz}) with a fifth-order Runge-Kutta integrator from an
arbitrary initial location in the propagation domain of the shell.
When reaching a boundary, the characteristic is reflected inside the
shell by switching the $\pm$ sign in equations (\ref{eq:dzds}) or
(\ref{eq:dsdz}).  Even though both equations are equivalent, their
denominator vanishes at different values of $s$ and $z$.  We therefore
toggle from one to the other, always using the smaller absolute value of
the derivative to compute the path of a characteristic.  This procedure
allows us to avoid numerical integration errors that arise when the
slope of characteristics is too large.

\subsection{Turning surfaces and mode classification}
\label{sec:classif}

Equations (\ref{eq:dzds}) and (\ref{eq:dsdz}) allow
us to predict the propagation domain of gravito-inertial waves within
the shell.  The quantity $\Delta$, given in equation (\ref{eq:delta}),
may change sign in the shell, hence dividing the shell into hyperbolic
($\Delta>0$) and elliptic ($\Delta<0$) domains. Oscillatory solutions of
equation (\ref{eq:poincare}) only exist in the hyperbolic domain, whereas
solutions are evanescent in the elliptic part of the shell.  These two
regions are separated by a turning surface where characteristics bounce.
This description is similar to that of inertial modes in the presence
of differential rotation \cite[][]{BR13}, gravito-inertial modes
with solid-body rotation \cite[][]{DRV99} or magneto-inertial waves
\cite[][]{friedl89}.  

We divide the modes into two categories, following
the classification introduced by \cite{BR13}:

\begin{itemize}
  \item Modes with no turning surface inside the shell, which we name H
  modes (H for Hyperbolic domain). For these modes, gravito-inertial waves
  can propagate in the whole shell ($\Delta>0$ everywhere in the shell).

  \item Modes exhibiting one or several turning surfaces (defined by
  $\Delta =0$) inside the shell, which we name HT modes (T for Turning
  surfaces).

\end{itemize}

From equations (\ref{eq:dzds}) and (\ref{eq:dsdz}) we see that, for
a given azimuthal wavenumber $m$, the parameter space is restricted
to only two dimensionless parameters, namely the surface \BV frequency $N$,
and the wave frequency $\op$.  We determine the regions occupied by H
and HT modes in this parameter space by checking whether the equation
$\Delta=0$ has a solution in the shell at given $N^2$ and $\op$.
Recasting equation $\Delta=0$ in terms of radius $r$ and colatitude
$\theta$, using $s=r\sin\theta$ and $z=r\cos\theta$, we get
\begin{equation}
  \label{eq:delta_sint}
  \opt^4 - (4\Omega^2 + N^2 r^2)\opt^2 + 4\Omega^2 N^2 r^2 = 2\Omega N^2r^2 \left(2\Omega - \opt^2\right) \sin^2\theta.
\end{equation}
This is an implicit non-linear equation that gives the shape of the
turning surfaces projected on a meridian plane. The general solution 
is difficult to obtain but one can get a rather detailed
view of the various solutions by solving equation (\ref{eq:delta_sint}) at 
specific points of the shell. We thus choose the poles and equator of 
the two bounding spheres, namely fixing $r=1$ or $r=\eta$,
and $\theta=0$ or $\theta=\pi/2$. Thus doing we find the following 
relation between $\omega$ and $N$:
\begin{eqnarray}
  &\bullet&\ \mathrm{for } \sin^2\theta(1)=0       : \op = -m \pm 2 \ \mathrm{or\ } \op = -m \pm N,
  \label{eq:transition1}\\
  &\bullet&\ \mathrm{for } \sin^2\theta(\eta)=0  : \op = -m \Omega(\eta) \pm N\eta \ \mathrm{or\ } \op = \left(-m \pm 2\right) \Omega(\eta),
  \label{eq:transition2} \\
  &\bullet&\ \mathrm{for } \sin^2\theta(1)=1       : \op = -m \pm \sqrt{4-N^2},
  \label{eq:transition3} \\
  &\bullet&\ \mathrm{for }\sin^2\theta(\eta)=1   : \op = -m \Omega(\eta) \pm \sqrt{N^2 \eta^2 \left(1-2\Omega(\eta) \right) + 4\Omega(\eta)^2}.
  \label{eq:transition4}
\end{eqnarray}
We now discuss these results according to the value of $m$.

\subsubsection{Axisymmetric modes}
\begin{figure}
   \includegraphics[angle=-90,width=\textwidth]{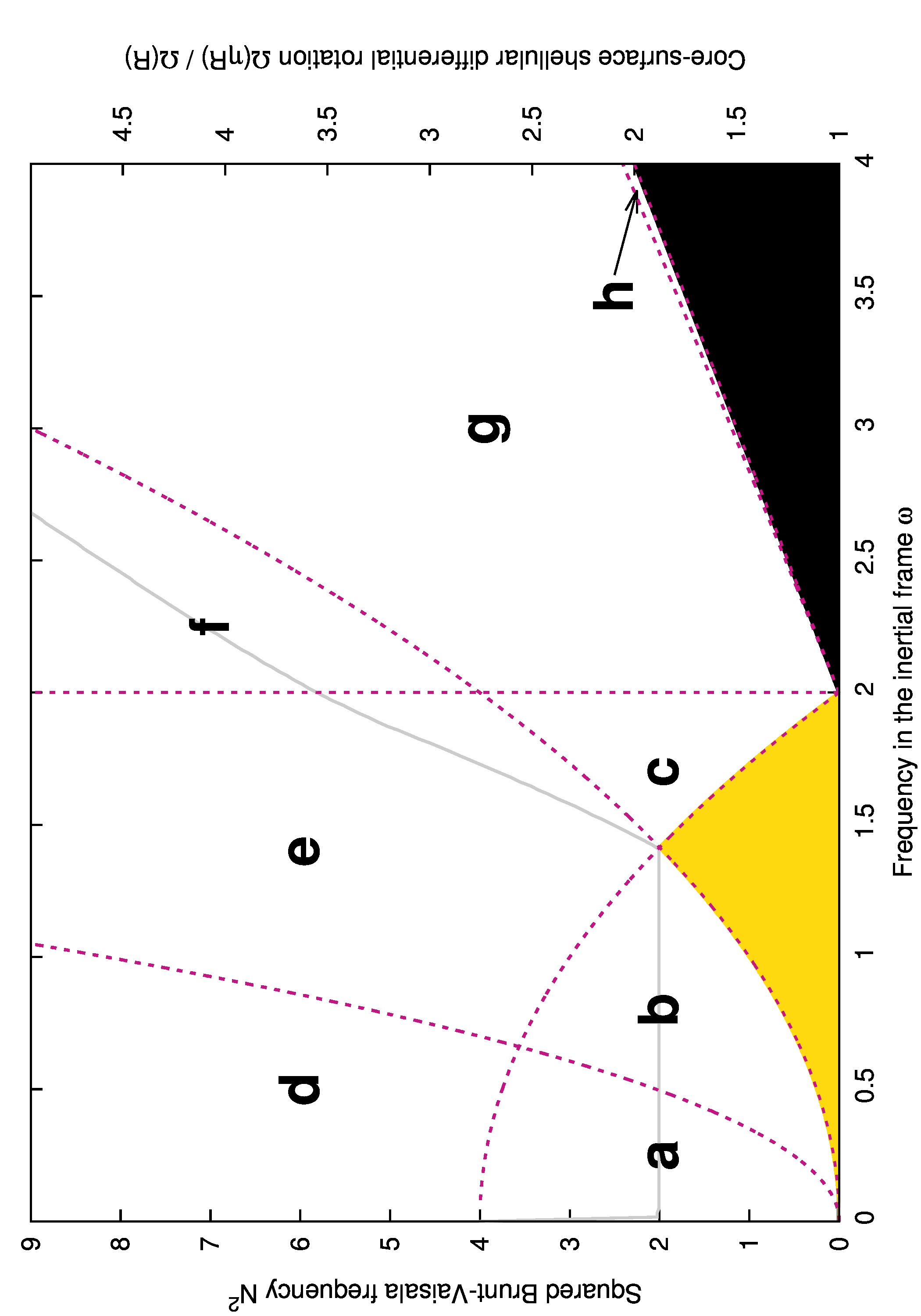}\\
\begin{center}
  \begin{tabular}{cccc}
   \includegraphics[angle=-90,width=0.25\textwidth]{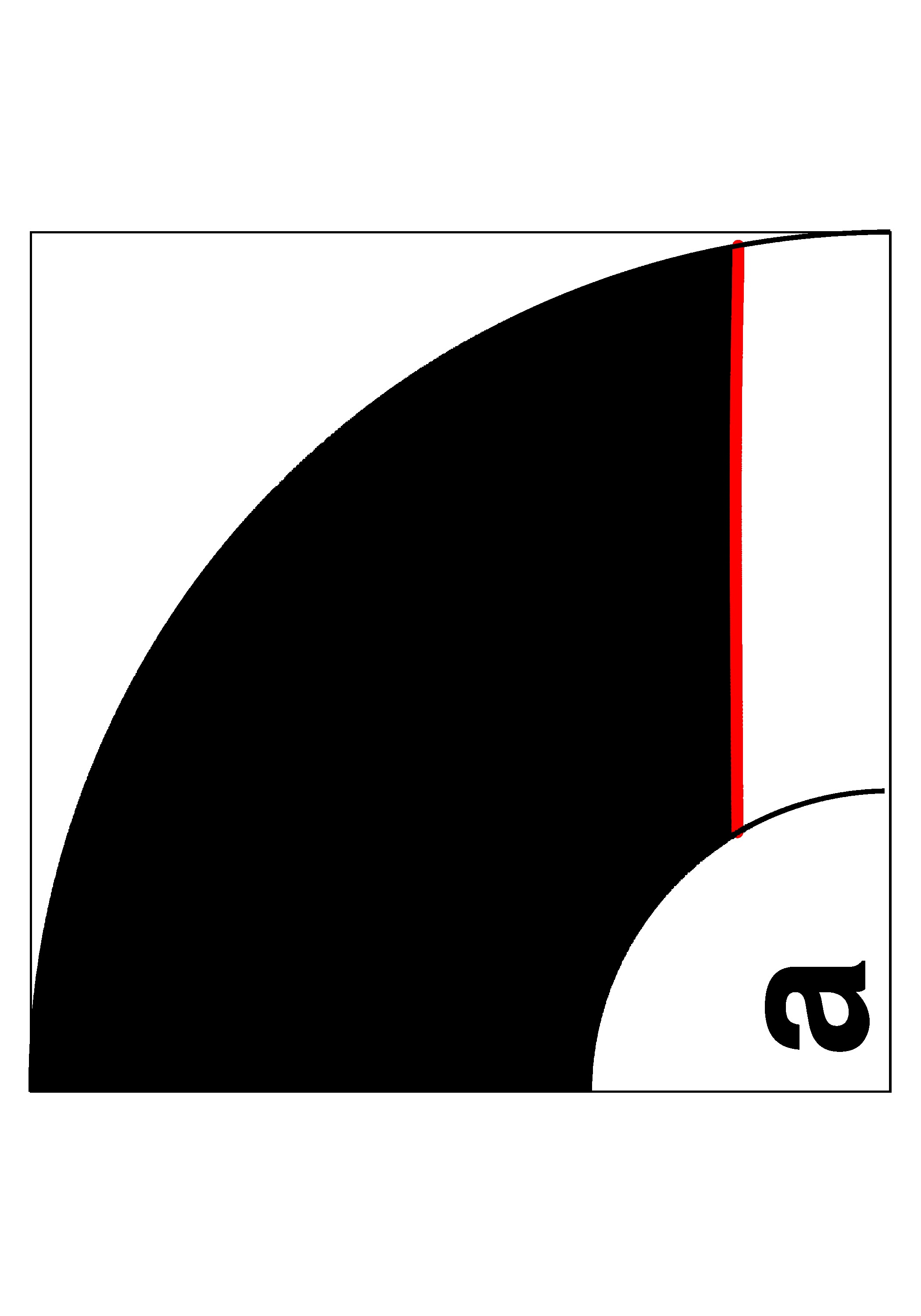}&
   \includegraphics[angle=-90,width=0.25\textwidth]{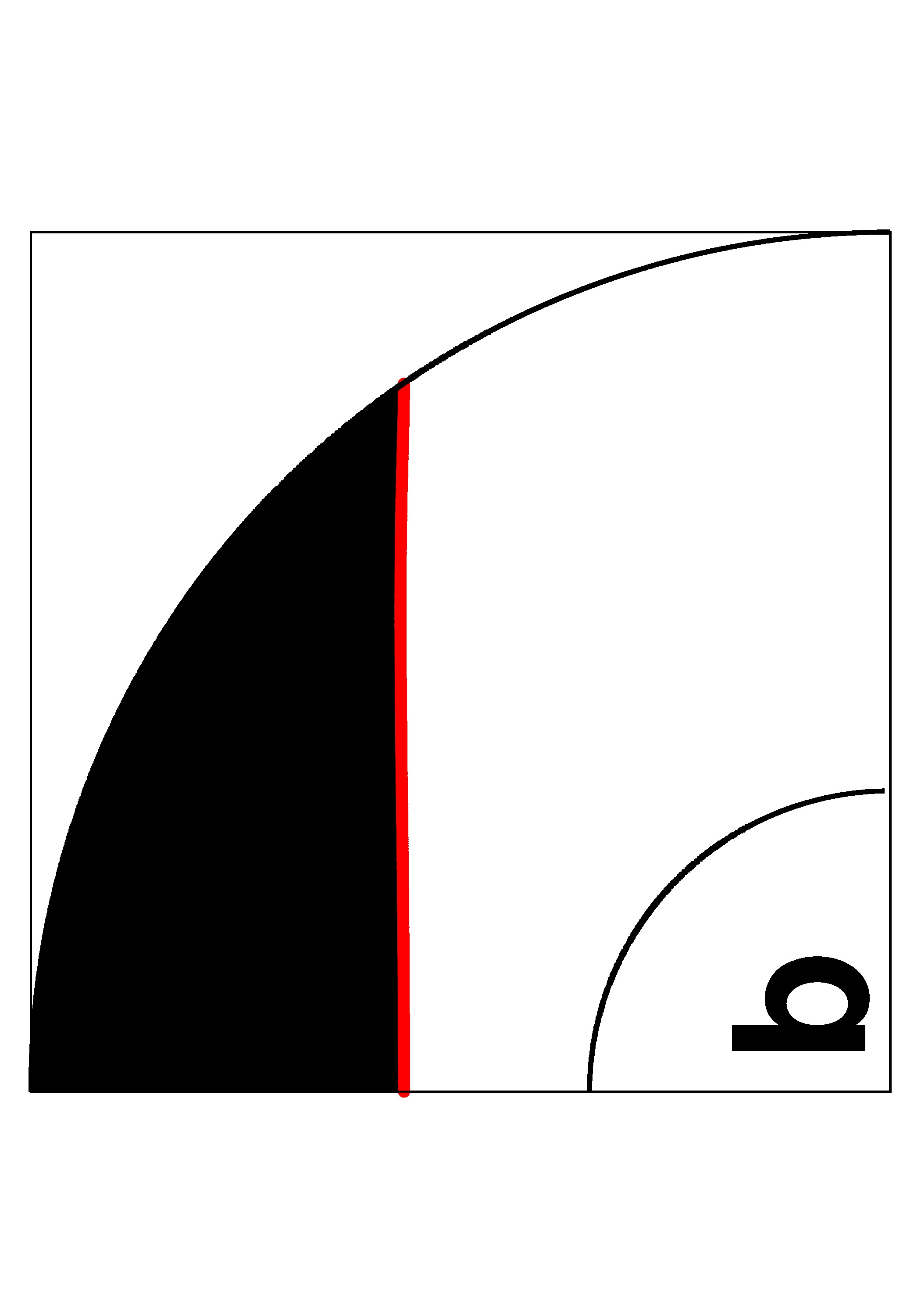}&
   \includegraphics[angle=-90,width=0.25\textwidth]{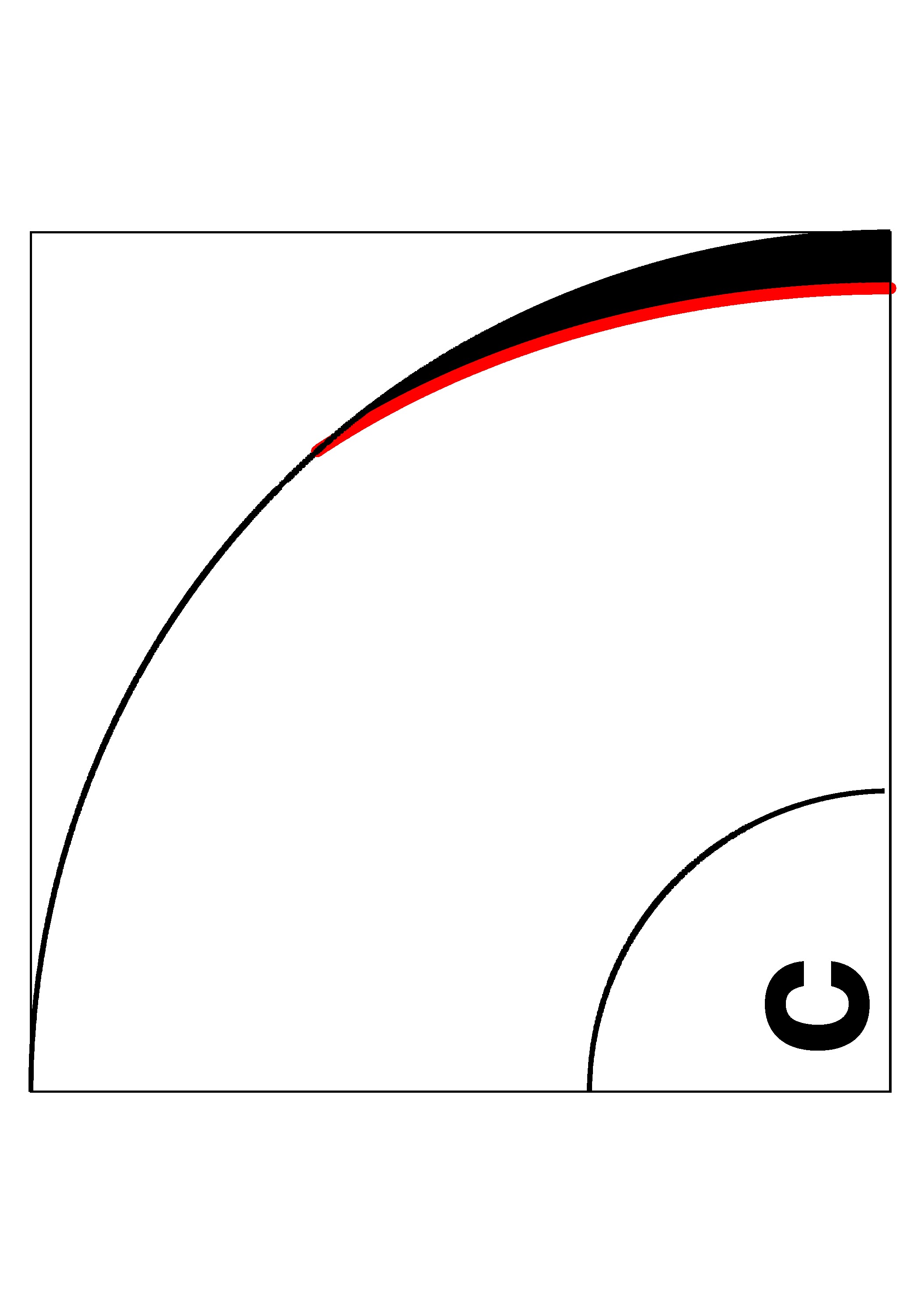}&
   \includegraphics[angle=-90,width=0.25\textwidth]{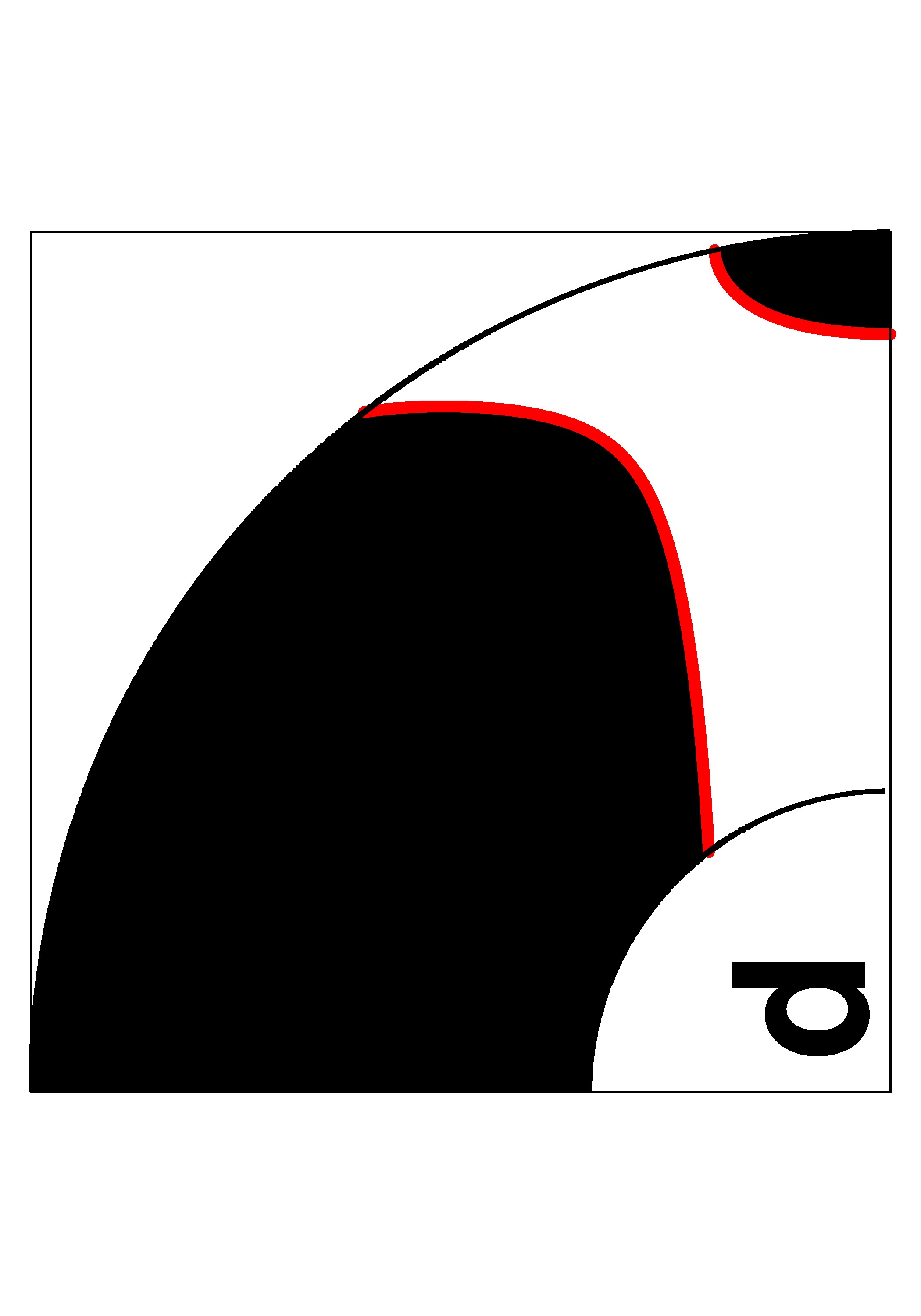}\\
   \includegraphics[angle=-90,width=0.25\textwidth]{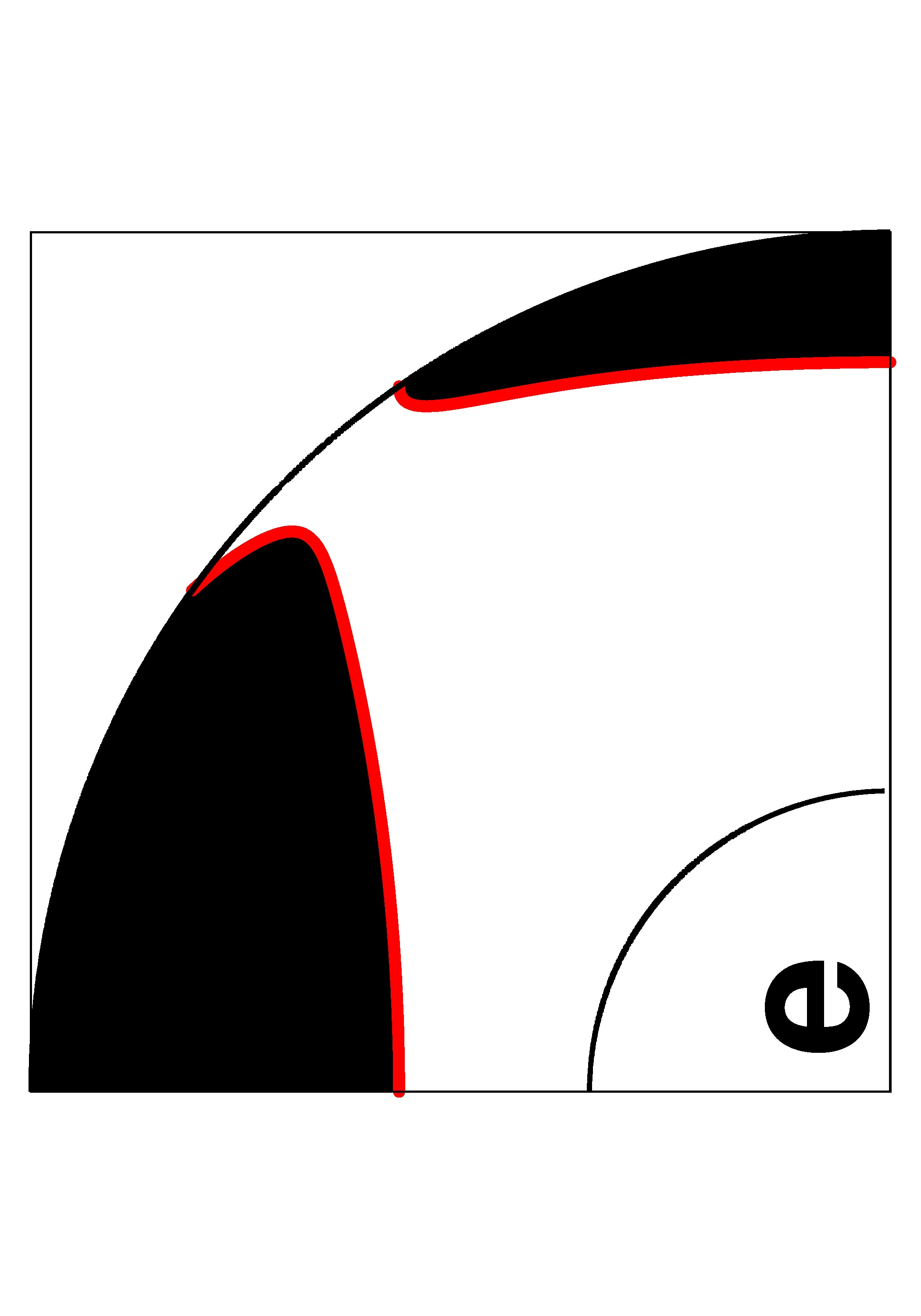}&
   \includegraphics[angle=-90,width=0.25\textwidth]{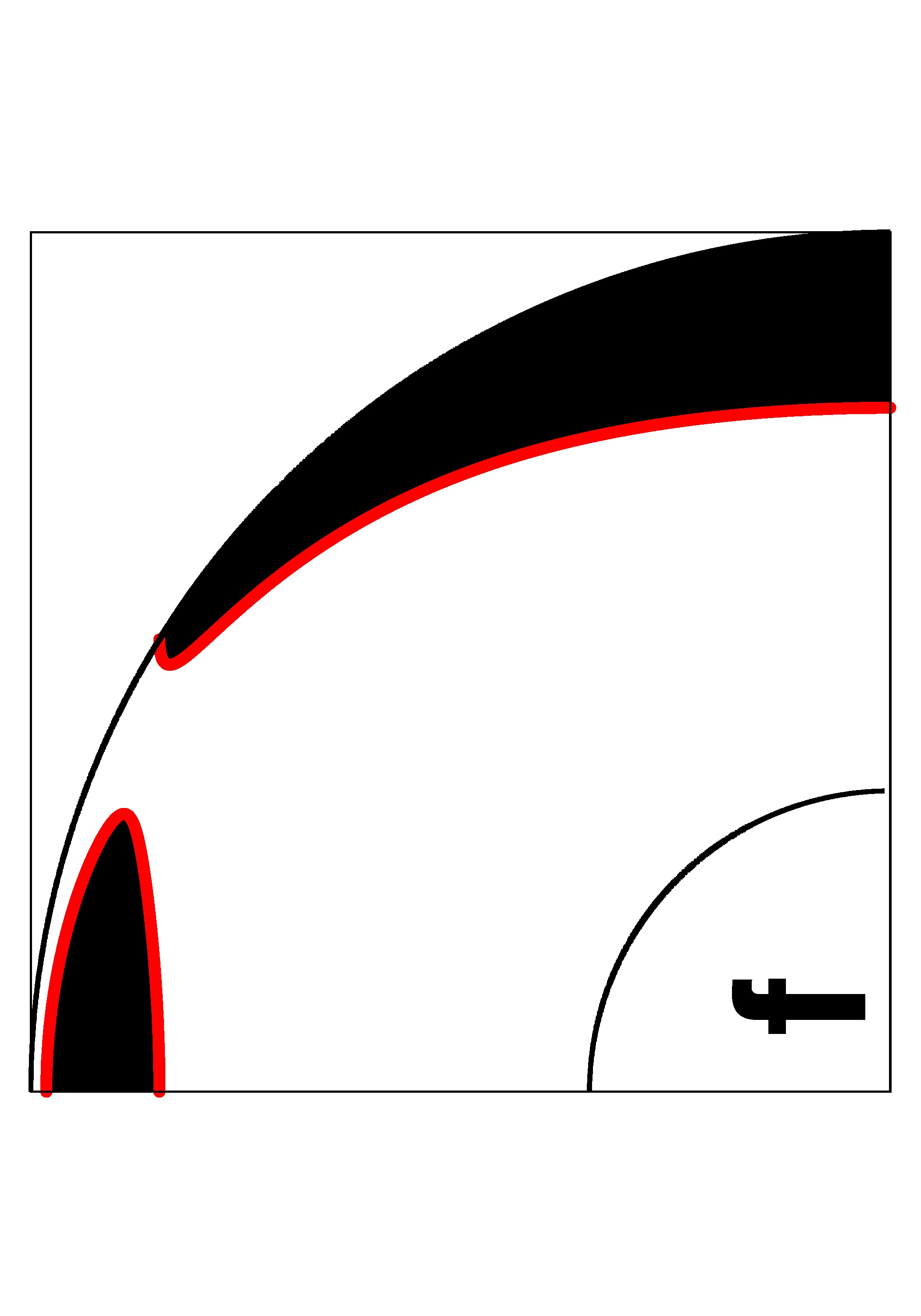}&
   \includegraphics[angle=-90,width=0.25\textwidth]{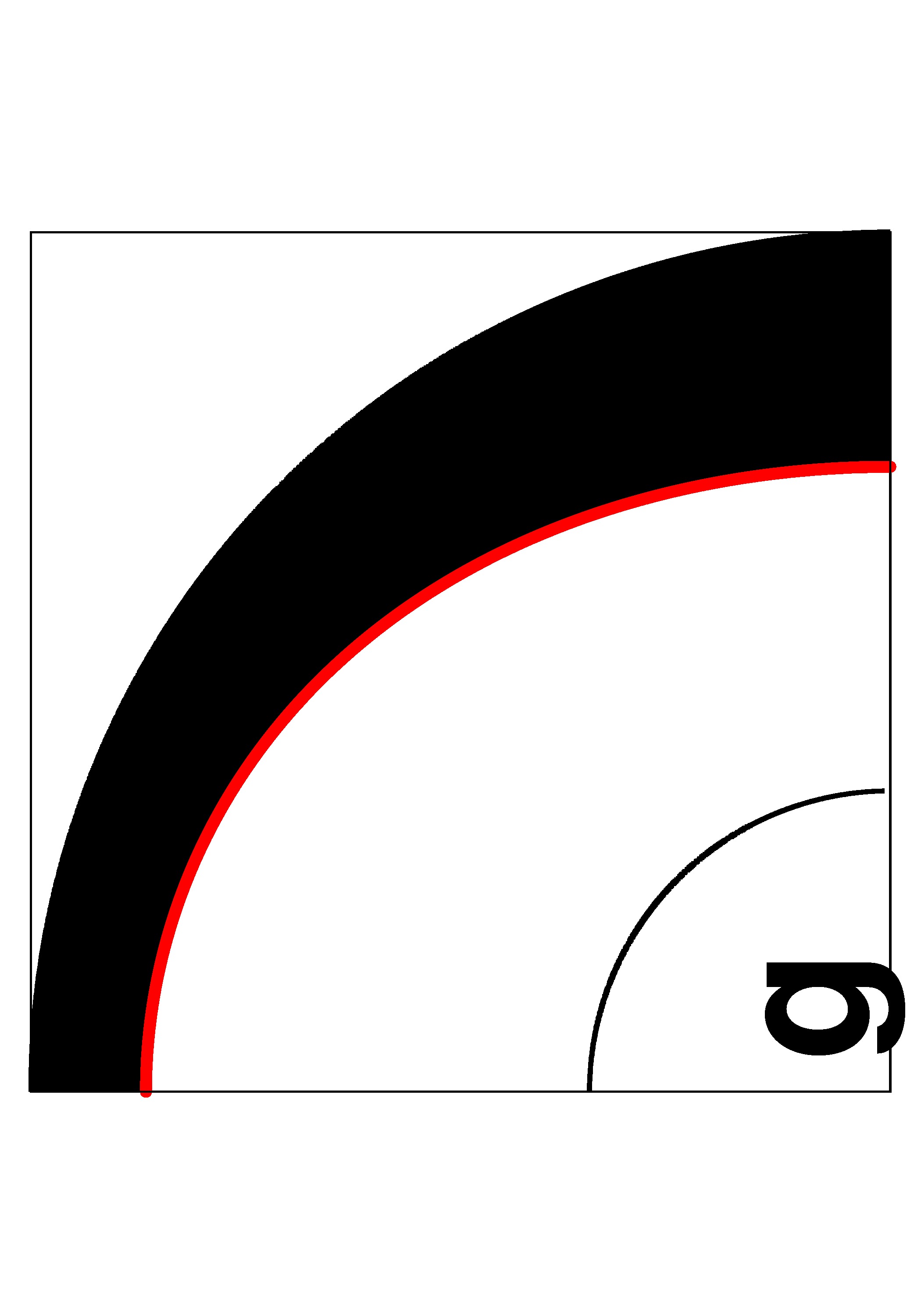}&
   \includegraphics[angle=-90,width=0.25\textwidth]{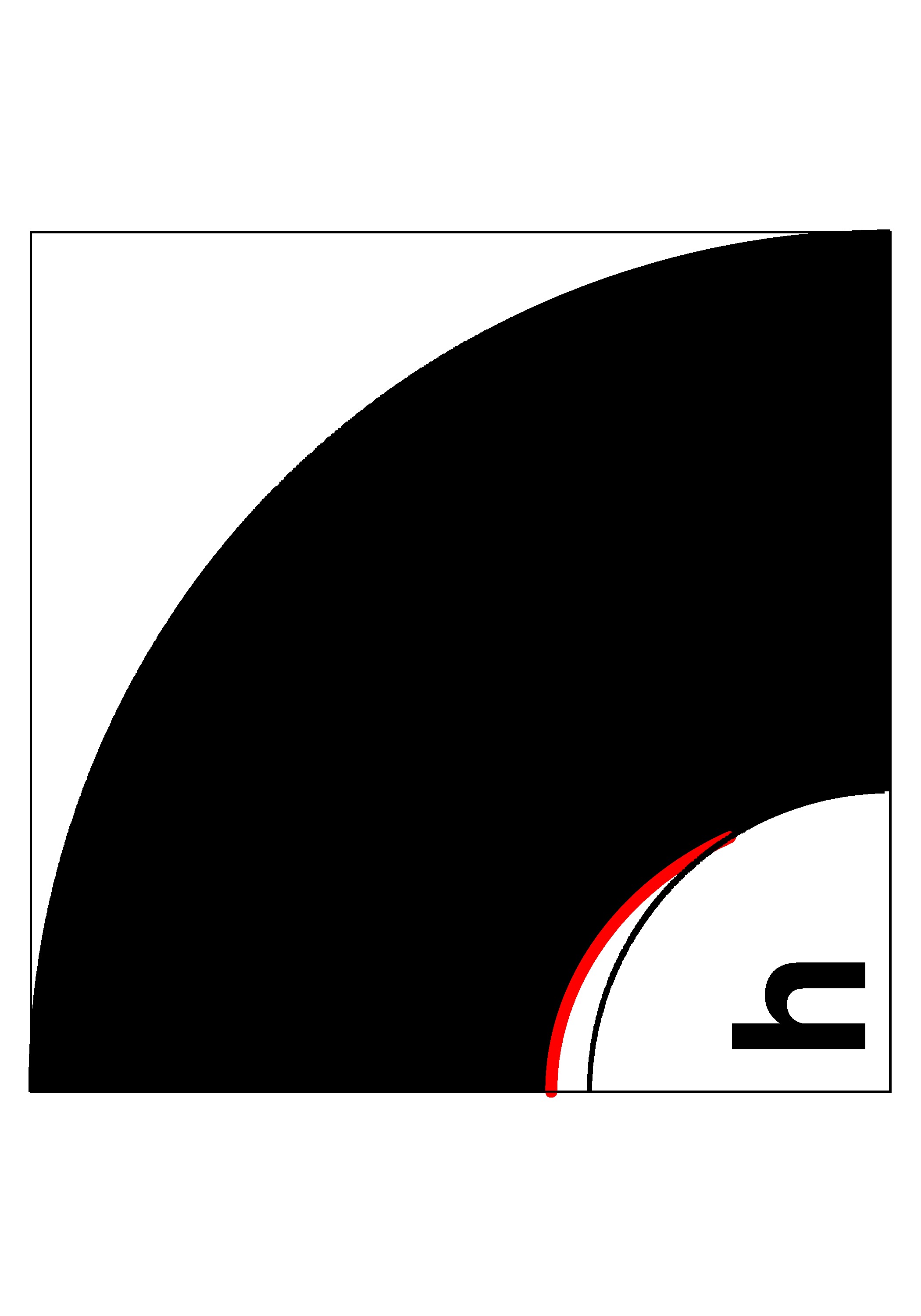}
  \end{tabular}
  \caption[]{
The top panel shows a map of the various classes of axisymmetric modes in
the $(\op,N^2)$ parameter space, for a a shell aspect ratio $\eta=0.35$.
  The yellow domain is the H-mode domain (corresponding to modes
  propagating in the whole shell), whereas the white areas show
  the HT-mode domains (corresponding to modes propagating in only a
  part of the shell, and bounded by turning surfaces). The black area
  corresponds to configurations where no modes exist.  Above the solid
  grey curve, modes are potentially destabilized by the ABCD instability
  for $\mathcal{P}=10^{-5}$.  The dashed purple curves correspond to the apparition 
  of a turning surface at $r=\eta, r=1$ and $\sin^2\theta = 0, 1$ and
  delimit the various HT-mode propagation regions.  For each subdomain,
  a letter indicates the geometry shown in the bottom subfigures, where
  the domain in which waves may propagate is in white and the evanescent
  region is in black. The red line is the $\Delta = 0$ turning surface
  on which characteristics bounce.}
  \label{fig:ax}
\end{center}
\end{figure}

Figure~\ref{fig:ax} shows the various regions of the parameter space,
for axisymmetric modes ($m=0$), for an aspect ratio of the shell
$\eta=0.35$\footnote[2]{$\eta=0.35$ is the aspect ratio of the liquid core
of the Earth, but this is also the aspect ratio of the radiative region
of a young massive star of about 40 solar masses.}.  The yellow domain
contains all H modes. There are no gravito-inertial modes in the black
area (the elliptic domain covers the whole shell). The white area is for
HT modes. We notice that frequency domains accessible to HT modes increase 
with $N^2$ and the subsequent differential rotation. As there are 
regions of the shell rotating faster than the average velocity, gravito-inertial
modes can exist with a higher frequency than in the solid-body rotation case.
When $N^2$ is increased, the H-mode domain gets smaller, and
there are no H modes at $N^2> 2$.

The various propagation domains for HT modes are shown in the miniatures.
The purple dotted lines mark the transition between the different
geometries of the modes, and are obtained by setting $m=0$ in equations
(\ref{eq:transition1})-(\ref{eq:transition4}).

We therefore find eight possible HT modes geometries with distinct
propagation properties for the characteristics. Remarkably,

\begin{itemize}
  \item geometries (a) and (b) are similar to the so-called H2 modes of
    \cite{DRV99}, whereas geometries (c) and (g) are somewhat reminiscent of
    their E2 modes, 
  \item geometries (d), (e) and (f) feature two turning surfaces in the shell, 
  \item geometries (a), (d) and (f) feature an acute angle between
    a turning surface and a boundary of the shell that may lead to the so-called
    wedge trapping of the modes \cite[][]{DRV99, Gerk08}.
\end{itemize}

Domain (h) is not studied any further due to its small extent in
the parameter space, and due to the small extent of the corresponding
propagation domain in the shell.  Note that, while the borders between the
HT-mode geometries generally depend on the shell's aspect ratio $\eta$,
the boundary of the H-mode region does not.

In Fig.~\ref{fig:ax}, we have delimited the region of parameter space
(located above the grey solid curve), where the waves may be unstable
according to criterion (\ref{eq:rayleigh}).  However, since this criterion
is local, this condition is not sufficient.

Finally, we deduce that the frequency range accessible to axisymmetric
modes is $[0, 2\Omega(\eta)]$, which may be as wide as $[0 ,
\frac{1}{\Omega_s}\left(2\Omega_s^2+N^2\right)]$ in our model for the
full sphere $\eta=0$.  This range is significantly larger than the
predicted range for gravito-inertial modes in a uniformly-rotating stratified sphere, 
that is $[0,\sqrt{4\Omega_s^2 + N^2}]$ \cite[][]{FS82a,DRV99}. Note that our
normalization differs from the one used by \cite{FS82a} and \cite{DRV99}
in solid-body rotation: the dimensionless numbers $E$ and $N^2$ differ by
a factor of 2, and their $\Omega$ is constant throughout the shell.

\subsubsection{Non-axisymmetric modes}
\begin{figure}
\begin{center}
   \includegraphics[angle=-90,width=\textwidth]{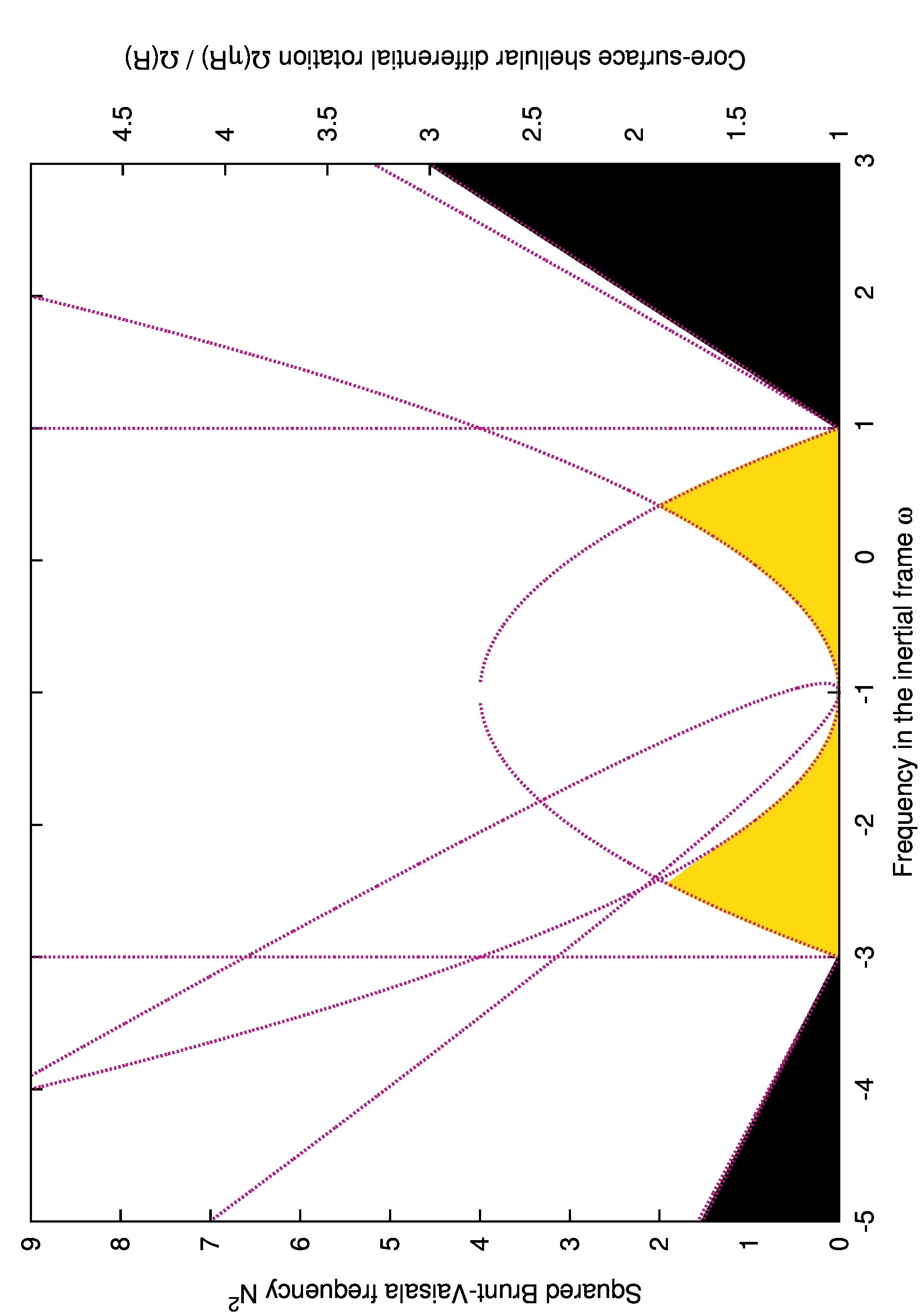}\\
     \includegraphics[angle=-90,width=0.55\textwidth]{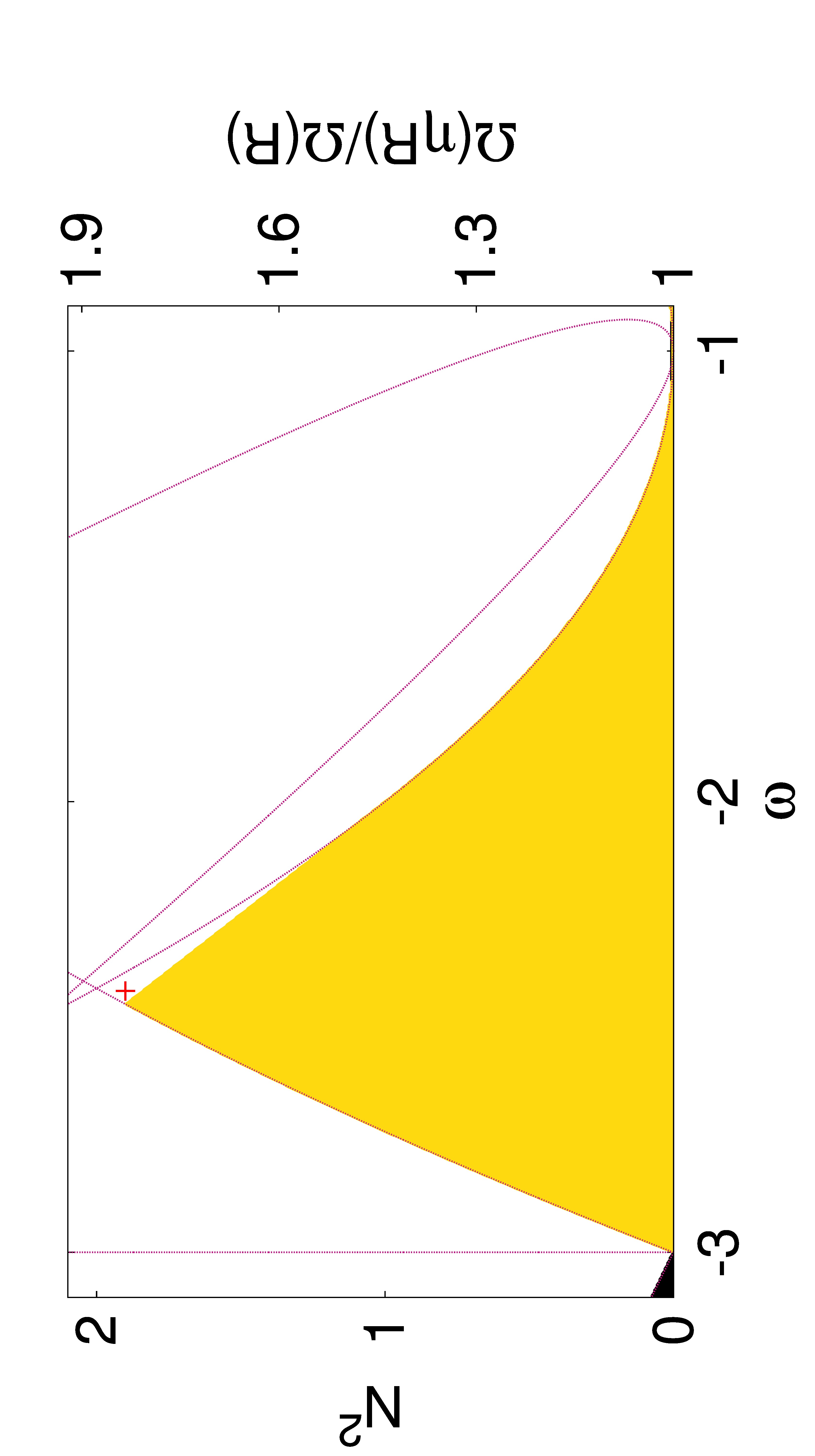}  
     \hfill \includegraphics[angle=-90,width=0.4\textwidth]{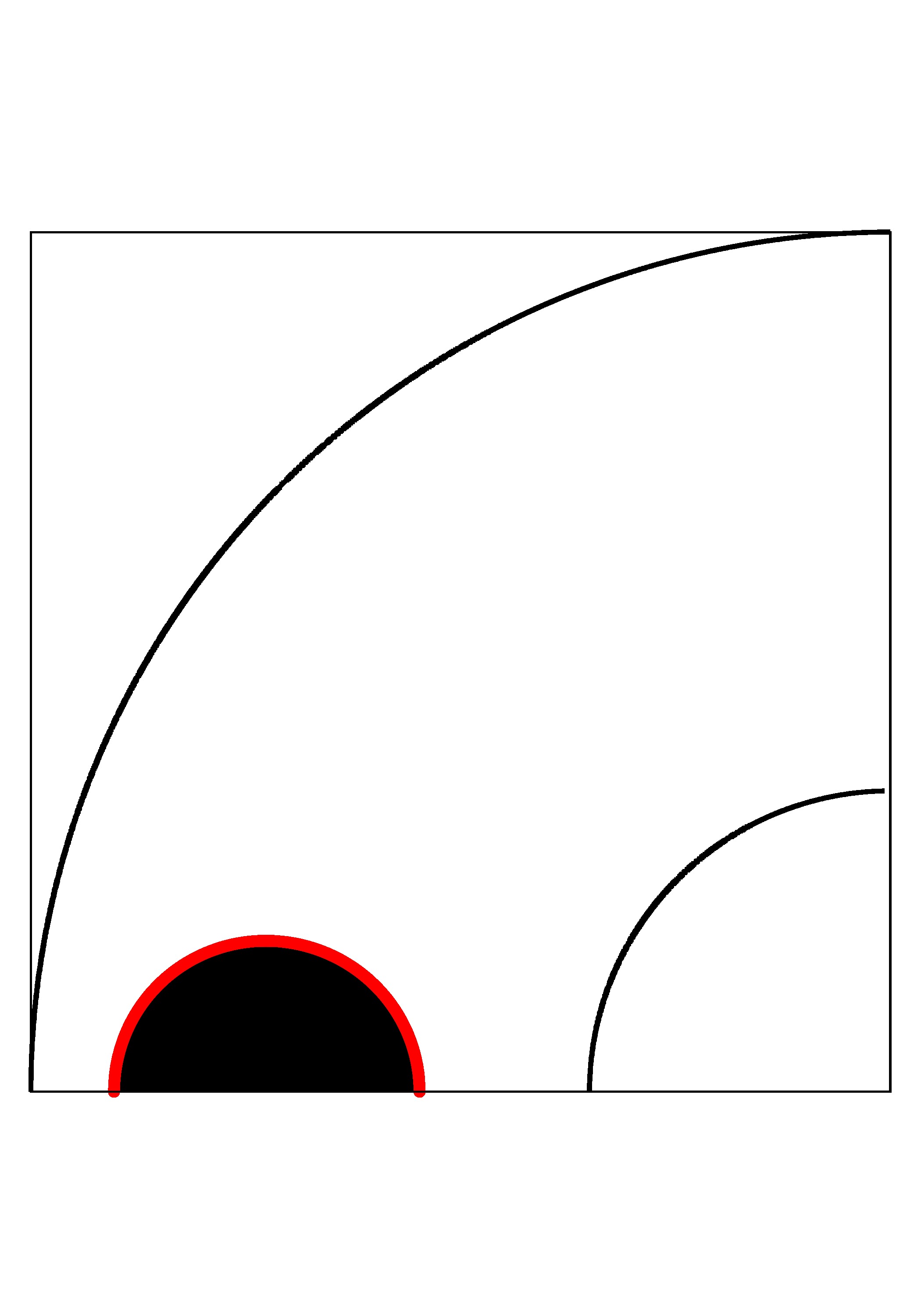}
  \caption[]{\small
    Top panel: same as Figure~\ref{fig:ax} but for $m=1$.
  Bottom left: extent of the H-mode domain (yellow) for $m=1$ retrograde 
  modes. The purple curves mark the expected transition between H and HT 
  modes (white). Bottom right: elliptic and hyperbolic domains across the 
shell for a mode at $m=1, N^2=1.9, \op=-2.42$, indicated by a red tick on the left plot.}
  \label{fig:DTfromaxis}
\end{center}
\end{figure}

For non-axisymmetric modes with positive $m$, the accessible frequencies
are grouped around $\op = -m$. The range is no longer symmetrical:
the accessible frequency domain extends to frequencies below $-m$, as H modes
progressively become HT modes as $m$ increases. 
This can be seen by comparing 
the top panels of figures~\ref{fig:DTfromaxis} and~\ref{fig:corot}.
From $m$ larger than two, the H-mode domain for $\op > -m$ also becomes smaller as $m$ increases. 
We find that the frequency range accessible to HT modes with $m\geq2$ is 
$[-(m+2)\Omega(\eta):-m\Omega_s+\max(2\Omega_s, N)]$.
As $\max(2\Omega_s, N)$ does not depend on $m$, we note that the extent of the 
parameter space for prograde modes is constant while the parameter
space for retrograde modes increases with $m$.

\subsubsection{Other turning surfaces}

The foregoing results show that the turning surfaces are not simple,
leading to numerous configurations for the hyperbolic domain. 
As warned above our method for solving equation (\ref{eq:delta_sint}) 
captures only part of the whole set of solutions.
Some turning surfaces cannot be calculated from
the set of equations (\ref{eq:transition1})-(\ref{eq:transition4}).
For our rotation profile, we note that elliptic domains can show up
around part of the rotation axis with the pole and the core still in
the hyperbolic domain.  This is illustrated in the lower panels of
figure~\ref{fig:DTfromaxis}, which show the distribution of H and HT
modes for $m=1$ retrograde modes at low stratification.  The small
white area enclosed between the H-mode domain (yellow area) and its
expected boundaries (purple curves) actually corresponds to HT modes.
For these specific HT modes, the turning surface does not encompass
one of the ``corners'' of the shell but appears on the rotation axis.
The bottom right panel of figure~\ref{fig:DTfromaxis} illustrates this 
case. Even though only a small part
of the parameter space is concerned, it indicates that the boundaries
between H and HT modes may not be determined analytically for more
complex rotation profiles.

\subsubsection{Corotation resonances}
\label{sec:corot}

For some non-axisymmetric modes, corotation resonances appear where
Doppler-shifted frequency vanishes, i.e. where 
$\opt = \op + m \Omega(r) = 0$. This resonance is also called a
critical layer.

From equation (\ref{eq:omega}), the domain of the parameter space $(\op,
N^2)$ where a corotation resonance exists in the shell is such that
\begin{equation}
  - m \left( 1 + \frac{N^2}{2} \left( 1 - \eta^2 \right) \right) \leq \op \leq -m.
  \label{eq:corot}
\end{equation}
We notice from (\ref{eq:corot}) that the range of frequencies 
at which corotation resonances exist gets larger as $m$ increases.  
It also increases with $N^2$ at a given $m$.

\begin{figure}
\begin{center}
   \includegraphics[angle=-90,width=\textwidth]{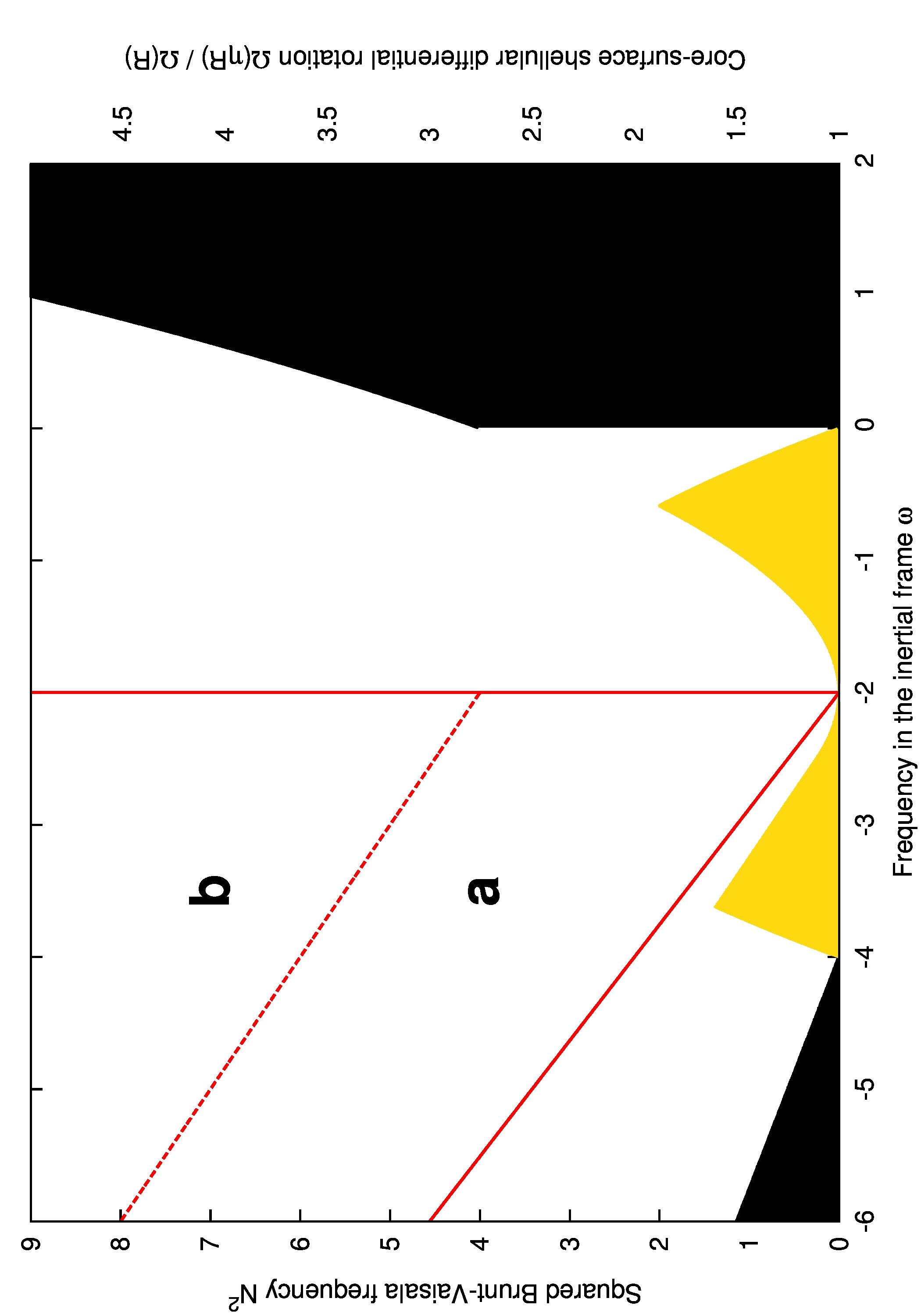}\\
   \centerline{\includegraphics[width=0.3\textwidth,angle=-90]{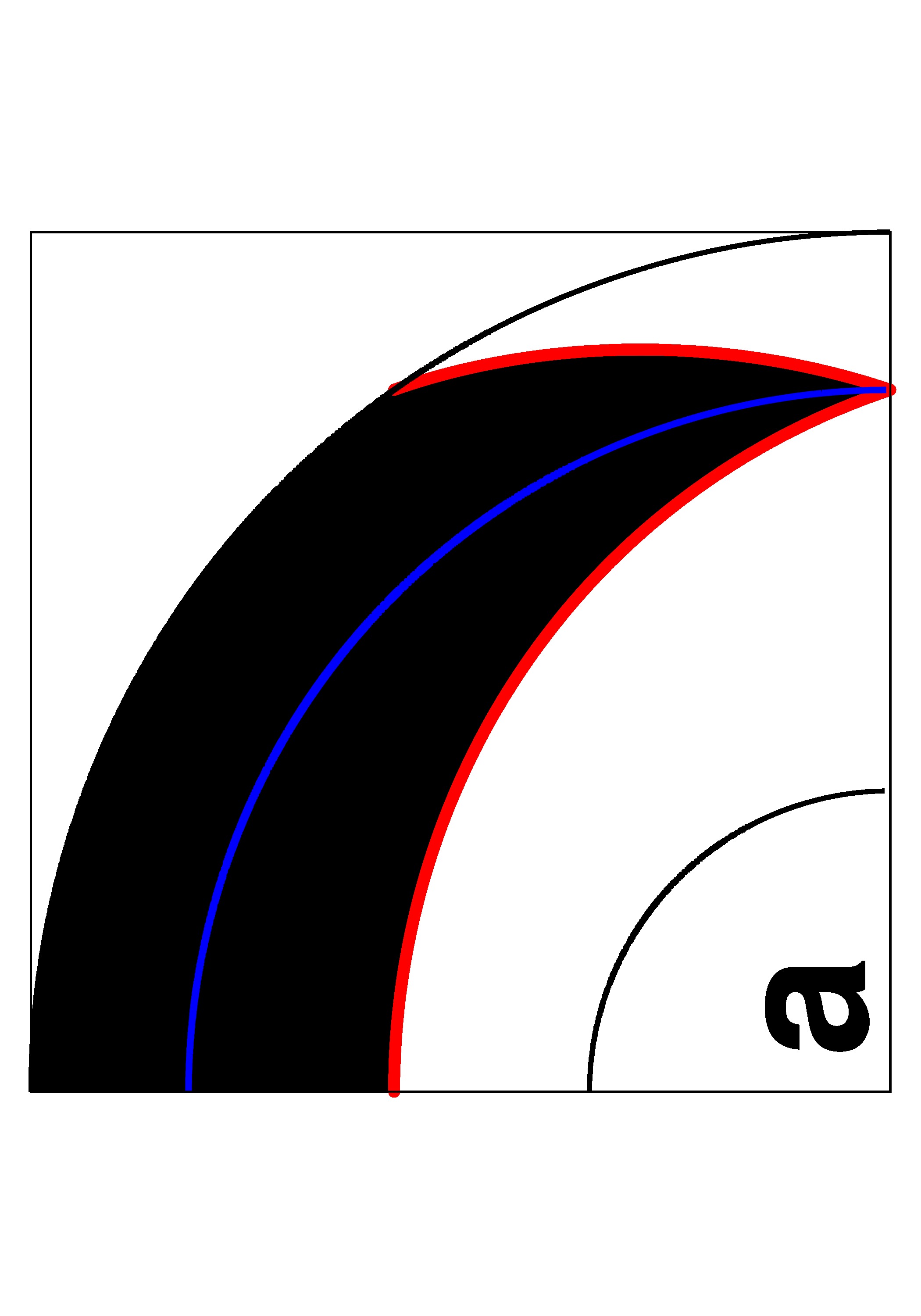}
     \hspace{0.2\textwidth}\includegraphics[width=0.3\textwidth,angle=-90]{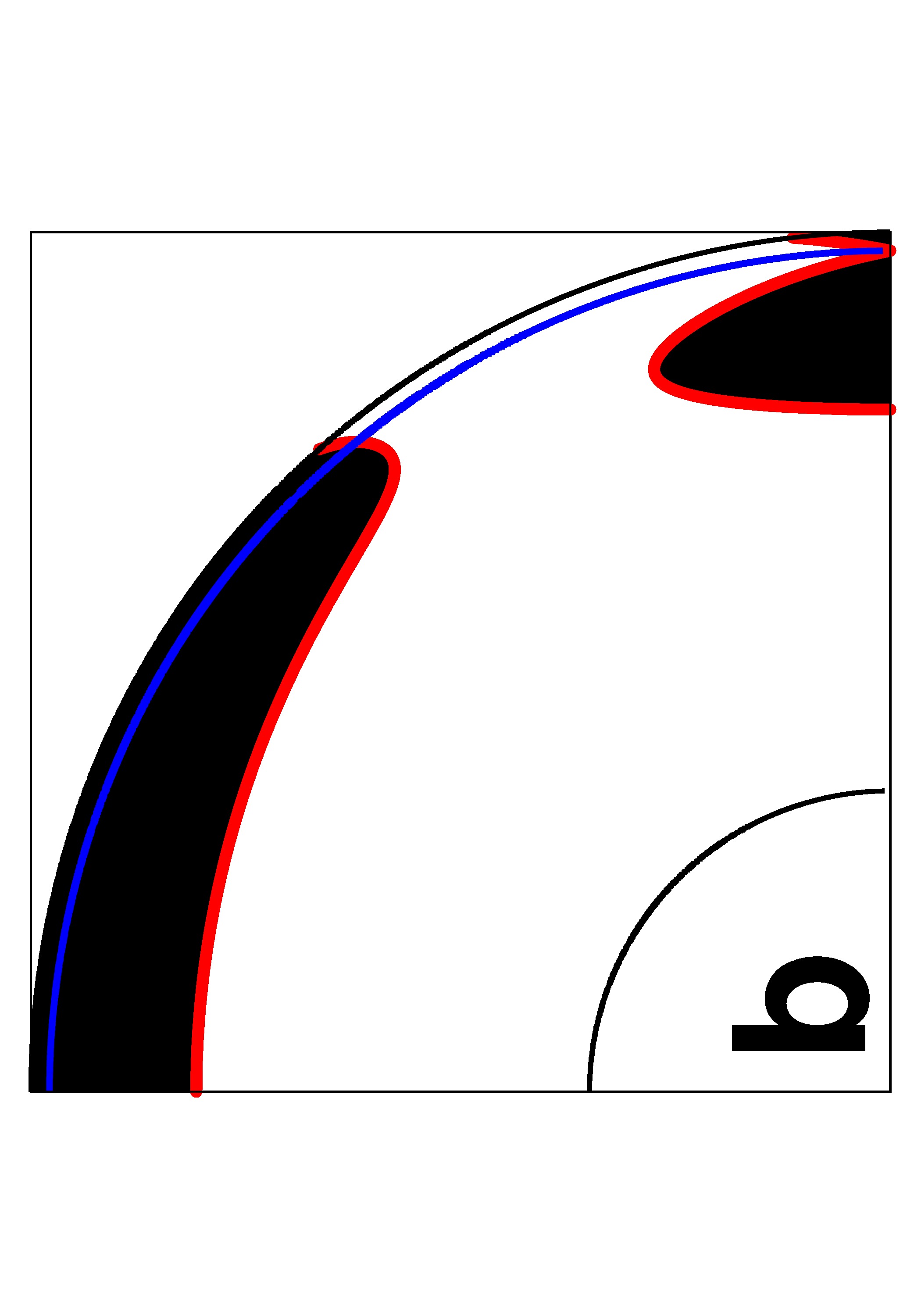}}  \caption[]{\small
  The top graph shows the occurrence of corotation resonances as a function of
  $\op$ (x-axis) and $N^2$ (y-axis), for a shell of aspect ratio $\eta=0.35$
  and azimuthal wavenumber $m=2$.  The yellow domain is the H-mode domain,
  whereas the white area is the HT-mode domain.  The black area corresponds to
  configurations where no modes exist.  Corotation resonances appear in the
  star for configurations between the two red solid lines.  The dashed line
  separates modes where the critical layer is wholly included in the elliptic
  part of the shell (subdomain a) and modes where the critical layer cuts
  through the hyperbolical domain (subdomain b).  The subfigures, labelled
  according to the subdomain they belong to, show how the corotation resonance
  (marked by the solid blue line) may cross the hyperbolic domain (white area).
  The black area is the elliptic domain and the red lines the turning surfaces.}
  \label{fig:corot}
\end{center}
\end{figure}

For configurations where corotation resonances exist, the corotation radius 
$r_c$ is given by
\begin{equation}
  r_c = \sqrt{1 + \frac{2}{N^2}\left(1 + \frac{\op}{m} \right)}.
  \label{eq:rc}
\end{equation}
We need to determine whether the corotation radius intersects the hyperbolic domain, 
that is, if $\Delta(r_c) > 0$.  Setting $r=r_c$ and $\opt =0$ in
equation (\ref{eq:delta}), we find that $\Delta(r_c)$ is positive if and only if
$N r_c > 2$, that is, using equation~(\ref{eq:omega}), if
\begin{equation}
  \label{eq:hypercorot}
  \op > m \left(1 - \frac{N^2}{2} \right).
\end{equation}
Figure~\ref{fig:corot} shows the parameter range in the $(\op, N^2)$
plane where corotation resonances exist in the shell for $m=2$. The domain
where critical layers exist in the shell is enclosed between the two solid
lines. We note that corotation resonances only exist for HT modes, as $\opt=0$
implies $\Delta = 0$ at the equator in equation (\ref{eq:delta_sint}).  We find
that the parameter domain for modes exhibiting corotation resonances gets
larger when $m$ increases, and most of the modes in the $\op<-m$ domain exhibit
corotation resonances.\\ We distinguish two possibilities, depending on whether
equation (\ref{eq:hypercorot}) is satisfied or not.  If $\op$ satisfies
(\ref{eq:corot}) but not (\ref{eq:hypercorot}), the corotation radius is inside
the elliptic domain, in which modes do not propagate, and the interaction between 
the corotation resonance and the modes is likely
small. Since $\opt=0$ and $z=0$ in (\ref{eq:delta}) implies $\Delta = 0$, we
find that the turning surface and the corotation radius intersect each other in
the equatorial plane only. In the top panel of figure~\ref{fig:corot}, this
happens in the subdomain (a) below the dashed line. A meridional cut of a mode
in the subdomain (a) is shown in the subfigure~\ref{fig:corot}a, where the
critical layer is denoted by a blue line.

Conversely, if (\ref{eq:corot}) and
(\ref{eq:hypercorot}) are both satisfied, the corotation resonance goes through
the hyperbolic domain and is expected to influence the propagation of the waves
and the associated characteristics either by introducing non-linear effects or
by limiting the propagation domain. This corresponds to the subdomain (b) in
the top panel of figure~\ref{fig:corot}, from the dashed line upwards. The
corresponding subfigure~\ref{fig:corot}b shows an example of such a mode where
the critical layer (blue line) crosses the hyperbolical domain (white area).
For this case, the calculation leading to (\ref{eq:hypercorot}) also shows that
the intersection between the hyperbolic domain and the corotation surface is
located at latitudes $\vartheta$ satisfying
\begin{equation}
  \sin \vartheta < \frac{2}{N r_c}.
\end{equation}
As we expect the critical layer to have an effect on the damping or
excitation of the mode \cite[e.g.][]{RTZL12}, the ability of the wave to
cross this layer needs to be investigated.  This is done by computing
the phase and group velocities (see below), and by solving
the fully-dissipative problem (section~\ref{sec:diss}).

\subsection{Dispersion relation, phase and group velocities}
\label{sec:disp}

Further insight into the propagation properties of gravito-inertial waves over
a shellular differential rotation may be obtained through the dispersion
relation of the waves.

To determine the interplay between paths of characteristics and corotation
resonances, we derive from equation (\ref{eq:poincare}) the following dispersion
relation
\begin{equation}
  \opt^2 = \frac{\mathcal{B}^2}{k^2} \ {\rm with}\ \mathcal{B} = \left[ N^2 \left(k_s z - k_z s\right)^2 + k_z^2 A_s - k_s k_z A_z\right]^{1/2},
  \label{eq:disp}
\end{equation}
where $\bb{k} = k_s \bb{e}_s + k_z \bb{e}_z$ is the wavevector, $k = ||\bb{k}|| = \sqrt{k_s^2 + k_z^2}$.
Equation (\ref{eq:disp}) reduces to (A.7) of \cite{BR13} in the case with no
stratification $(N^2=0)$.

This equation yields the phase and group velocities in the corotating frame,
$\vv_p$ and $\vv_g$ respectively,
\begin{equation}
  \vv_p = \pm \frac{\mathcal{B} \bb{k}}{k^3},
\end{equation}
and
\begin{equation}
  \vv_g = \pm \frac{1}{k^3}\left[ \left( k^2 \frac{d \mathcal{B}}{d k_s} - k_s \mathcal{B} \right)  \bb{e_s} +
                                         \left(k^2 \frac{d \mathcal{B}}{d k_z} - k_z \mathcal{B} \right)  \bb{e_z} \right].
\end{equation}

The dispersion relation indicates the behaviour of a wave near the
corotation radius.  Upon approaching the corotation resonance, $\opt
\rightarrow 0$ which implies that either $k\rightarrow \infty$ or $\mathcal{B}
\rightarrow 0$.

\begin{itemize}
\item If $k\rightarrow \infty$ at finite $k_z$, then $\vv_p = 0$ and
$\vv_g = 0$, the gravito-inertial waves do not cross the corotation radius. If
  corotation is in the elliptic domain but touches the turning surface
  on the equator $z=0$, then the characteristics become more and more
  vertical (namely parallel to the rotation axis) as they approach the
  corotation, but the corotation is never reached. The wave is likely
  dissipated there.

\item If $\mathcal{B} \rightarrow 0$, then $\vv_p \rightarrow 0$ and
$\vv_g \ne 0$ in the general case.
  In this situation, the waves can go through the corotation resonance
  in the hyperbolic domain, but non-linear effects are expected
  \cite[][]{BO2010}.
\end{itemize}

We only see the occurrence of this second possibility in our simulations
(section~\ref{sec:diss}):
as shown in figure~\ref{fig:hypercorot}, the eigenmodes we compute are
able to cross the corotation radius.

\subsection{Critical latitudes}

Critical latitudes are another type of singularities that may affect
waves in a spherical shell~\cite[e.g.][]{RGV01}.  As all singularities,
they are important to understand the dynamics of tidally interacting
bodies, as they generate dissipative structures \cite[][]{GL09, O09,
RV10, FBBO14}.

A critical latitude is a latitude $\vartheta$ on the bounding spheres where the
characteristics are tangent to the boundary. They appear when $|\opt(r)|
\le 2\Omega(r)$, that is
\begin{equation}
\sin\vartheta = \frac{\opt(r)}{2\Omega(r)},\quad  {\rm \ with\ }\quad r= \eta \quad {\rm\ or\ }\quad r=1.
\end{equation}
The inner and outer critical latitudes, $\theta_i$ and $\theta_o$, are given by
\begin{equation}
  \sin\vartheta_o = \frac{\op+m}{2}, \qquad \sin\vartheta_i = \frac{\op}{2\Omega(\eta)}+ \frac{m}{2}.
\end{equation}
There is a critical latitude on the inner boundary if $0 < \sin\vartheta_i < 1$, that is when
\begin{equation}
  -m \Omega(\eta) < \op < (-m+2) \Omega(\eta),
\end{equation}
and similarly, there is a critical latitude on the outer boundary if
\begin{equation}
  -m < \op < -m +2.
\end{equation}

In the frequency ranges where critical latitudes exist, some modes may be
associated with shear layers emitted at these latitudes \cite[][]{RV10}.  While
it is not necessarily the case for free modes of oscillations, tidally-forced
flows in a spherical shell seem to always excite the shear layers associated with the 
inner critical latitudes when they exist \cite[as shown by][]{O09, GL09}.
For other geometries, concave critical latitudes may be excited 
\cite[see, e.g.][]{Swart10}.

\subsection{Lyapunov exponents}

The paths of characteristics computed using equations
(\ref{eq:dzds})-(\ref{eq:dsdz}) often tend towards a
short-period limit cycle, known as an attractor. When such a
structure exists, the kinetic energy of an eigenmode is generally
concentrated around the attractor.  These modes are expected to be
strongly damped due to the singular nature of the attractor at
vanishing viscosities. It is therefore of interest to assess the 
presence and the strength of the focusing towards attractors, and 
for this we compute the Lyapunov exponent of the characteristic
trajectories. It quantifies whether characteristics get closer to
each other after multiple reflections onto the boundaries of the
hyperbolic domain. Since our problem can be studied in a
meridional quarter-plane of the shell, we can conveniently use the
rebounds on the rotational and equatorial axes. From the
distance between two consecutive rebounds on the rotation axis, or on the
equatorial plane, $\delta s_k$ or $\delta z_k$ respectively, we derive the
two following Lyapunov exponents, $\Lambda_s$ or $\Lambda_z$ respectively:
\begin{equation}
\Lambda_s = \lim_{N_s\rightarrow \infty} \frac{1}{N_s}\sum\limits_{k=1}^{N_s} \ln \left| \frac{\delta s_{k+1}}{\delta s_k} \right| \quad {\rm and} \quad
  \Lambda_z = \lim_{N_z\rightarrow \infty} \frac{1}{N_z}\sum\limits_{k=1}^{N_z} \ln \left| \frac{\delta z_{k+1}}{\delta z_k} \right|.
  \label{eq:lambda}
\end{equation}
There are some instances where an attractor may feature more than
one rebound on a given axis where the Lyapunov exponent is computed.
When that happens, our numerical procedure finds the coordinate of each
convergence point and computes the exponent selecting the rebounds
that tend towards that convergence point.

Formally, Lyapunov exponents on both axes should
have the same value.  We compute the Lyapunov exponents with several pairs
of characteristics starting in various parts of the shell.  Since this
calculation is sometimes difficult numerically, we use
the average value of $\Lambda_s$ and $\Lambda_z$ to derive the Lyapunov
exponent.

A negative Lyapunov exponent means that
the two characteristics get closer to each other, and end up converging towards 
an attractor. The more negative the exponent, the faster the convergence.

\begin{figure}
  \centerline{ \includegraphics[width=0.7\textwidth]{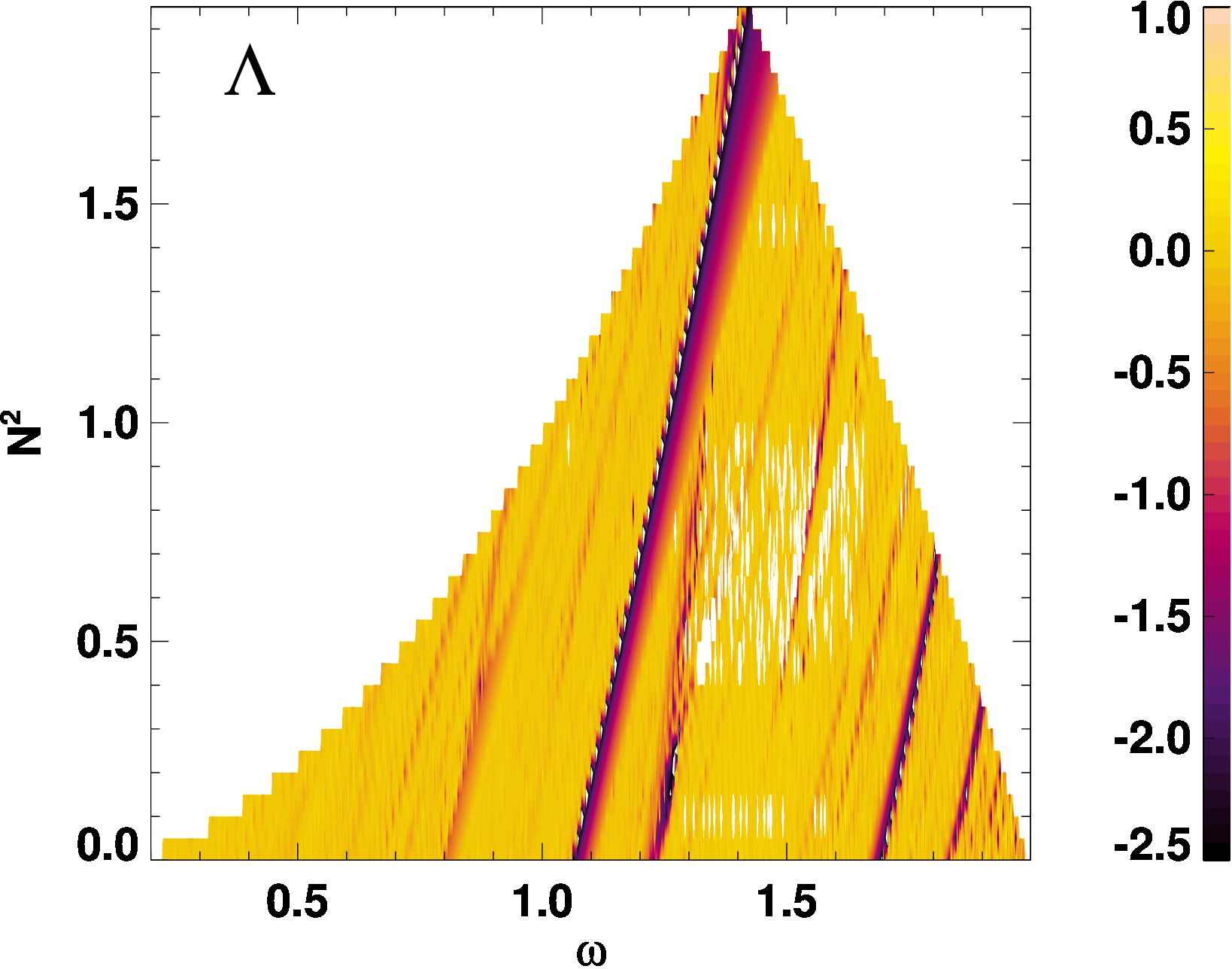}}
  \caption{
Lyapunov exponent for H modes for $\eta=0.35$. The dark ridges are modes
focused on a short-period attractor.
  \label{fig:Lyap_D}}
\end{figure}

In figure \ref{fig:Lyap_D}, we show the Lyapunov exponent in the
$(N^2,\op)$ plane, for axisymmetric H modes. We see ridges where the Lyapunov
exponent is very negative, which implies the presence of a strong
attractor in the shell. Each ridge of negative
Lyapunov exponent consists of the same attractor progressively distorted 
by the
differential rotation: as $N^2$ and the subsequent differential rotation
increase, the frequency varies but the overall shape of the attractor
is conserved, as will be illustrated in figure~\ref{fig:suiviN2}. 
We note that the
strength of the focussing for a given attractor is constant: the Lyapunov
exponent is constant along a ridge of negative value. These modes are
expected to be strongly damped by dissipative processes.  However,
most of the $(N^2,\op)$ plane corresponds to long-period attractors
compatible with low dissipation, and these modes are therefore more
prone to yield observable stellar pulsations.  We note that by taking a
horizontal slice in figure~\ref{fig:Lyap_D} at $N^2=0$, we retrieve the
Lyapunov exponents computed for inertial modes in a solid-body rotating
shell by \cite{RGV00}, \cite{RGV01}.

For non-negative values of the Lyapunov exponent, there is no convergence
towards a limit cycle. Because the sums in equation (\ref{eq:lambda}) could not
necessarily be carried out with as many iterations as formally required,
they do not cancel out.  For some values (white regions inside the graph),
the computation has not converged, but the Lyapunov exponents are likely
very small in absolute value.

As will be shown in section \ref{sec:diss}, the kinetic energy of
eigenmodes is not necessarily distributed along the predicted attractor: other
kinds of singularities like the critical latitude singularity or
corotation resonances may also affect the kinetic energy
distribution.
Due to turning surfaces, computing the Lyapunov exponents of HT modes
(whether they are focused on a short-period attractor or in a wedge)
is much more difficult, and no significant map could be computed.

\section{Dissipative problem}
\label{sec:diss}
After the foregoing study of the properties of the eigenmodes
in the non-dissipative limit, we now investigate the role of
dissipative processes on the structure and damping (or growth) rates of
gravito-inertial modes.  In this section we describe the numerical method
we used, show representative examples of axisymmetric and non-axisymmetric
modes, and discuss the physical implication of the results.

\subsection{Numerical method}
\label{sec:numerics}

In order to study the full dissipative eigenvalue
problem, that consists of equations (\ref{eq:ns_a}) to (\ref{eq:heat_a})
with related boundary conditions, we now solve the problem numerically
using a spectral method similar to that used in \cite{BR13}.  Equations
are projected onto spherical harmonics \cite[see][]{R87}, expanding the
velocity and temperature perturbations as
\begin{equation}
  \vv(r,\theta,\phi) = \sum\limits_{\ell=0}^\infty \sum\limits_{m=-\ell}^\ell \ulm(r) \rlm + \vlm(r) \slm + \wlm (r) \tlm, 
   \quad T= \sum\limits_{\ell=0}^\infty \sum\limits_{m=-\ell}^\ell t^\ell_m \ylm,
\end{equation}
where $\rlm = \ylm(\theta,\phi) \bb{e_r}, \slm = \del \ylm(\theta,\phi),
\tlm =\del\times \ylm(\theta,\phi)$ and where $\ylm$ denotes
the normalized spherical harmonics.  In the radial direction,
equations are discretised on the Gauss-Lobatto grid associated with
Chebyshev polynomials.  The equations are truncated at order $L$ in the
spherical harmonics expansion and at order $N_r$ on the Chebyshev grid.
Appropriate values of $N_r,L$ depend on the mode properties and the
Ekman and Prandtl numbers.  We use the Arnoldi-Chebyshev algorithm to
compute eigenvalues $\op$ and their associated eigenvectors.  This method
allows us to find the frequencies corresponding to the least-damped modes
around a given initial guess.  By changing slightly the guess, we can
test for round-off errors \cite[for more details on the numerical method, 
see ][]{VRBF07}.  All modes are computed assuming symmetry with respect
to the equatorial plane, and using stress-free boundary conditions at
both the inner and outer radial boundaries of the shell.

Projecting the
momentum equation (\ref{eq:ns_a}) on $\rlm$ and $\tlm$
and the heat equation (\ref{eq:heat_a}) on $\ylm$ we get
\begin{eqnarray}
  \label{eq:hs2}
  E\Delta_\ell\wlm &+& \left[\frac{2im\Omega}{\ell(\ell+1)}- i\opt\right]\wlm \nonumber\\
                         &=& -2\Omega\left[A(\ell,m) r^{\ell-1} \prt_r\left(r^{-\ell+2} u_m^{\ell-1} \right) 
                              + A(\ell+1,m) r^{-\ell-2}\prt_r\left(r^{\ell+3}u_m^{\ell+1} \right) \right]\nonumber\\
                         &+& r\prt_r\Omega \left[-(\ell+1) A(\ell+1,m) u_m^{\ell+1} + \ell A(\ell,m) u_m^{\ell-1} \right],
\end{eqnarray}

\begin{eqnarray}
  \label{eq:hs3}
  E \Delta_\ell \Delta_\ell (r\ulm) &+&  \left[\frac{2im\Omega}{\ell(\ell+1)}- i\opt\right] \Delta_\ell(r\ulm)= - 2im\prt_r\Omega \left(\ulm+ \vlm\right) - imr\prt^2_{rr}\Omega \ulm \nonumber\\
                                          &+& 2\Omega\left[B(\ell,m) r^{\ell-1} \prt_r\left(r^{1-\ell} w_m^{\ell-1} \right) 
                                              + B(\ell+1,m) r^{-\ell-2} \prt_r(r^{\ell+2} w_m^{\ell+1}) \right]\nonumber\\
                                          &+& 2\prt_r\Omega\left[B(\ell,m) w_m^{\ell-1} +  B(\ell+1,m) w_m^{\ell+1} \right] -N^2\ell(\ell+1) \thlm,
\end{eqnarray}

\begin{equation}
  \label{eq:hs1}
  i\opt \thlm - r\ulm = \frac{E}{\mathcal{P}} \Delta_\ell \thlm, 
\end{equation}

where we defined
\begin{eqnarray}
  \Delta_\ell = \ddnr{} + \frac{2}{r}\dnr{} &-& \frac{\ell(\ell+1)}{r^2} ,\\ 
  A(\ell,m) = \frac{1}{\ell^2} \left( \frac{\ell^2 - m^2}{4\ell^2-1} \right)^{1/2},&\ &
  B(\ell,m) = \ell^2(\ell^2-1) A(\ell,m).
\end{eqnarray}
In equations (\ref{eq:hs2}) -- (\ref{eq:hs3}), we used mass conservation
(\ref{eq:masscons_a}) which yields
\begin{equation}
  \label{eq:subs}
  \vlm = \frac{1}{\ell(\ell+1)r} \frac{\prt r^2 \ulm}{\prt r}.
\end{equation}

Setting to zero the stratification and the subsequent differential
rotation, these equations reduce to equations (2.2) of \cite{RV97} for
inertial modes in solid-body rotation. Removing only stratification,
these equations reduce to equations (4.2) of \cite{BR13}.  Finally,
taking only stratification into account without differential rotation,
this set of equations reduces to equations (2.5) of \cite{DRV99} for
gravito-inertial modes.

\subsection{Axisymmetric modes: illustrative cases}
\label{sec:axisym}
Let us first describe a few axisymmetric ($m=0$) modes obtained
for various stratifications and associated differential rotations.
We illustrate the various geometries for H and HT modes by comparing the
solutions to the fully dissipative equations with their non-dissipative
counterparts.  For all the modes that we have calculated, we show a
meridional slice of kinetic energy, normalized at its maximum value,
in a quarter-plane. We recall that all computed modes are symmetric with
respect to the equatorial plane.

\begin{figure}
\includegraphics[width=0.5\textwidth]{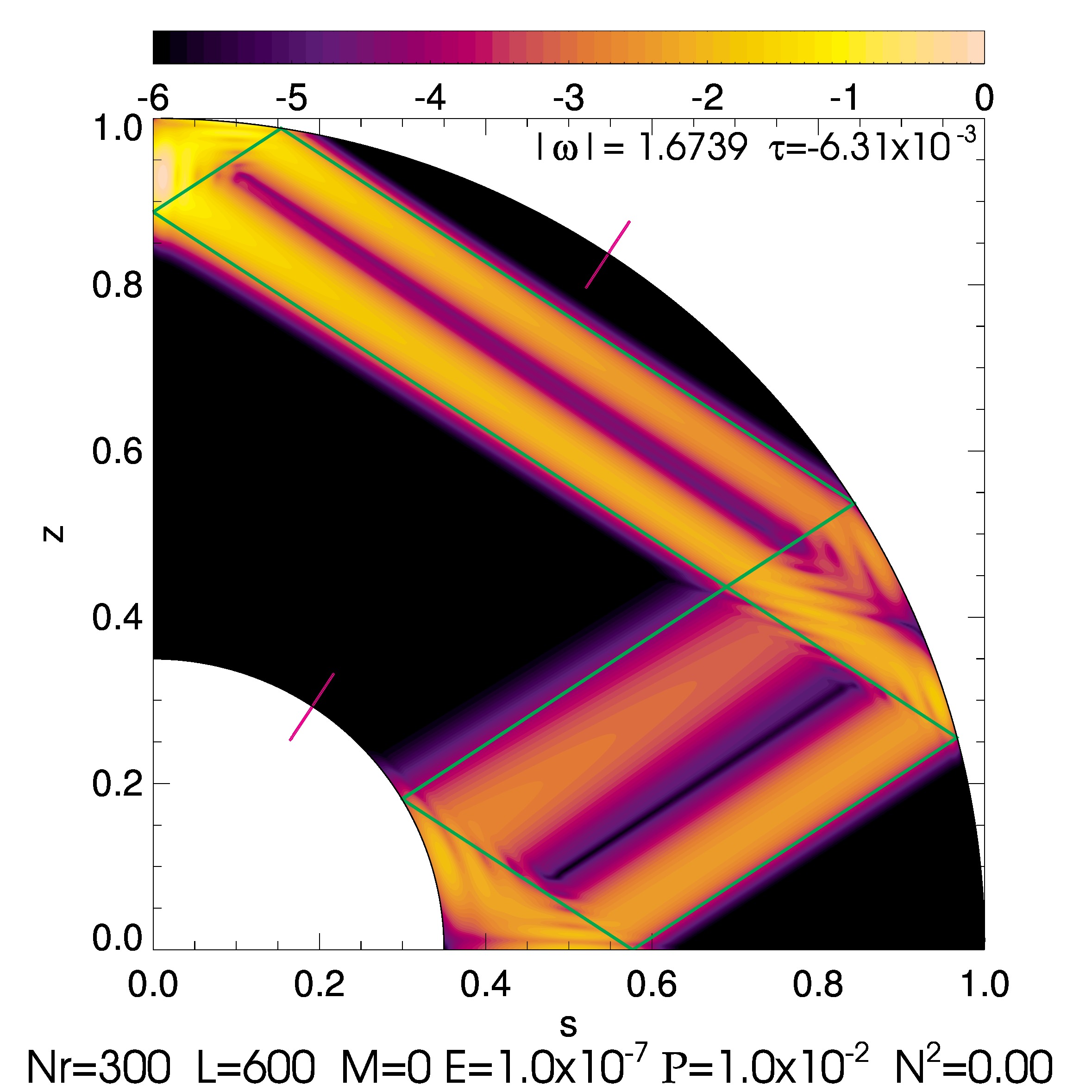}
\includegraphics[width=0.5\textwidth]{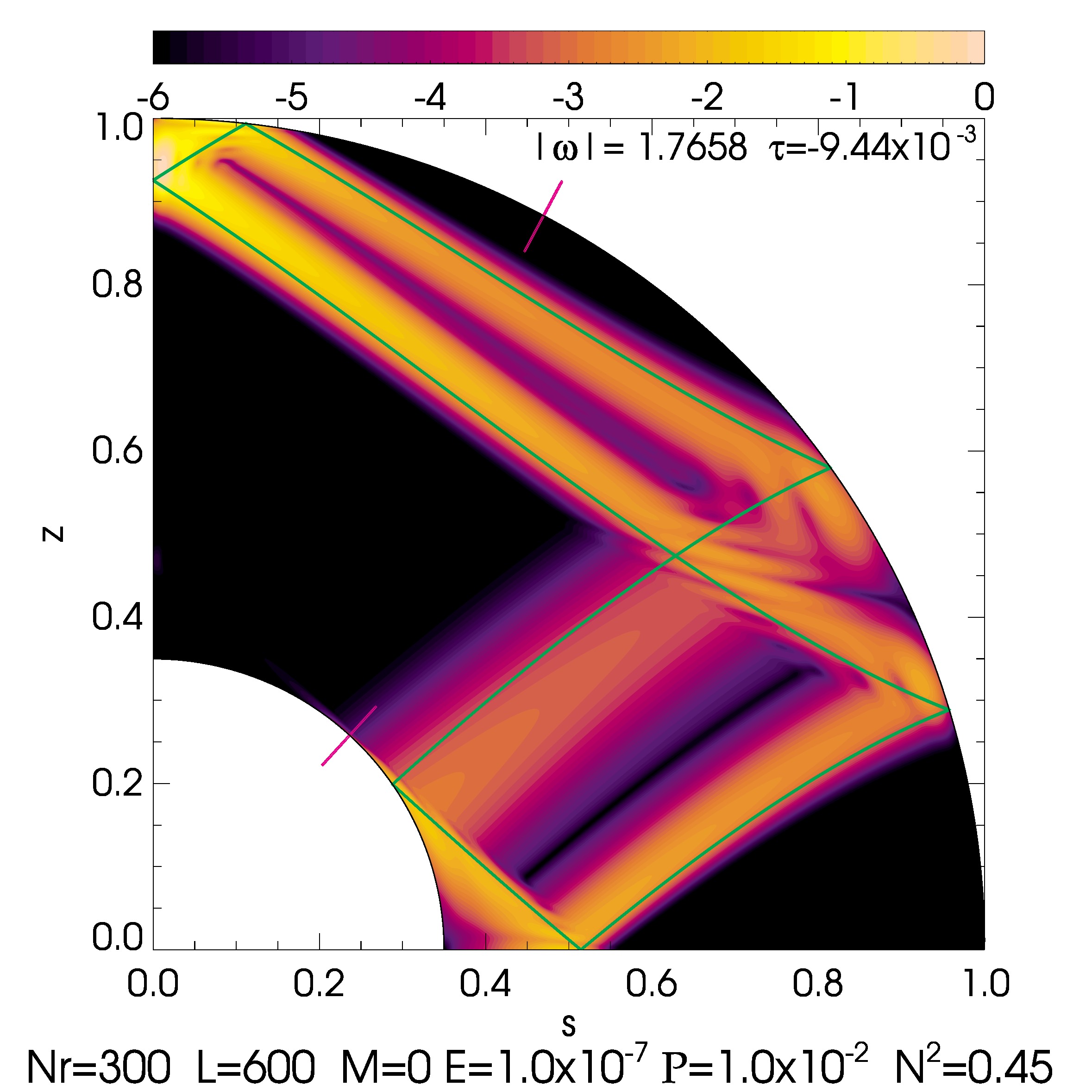}\\
\includegraphics[width=0.5\textwidth]{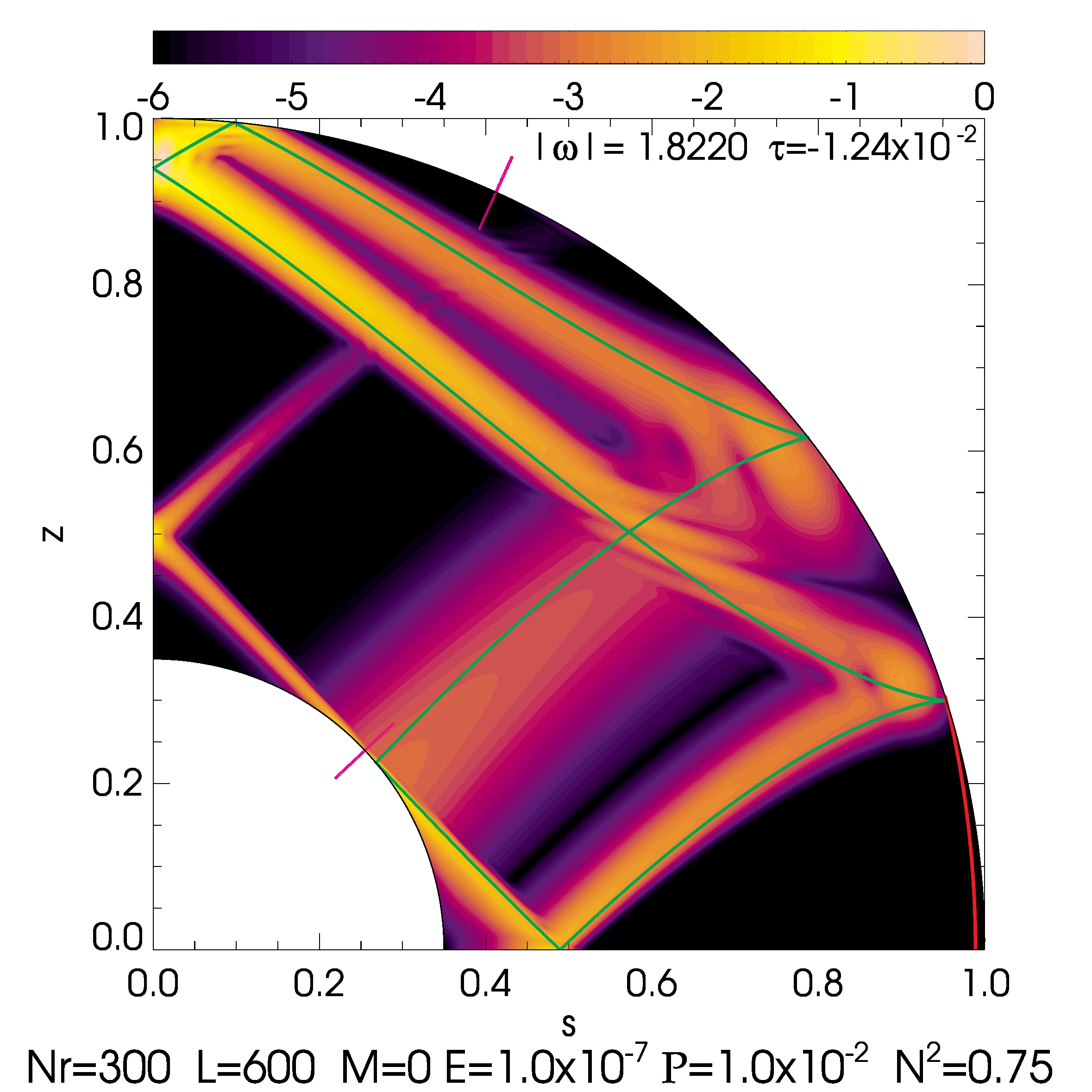}
\includegraphics[width=0.5\textwidth]{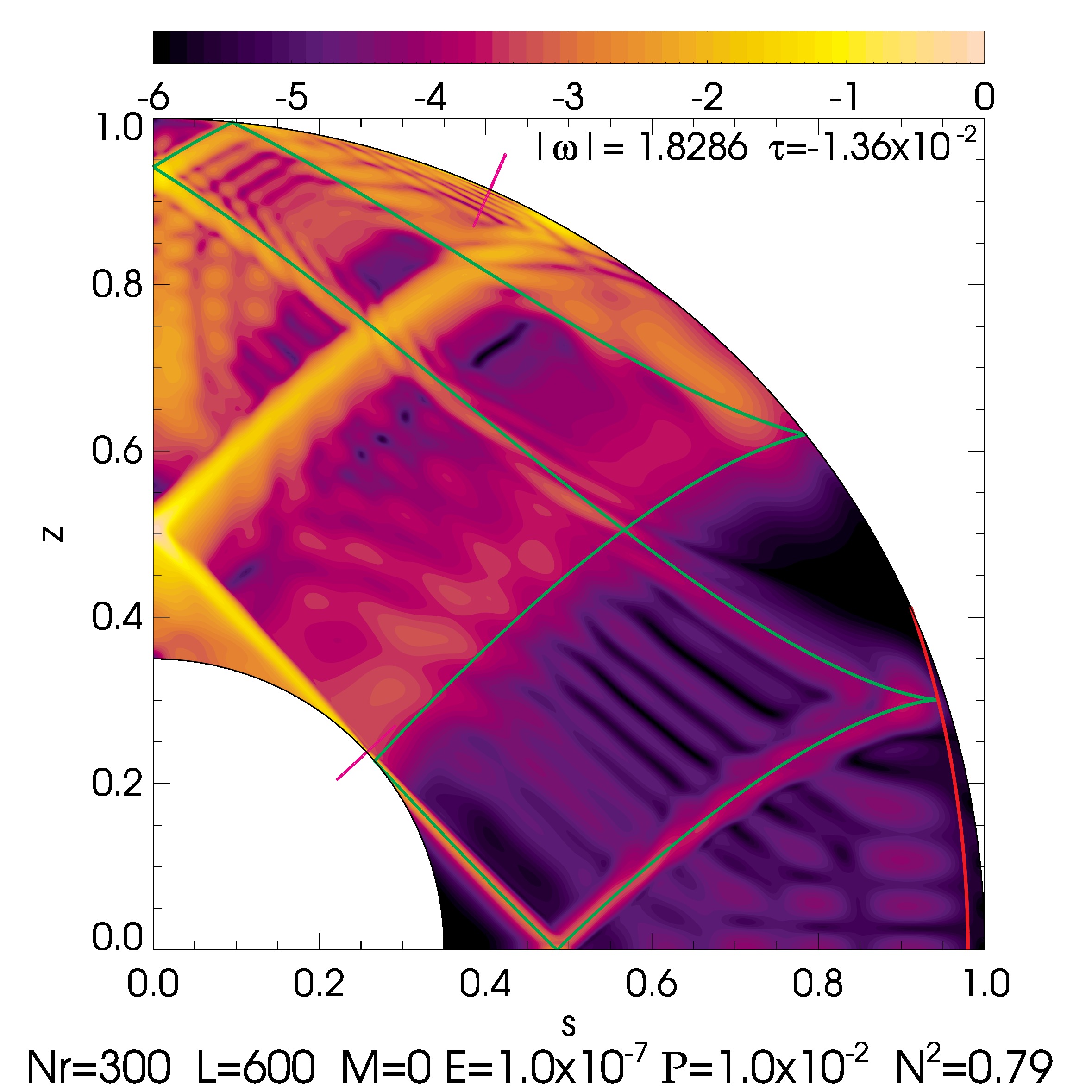}
   \caption{
     Meridional slices of kinetic energy (normalised to its maximum value, in logarithmic scale)
     obtained by solving the dissipative linearised hydrodynamics equations,
     with attractors (green) and turning surfaces (red) overplotted.
     The pink ticks at the inner and outer borders denote the critical latitudes.
     From top left to bottom right, the \BV frequency is increased for the same 
     mode ($\eta=0.35, E=10^{-7}, \mathcal{P}=10^{-2}$, starting from $\op = 1.67, N^2=0$). 
     The mode is distorted until the structure is suddenly changed when the shear 
     layer crosses the critical latitude (see main text).
     }
   \label{fig:suiviN2}
\end{figure}
As a first illustration of the impact of the stratification and the associated
differential rotation, we follow a pure inertial mode while increasing the \BV
frequency. To do so, we increase $N^2$ by small increments using the
eigenfrequency obtained for a given mode as an initial guess for the next
computed mode.
Figure~\ref{fig:suiviN2} shows the kinetic energy in a meridional
plane for increasing values of the \BV frequency at the surface of the shell.  
We recall that as the surface \BV frequency increases, so do the linear \BV 
frequency gradient and the rate of shellular differential rotation.  In
each panel, we find a singular shear layer that closely follows an attractor 
of characteristics (in green).

The top left panel is for $N^2=0$ and thus solid-body rotation. 
The characteristics and the shear layers then follow
straight lines in the shell.  The top right panel ($N^2 = 0.45$) shows the
distortion induced by the stratification and the consequent differential
rotation: the characteristics are now curved, even though the overall shape of
the attractor is conserved.  By increasing the stratification further (bottom
left panel, $N^2=0.75$), the critical latitude singularity at the inner core
gets excited. Owing to the fluid's viscosity, part of the energy is then
focused on the shear layer tangent to the core.  The mode has now changed to a
HT mode, as a turning surface appears near the equator (shown by a red curve in
the panels).
As shown in the bottom right panel, when increasing the stratification even
further ($N^2=0.79$), the shear layer emitted at the critical latitude
singularity dominates the kinetic energy distribution inside the shell. 

\begin{figure}
\includegraphics[width=0.5\textwidth]{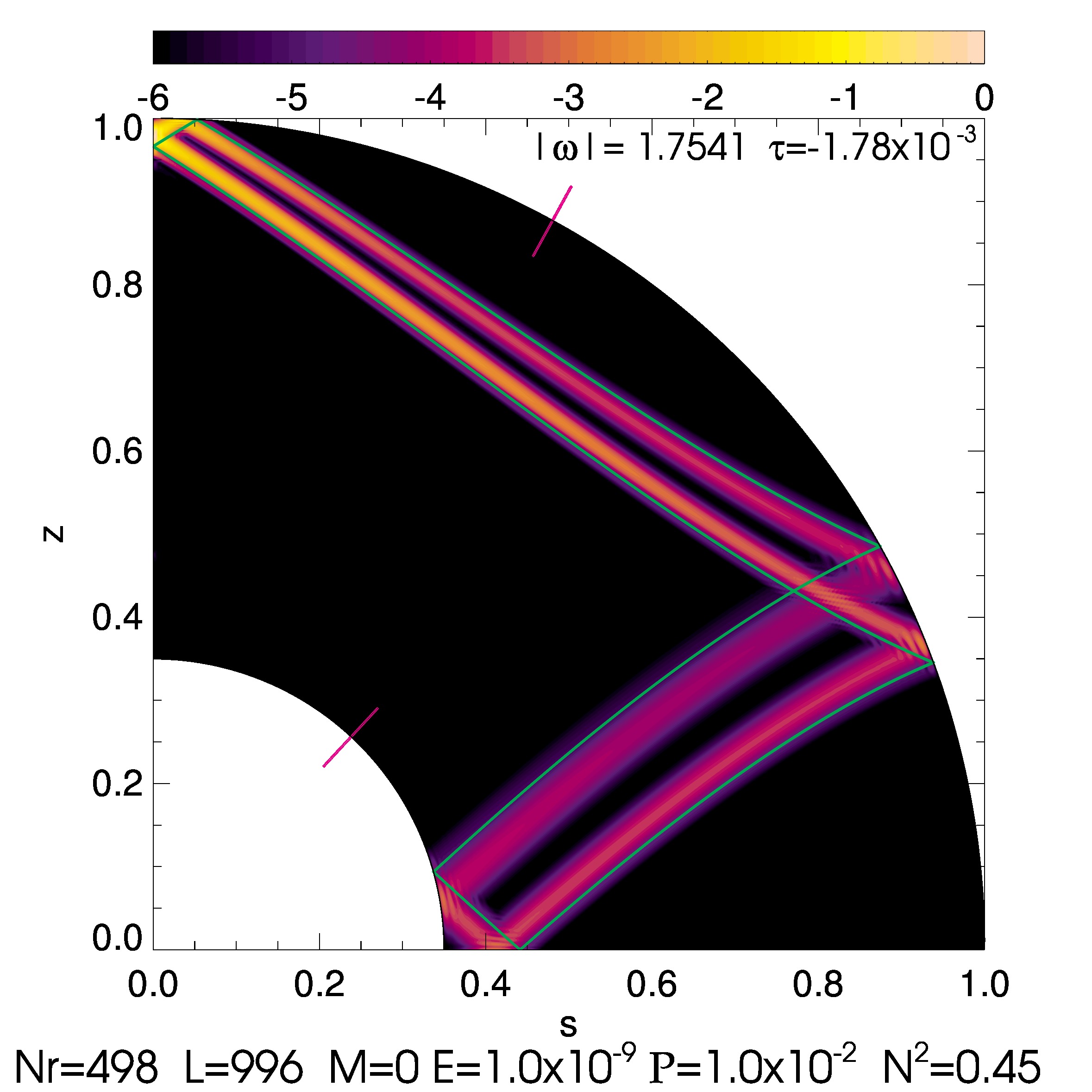}
\includegraphics[width=0.5\textwidth]{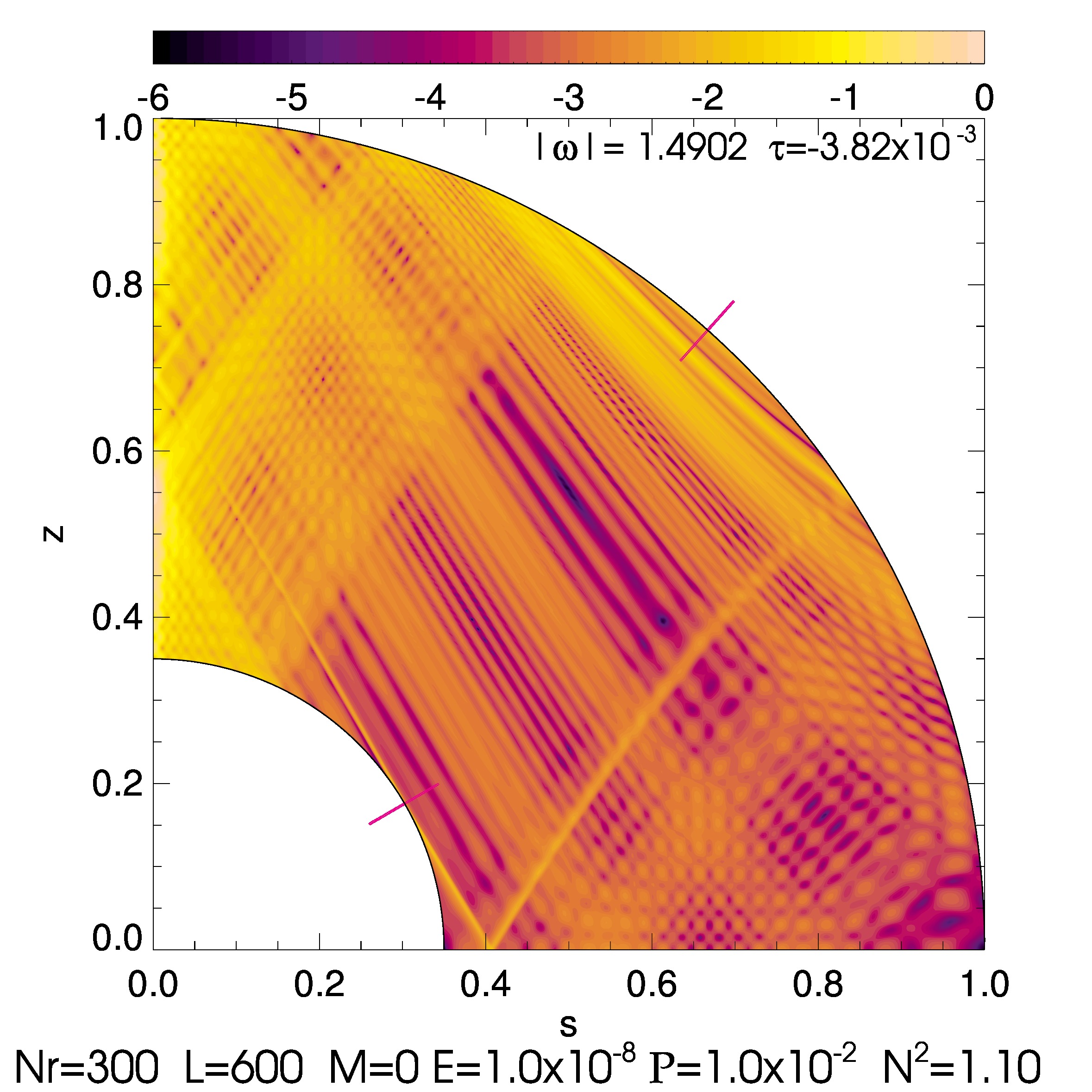} \\
\centerline{\includegraphics[width=0.5\textwidth]{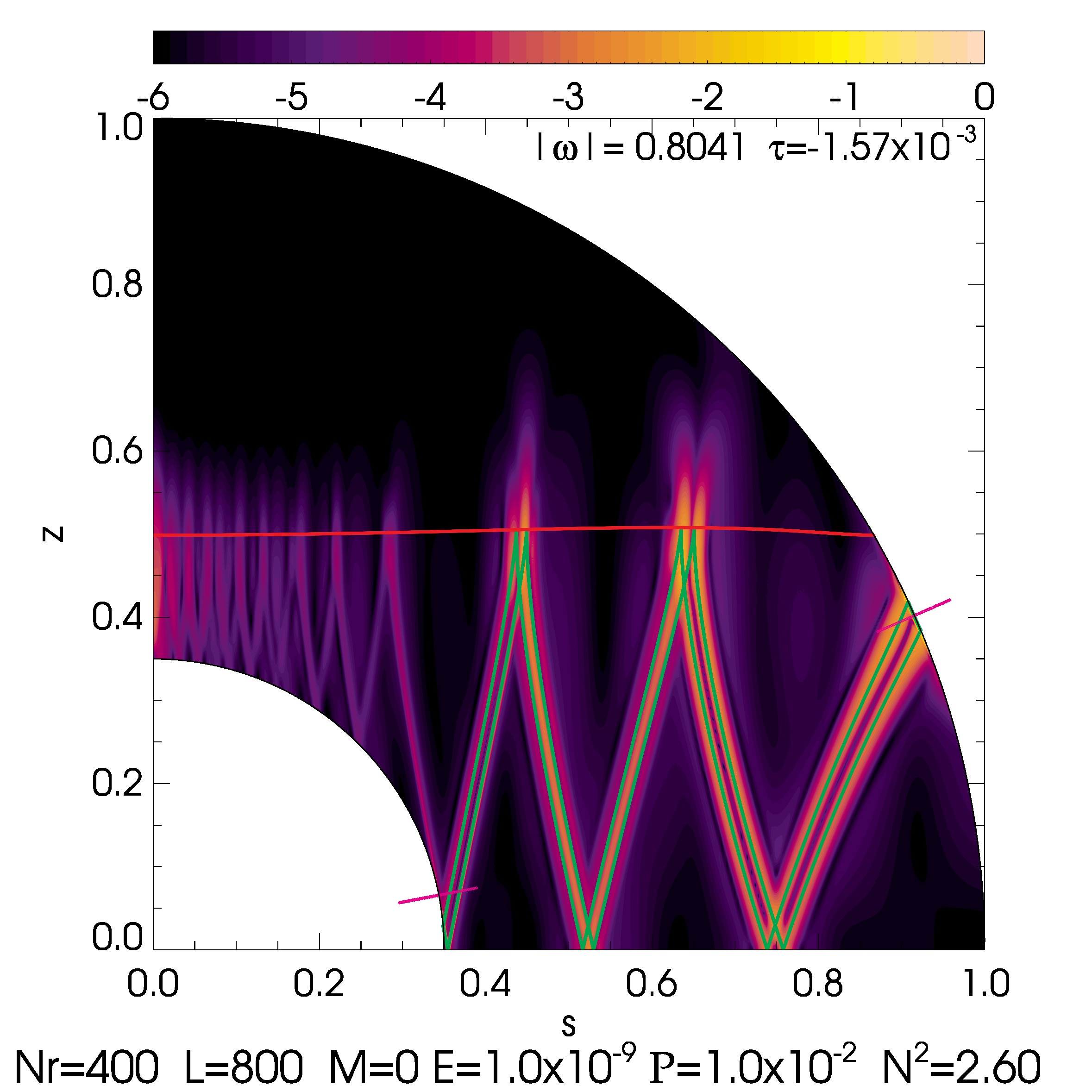} }
   \caption{
     Meridional slices of kinetic energy for a sample of modes (relative to maximum, in log scale).
     Attractors (green), turning surfaces (red) and critical latitudes (pink ticks) overplotted. 
     The various geometries are discussed in the main text.
     }
     \label{fig:ex_ax}
\end{figure}

Figure~\ref{fig:ex_ax} illustrates the diversity of the H and HT
modes predicted in our model.  The top row shows two H modes. The mode
displayed in the top-left panel is obtained by a follow-up of the mode
displayed in the top-right panel of figure~\ref{fig:suiviN2} by lowering
the Ekman number.  The energy is neatly focused on a shear layer whose
location is well described by the associated attractor.
The mode in the top-right panel is associated with a quasi-periodic
trajectory of characteristics.  The kinetic energy is distributed nearly
uniformly over the whole shell.
The mode in the bottom panel is a HT mode, which belongs to the subdomain 
(b) in figure~\ref{fig:ax}.
The predicted turning surface (red curve) nicely delimits the propagation
domain, even though the kinetic energy penetrates somehow in the
evanescent region.  The extent of the eigenfunction in the elliptic domain
is directly related to the diffusion coefficients $\nu$ and $\kappa$.
Attractors matching the shear layers (green curves) are obtained from
characteristics spanning the propagation domain and bouncing on the
turning surface.  This mode simultaneously focusses most of its energy
on an attractor and excites the critical latitude singularity, letting
a small fraction of its energy propagating towards the rotation axis. 
The turning surface may also introduce singularities,
such as the so-called wedge trapping, which is displayed in Figure~\ref{fig:wedgeinst} 
and discussed in the next section.

\subsection{Axisymmetric modes: dissipative properties}
\begin{figure}
\centerline{  \includegraphics[angle=-90,origin=cc,width=0.48\textwidth]{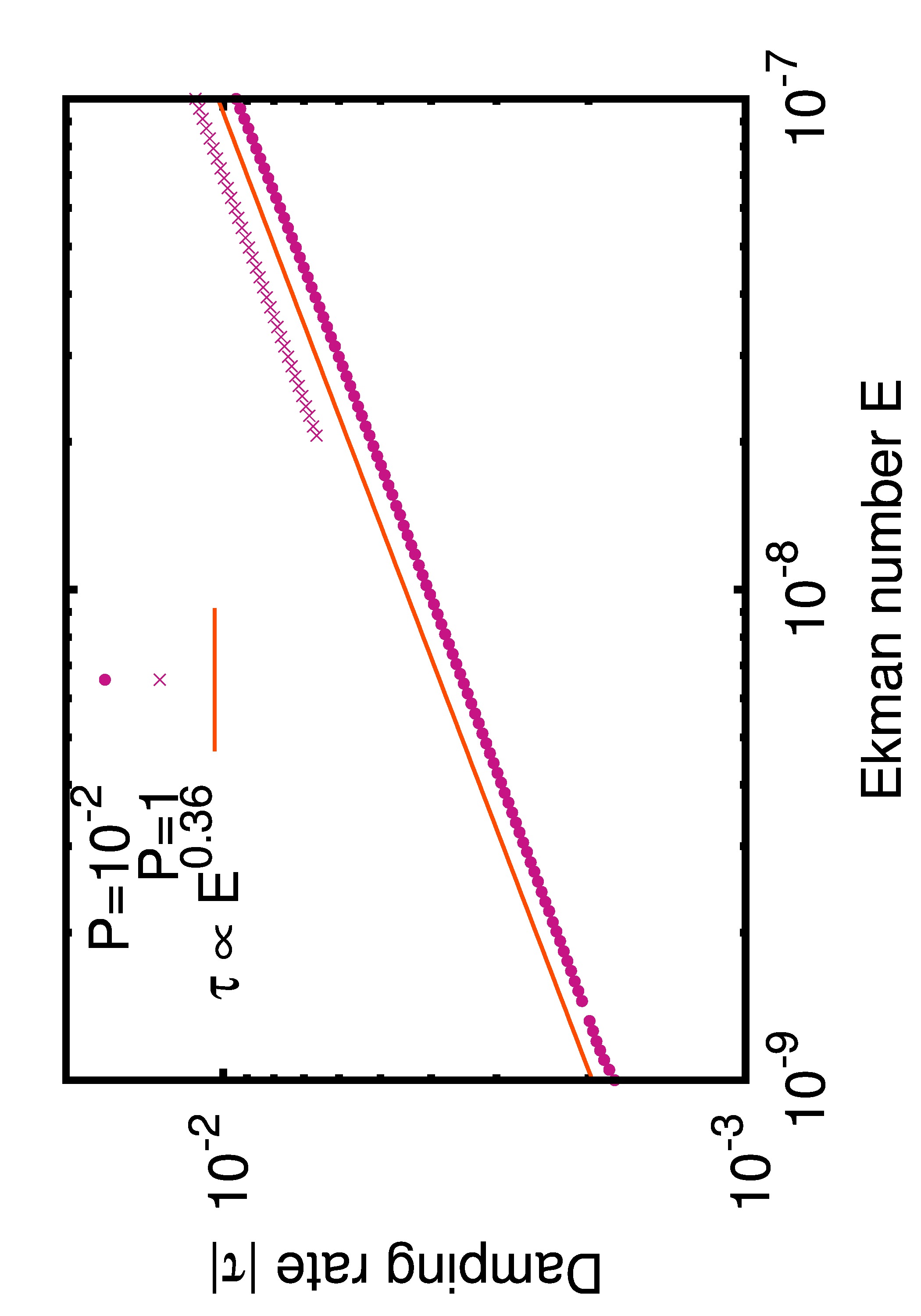} 
              \includegraphics[angle=-90,origin=cc,width=0.48\textwidth]{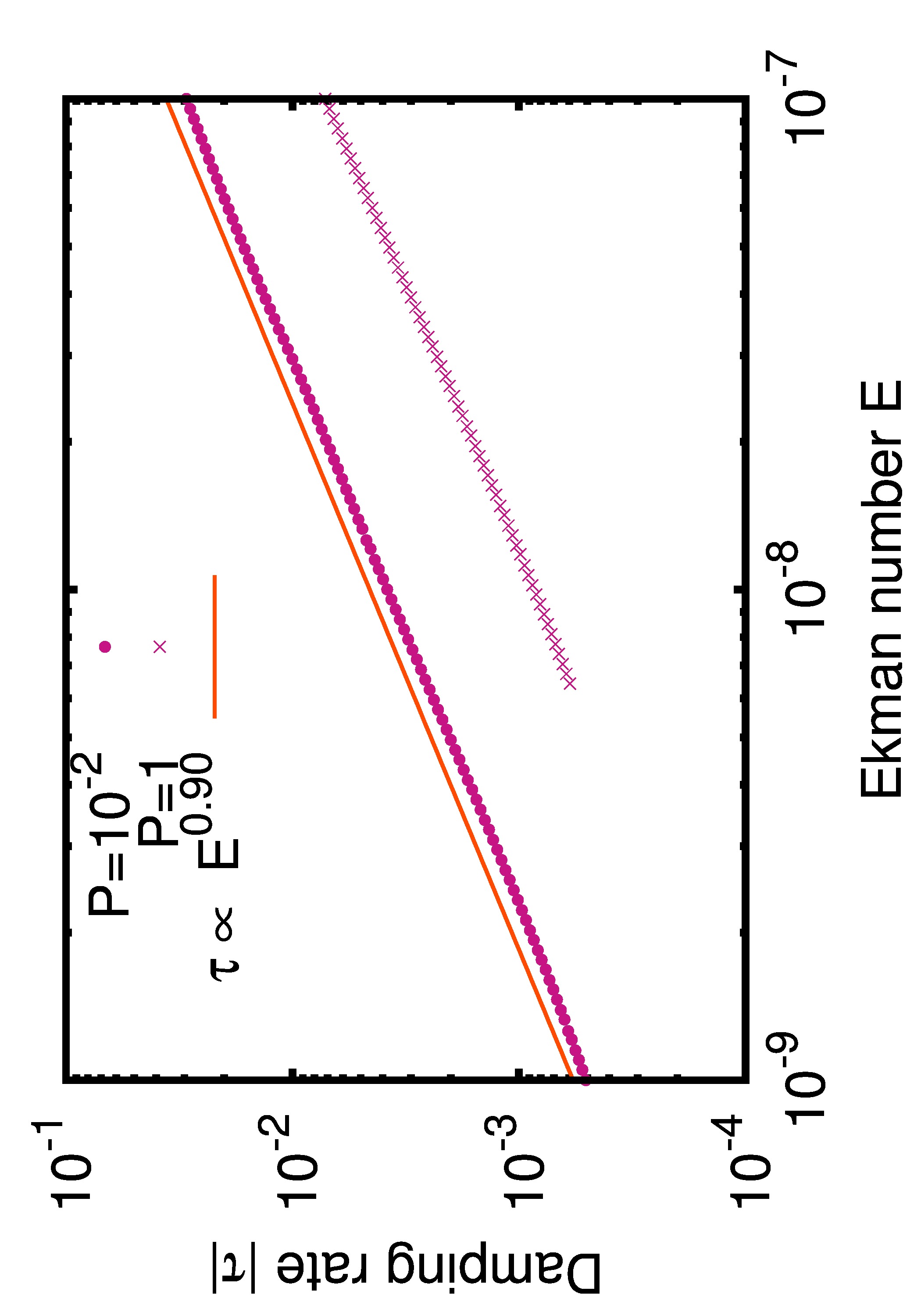} }
\caption{Damping rate as a function of the Ekman number for the H modes
  displayed in the top row of figure~\ref{fig:ex_ax}.  The left plot
  corresponds to the mode focused on a short-period attractor and the right
  plot to the quasi-periodical mode.  The data points are obtained by following
  a given mode with decreasing the Ekman number, for two different values of the
  Prandtl number: $10^{-2}$ (filled circles) and $1$ (crosses), while the solid
  line is the best fit trend at $E\rightarrow 0$.} 
  \label{fig:Dsuivi}
\end{figure}

We examine in this section the dissipative properties of axisymmetric modes
with shellular rotation resulting from the stratification.  Because of the very
small values of the Ekman and Prandtl numbers in stars or planets, we are
interested in the asymptotic behaviour of the modes and their eigenfrequency
when $E$ and $\mathcal{P}$ vanish (but verifying the constraint $E \ll \mathcal{P}$).  As
astrophysical parameters are beyond reach, we first try to infer asymptotic scaling
laws of damping rates for various modes. 

Figure~\ref{fig:Dsuivi} shows the damping rate $|\tau|$ as a function of the
Ekman number for two H modes at two different Prandtl numbers.  In both panels,
$E$ decreases from $10^{-7}$ to $10^{-9}$ for $\mathcal{P}=1$ and $\mathcal{P}=10^{-2}$.  The left
panel corresponds to the H mode in the top left panel of
figure~\ref{fig:ex_ax}, which is focused around a short-period attractor.  We
find that the decrease of $|\tau|$ is well reproduced by $E^{0.36}$.
This dependency on the Ekman number is reminiscent of the scaling laws obtained
by \cite{RGV01} for inertial modes focussed on short-period attractors.  For
these modes, the expected scalings are $|\tau| \propto E^\alpha$ when $E\rightarrow 0$,
with $\alpha \in [1/3,1/2]$. 
The right panel corresponds to the quasiperiodic H-mode in the top right panel
of figure~\ref{fig:ex_ax}.  We find the scaling of $|\tau| \propto
E^{0.9}$ for both Prandtl numbers. 

\begin{figure}
  \centering
\begin{tabular}{MM}
  \includegraphics[width=0.48\textwidth]{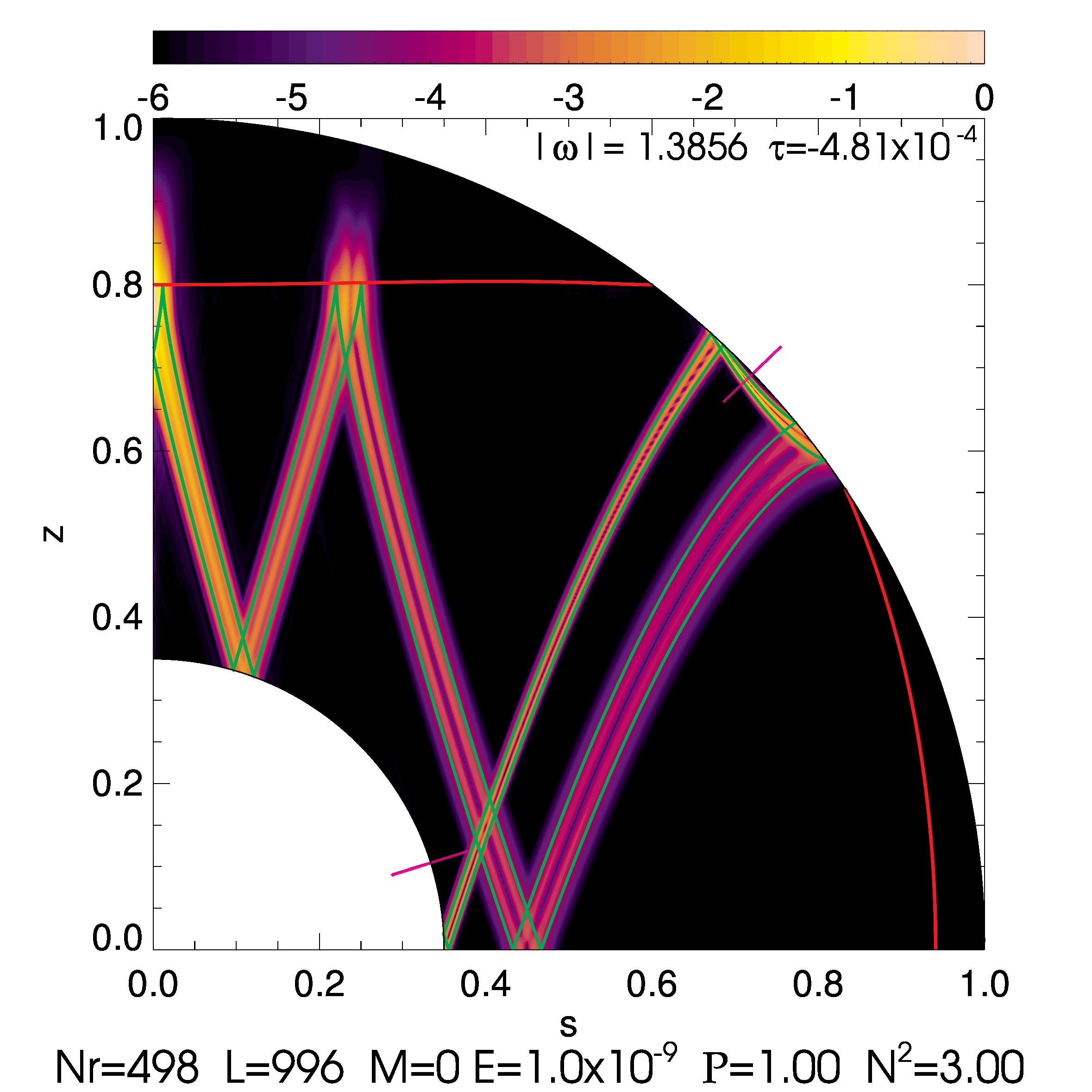} & 
  \begin{tabular}{c}
    \includegraphics[angle=-90,origin=cc,width=0.48\textwidth]{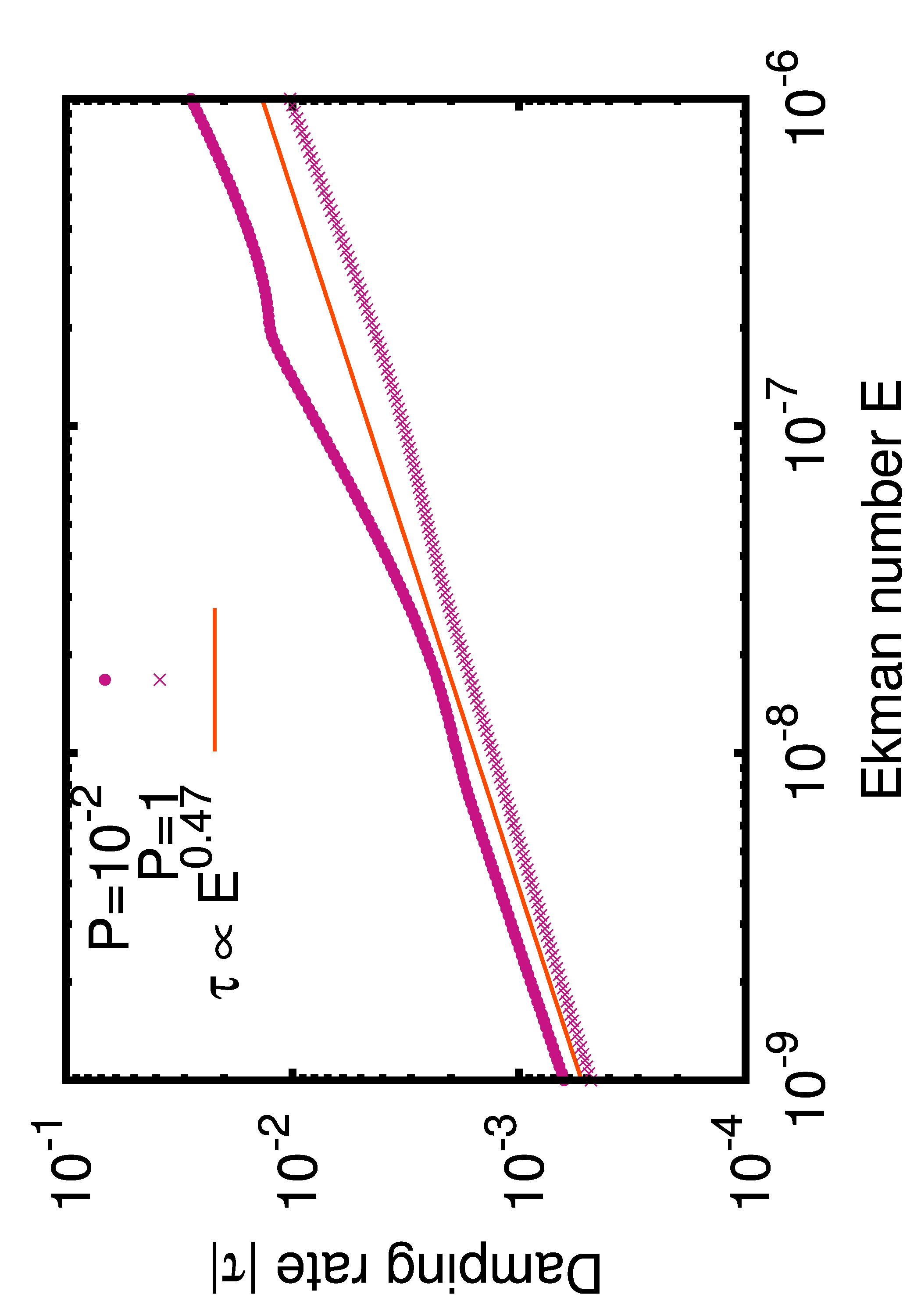}\\
    \includegraphics[angle=-90,origin=cc,width=0.48\textwidth]{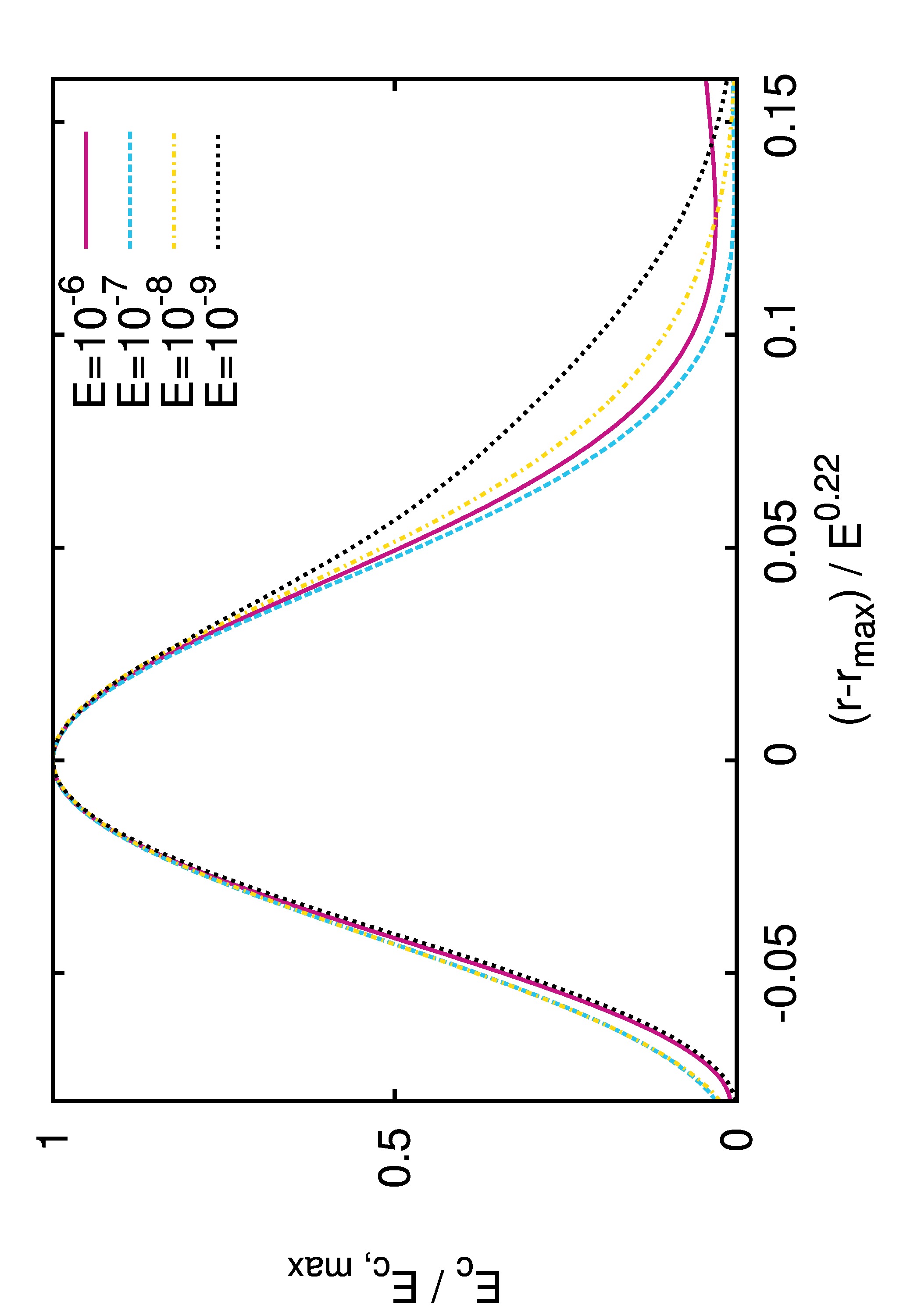}
  \end{tabular}
\end{tabular}
   \caption{
     Left: Kinetic energy for a HT mode featuring two turning surfaces, at
     $\op= 1.39, \eta=0.35, E=10^{-9}, \mathcal{P}=1$ and $N^2=3$ (relative to maximum, in log scale).  
     Top right: Damping rate
     as a function of the Ekman number.  The data points are obtained by
     following the mode with decreasing the Ekman number, for two different
     values of the Prandtl number: $10^{-2}$ (filled circles) and $1$
     (triangles), while the solid line is the best fit trend at $E\rightarrow 0$.
     Bottom right: Width of the shear layer over four decades in the Ekman number, rescaled by the best fit obtained. } 
   \label{fig:DTmode}
\end{figure}

Figure~\ref{fig:DTmode} shows the same follow-up calculation for an
HT mode with two turning surfaces and a short-period attractor.  The top right 
panel shows the evolution of the damping rate $|\tau|$ when decreasing the Ekman
number from $10^{-6}$ to $10^{-9}$, at $\mathcal{P}=10^{-2}$ and $ 1$.  
We find that $|\tau|$ approximately scales with $E^{0.47}$ at low $E$ and constant $\mathcal{P}$.
  Similarly, we follow this mode decreasing $\mathcal{P}$ from $1$ to $10^{-3}$ at $E=10^{-7}$ (not shown here),
  and find $|\tau| \propto \mathcal{P}^{-0.5}$ at low $\mathcal{P}$ at constant $E$.
We also measure the width of the shear layer while decreasing $E$, 
and find that the width of the shear layer scales with $E^{0.22}$ at fixed $\mathcal{P}$. 
This is illustrated in the bottom right panel of Figure~\ref{fig:DTmode}, where the kinetic energy profiles
in the shear layer at various Ekman number values, once rescaled by $E^{0.22}$, overlap neatly (this plot
is similar to that of Figure 13b in \citealt{DRV99}).
The obtained scaling is reminiscent of the $E^{1/4}-$scaling derived for the shear layer width by \cite{DRV99}.

\begin{figure}
  \centering
   \includegraphics[width=0.5\textwidth]{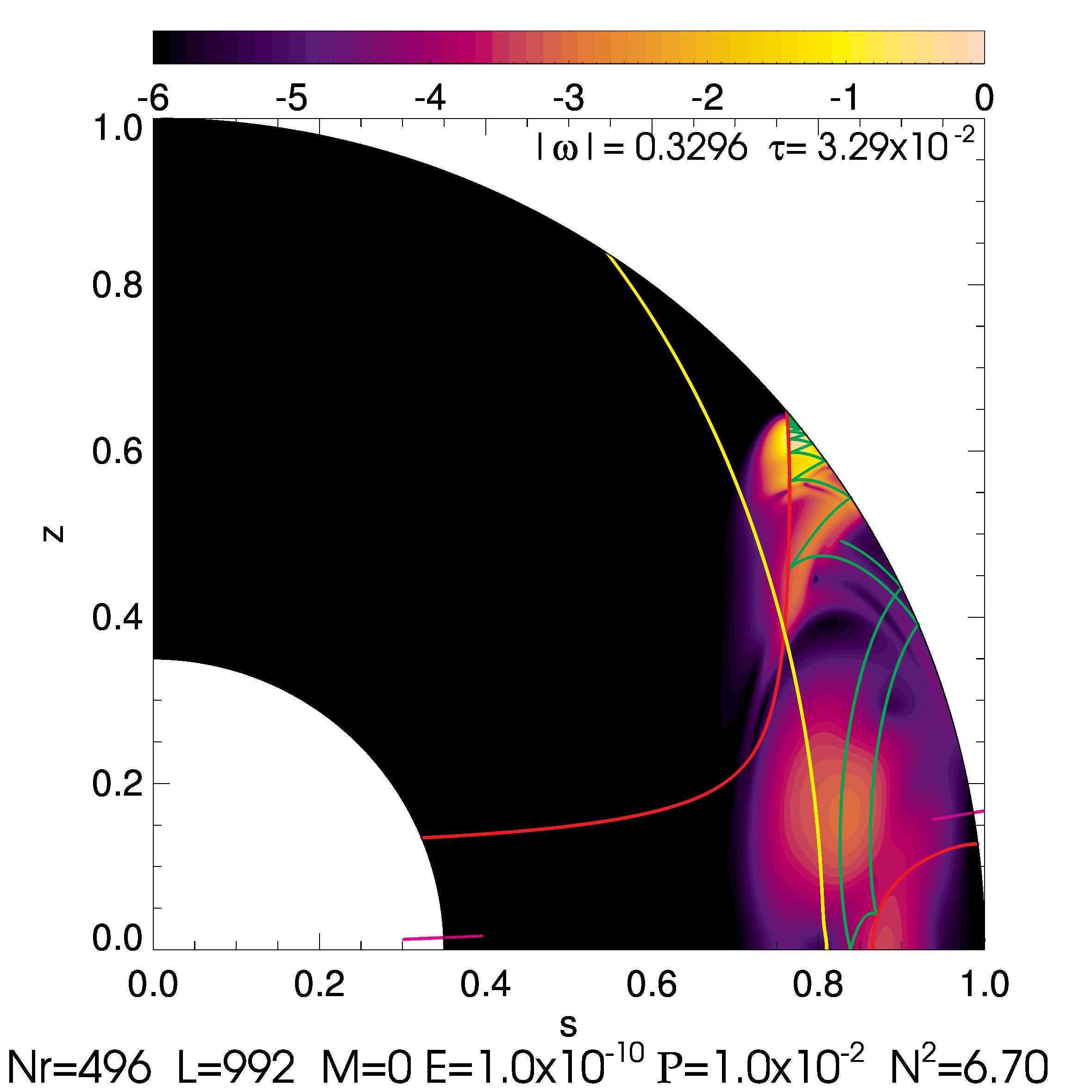}  
   \includegraphics[angle=-90,origin=cc,width=0.48\textwidth]{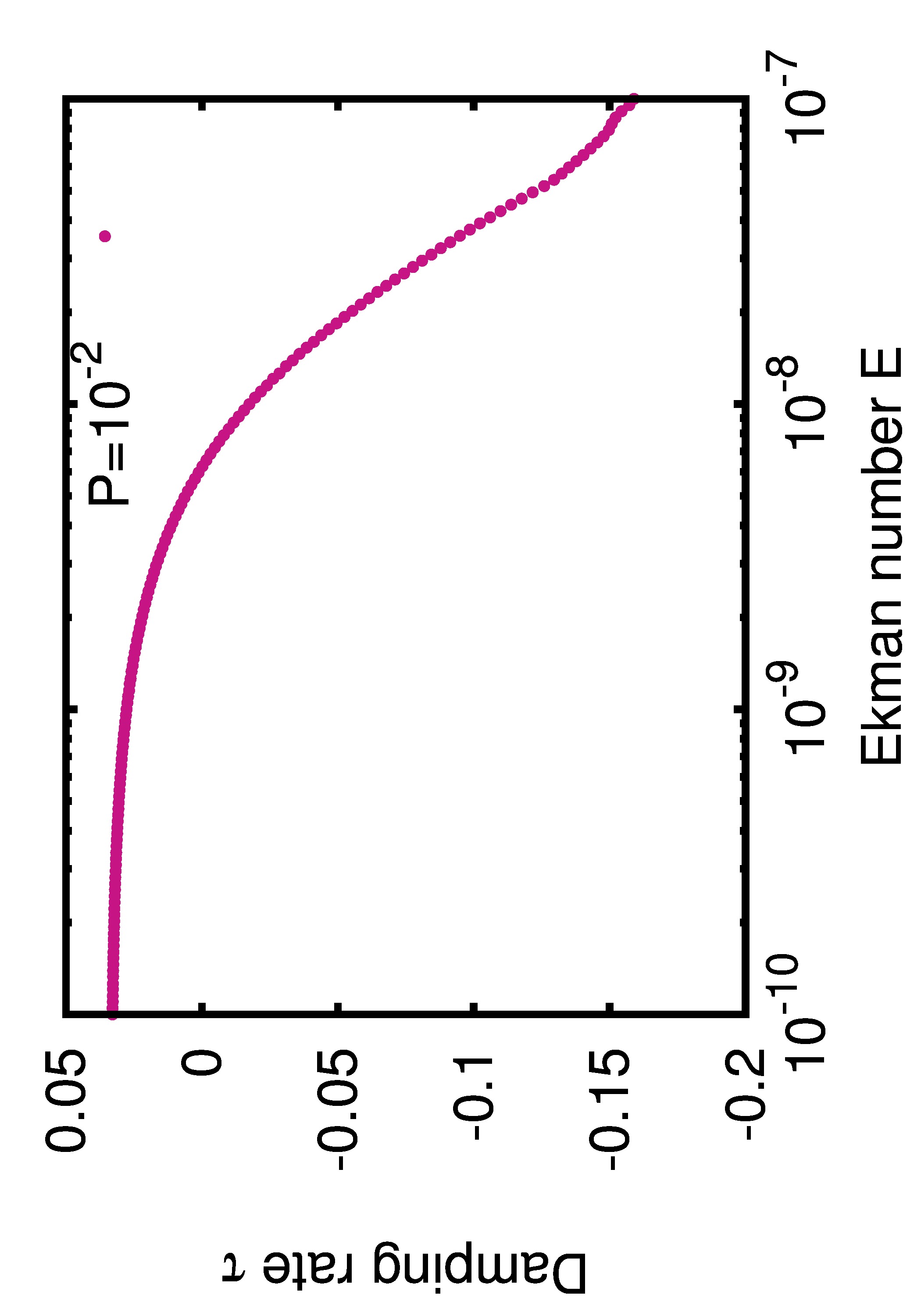} 
   \caption{
     Left: Kinetic energy for a HT mode in which a wedge trapping occurs, at
     $\op= 0.33, \eta=0.35, E=10^{-10}, \mathcal{P}=10^{-2}$ and $N^2=6.7$ (relative to maximum, in log scale).  The turning
     surfaces are shown in red and the characteristics in green, while the area
     outside the yellow curve is ABCD-unstable (see section~\ref{sec:abcd}).
     Right: Damping rate as a function of the Ekman number. The data points are
     obtained by following a mode with decreasing the Ekman number, the solid
     line is the best fit trend at $E\rightarrow 0$.
     }
     \label{fig:wedgeinst}
\end{figure}

The left panel of figure \ref{fig:wedgeinst} shows a HT mode with
two turning surfaces bounding the hyperbolic domain, corresponding to the
subdomain (d) of figure~\ref{fig:ax}. 
As alluded at the end of section \ref{sec:axisym}, the turning surfaces and
the outer surface of the shell form an acute angle in which the kinetic energy is focused.
This focussing, known as wedge trapping, is also described by
the characteristics which show that rays should converge towards the
intersection between the shell's surface and the higher-latitude turning surface.
Our calculation confirms that most of the mode’s kinetic energy is
indeed confined in that location. Wedge trapping is possible in the
subdomains (a), (d) and (f) of figure~\ref{fig:ax}.
The right panel of Fig.~\ref{fig:wedgeinst} shows that, when decreasing the Ekman number, 
this mode becomes unstable and its growth rate becomes independent of $E$. 
The critical Ekman number below which instability sets in is $E_c \sim 7\times 10^{-9}$.  
Most probably, this destabilization is due to the ABCD instability.  Indeed,
the part of the shell where the local ABCD instability criterion (\ref{eq:rayleigh}) is satisfied 
corresponds to the place occupied by the eigenfunction (see figure~\ref{fig:wedgeinst}).
The border of the ABCD-unstable domain is only slightly changed around the
equator when considering astrophysically-relevant Prandtl numbers such as $\mathcal{P}=10^{-5}$.
To be complete, the other possible driving of the instability, namely the shear instability (see
criterion \ref{eq:richardson}) is only satisfied along the inner boundary (at $\theta >
40^\circ$, $r<0.45$) and can thus be eliminated. We give further insight on the driving of the 
instability in section~\ref{sec:scaling}.


\begin{figure}
  \centering
  \includegraphics[width=0.5\textwidth]{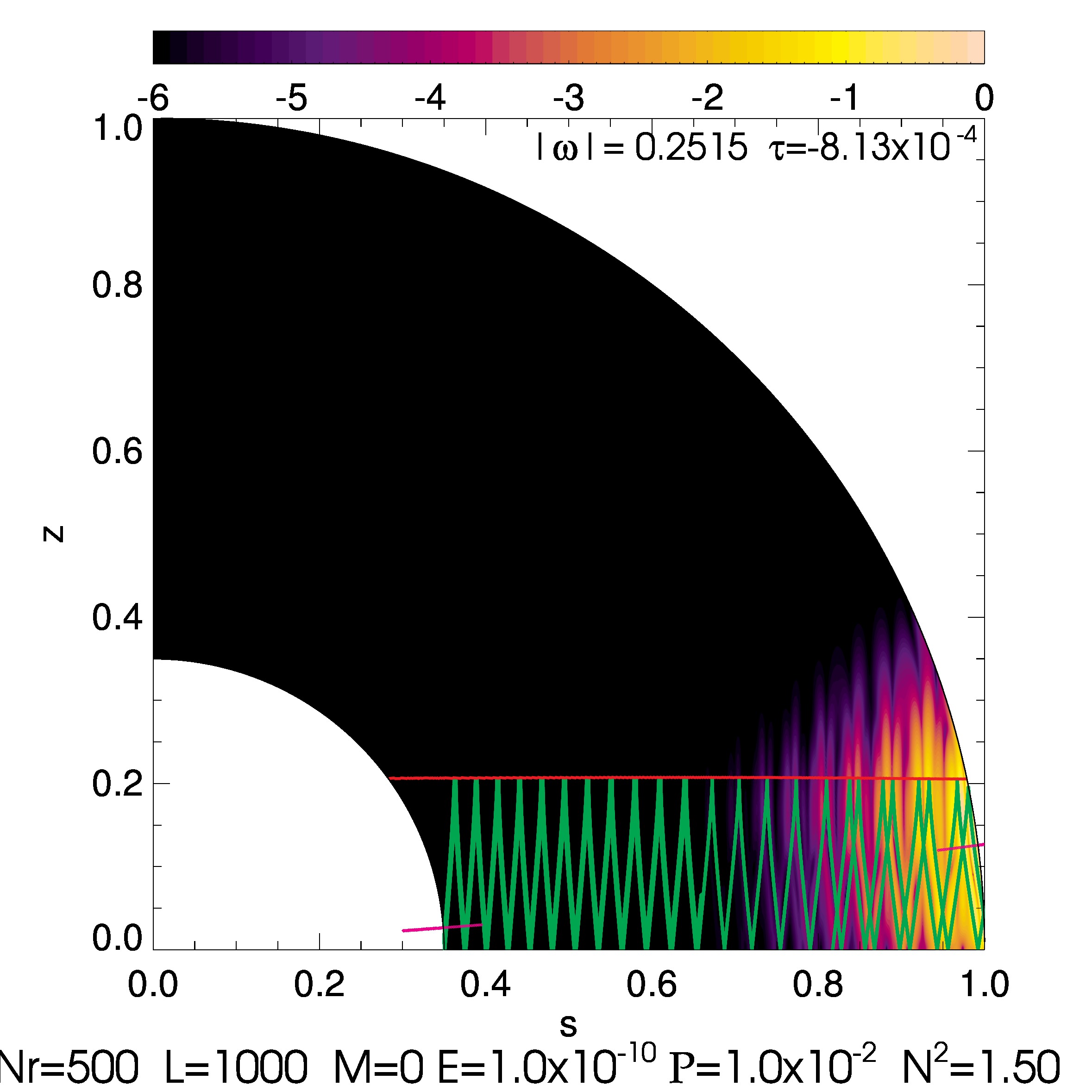} 
   \caption{
     Kinetic energy for a HT mode at $\op= 0.25, \eta=0.35, E=10^{-10}, \mathcal{P}=10^{-2}$ and $N^2=1.5$ (relative to maximum, in log scale). 
     For this mode, a wedge trapping is possible, but does not occur (see main text).
     The turning surface (red curve), the path of characteristic (green curve) and the critical latitudes (pink ticks) are overplotted.}
     \label{fig:wedgest}
\end{figure}

In figure~\ref{fig:wedgest}, we show another mode in which a wedge is formed at
the inner boundary.  We notice that the kinetic energy is not focused in
the wedge.  This is due to the presence of the critical latitude singularity: characteristics
hitting the core at latitudes $\theta > \theta_c$ converge towards the wedge in
a singular point, while characteristics hitting the core at $\theta < \theta_c$
form a periodic limit cycle.  When this dichotomy appears, the kinetic energy
seems to be always focussed on the periodic attractor, and the presence of a 
wedge does not impact the shape of the eigenmode.

\subsection{Non-axisymmetric modes: illustrative cases}
To complete the foregoing picture given by axisymmetric modes, we now focus on
non-axisymmetric ones, emphasizing the differences introduced by a non-zero
azimuthal wave-number $m$.  As explained in section~\ref{sec:classif}, the main
feature of non-axisymmetric modes is the presence of corotation resonances,
where the Doppler-shifted frequency vanishes, that is places where $\op + m\Omega(r) = 0$.

\subsubsection{Corotation resonances}

We have seen in section~\ref{sec:corot} the conditions for a corotation
resonance (or a critical layer) to occur in the spherical shell, its location
is given by equation (\ref{eq:rc}). 
Either the corotation resonance threads the hyperbolic domain, in which case we
expect the resonance to have a significant impact on the dissipation
properties of the mode, or it stays in the elliptic domain, in which
case its impact on the mode’s dissipation should be marginal.


\begin{figure}
  \centering
   \includegraphics[width=0.48\textwidth]{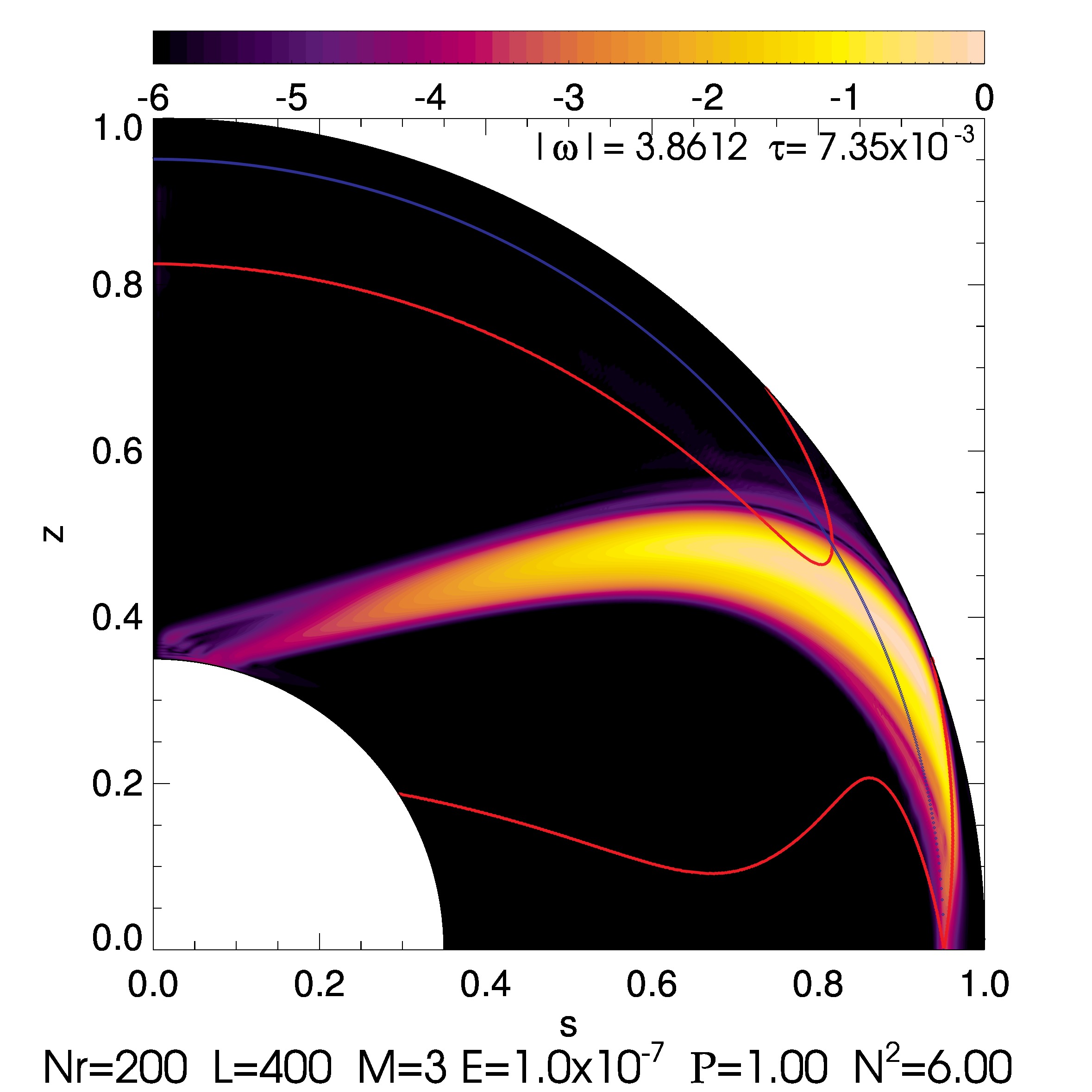}
   \includegraphics[width=0.8\textwidth]{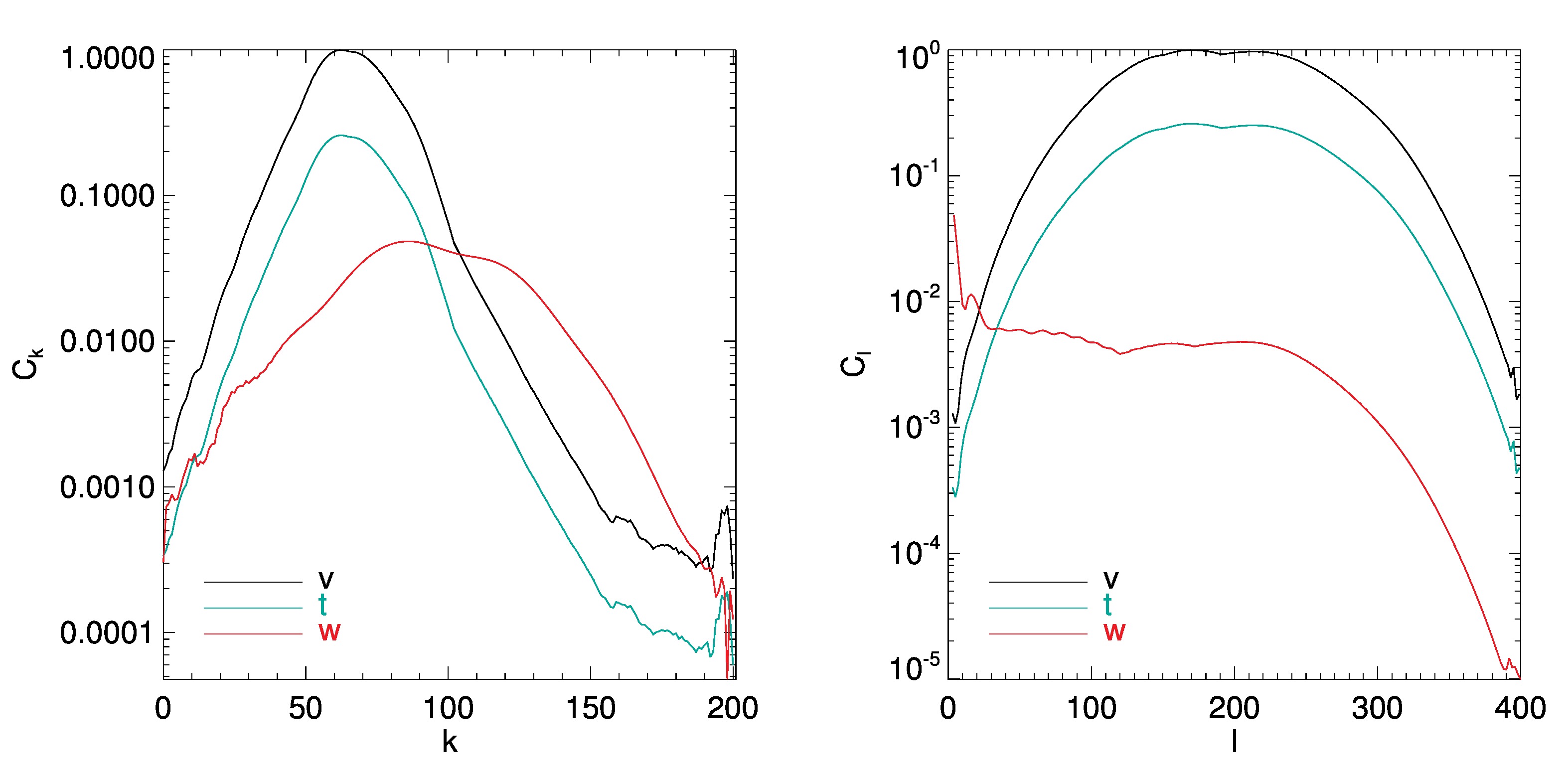}
   \caption{
     Top: Kinetic energy for a HT mode exhibiting a corotation resonance in the hyperbolical domain, 
     at $\op= -3.86, m=3, \eta=0.35, E=10^{-7}, \mathcal{P}=1$ and $N^2=6$ (relative to maximum, in log scale).
     This mode belongs to subdomain (b) in figure \ref{fig:corot}.
     Bottom: Spectral decomposition on the Chebyshev (left) and the spherical harmonics (right) bases.
     }
     \label{fig:hypercorot}
\end{figure}
An example of the former case is shown in the top panel of
figure~\ref{fig:hypercorot}.  We see that the energy is mostly focused on a
structure tangent to the corotation layer (blue radius), between the three
turning surfaces (red curves).  To avoid round-off errors, we had to use a
moderate resolution.  To reach spectral convergence with this resolution (as
shown in the bottom panels), we also had to use a relatively high Ekman number.
Many of these modes appear to be unstable, though we were not able to follow
any of these modes in a wide range of parameters $E$ and $\mathcal{P}$, owing to constraints from
spatial resolution. The mode shown in figure~\ref{fig:hypercorot} is unstable.
Its positive growth rate can be explained either by the baroclinic instability
or by a shear instability, but this latter possibility can actually be excluded:
the local instability criterion (\ref{eq:richardson}) is only met in two small
areas around the inner shell ($\theta>40^\circ,r<0.45$) and on the equator
($\theta>65^\circ,r>0.97$). The non-axisymmetric baroclinic instability
criterion (\ref{eq:zahn}) is met in most of the shell ($z>1/6$), and is likely at
the origin of the positive growth rate. As mentioned in section~\ref{sec:corot}, 
these corotations are expected to trigger non-linear effects \cite[see ][]{Maslowe86}.

When the corotation resonance is wholly included in the elliptic part of the
shell, we were not able to compute any mode: a satisfactory trade-off between 
spectral convergence and round-off errors could not be reached (because a high resolution,
needed to describe this type of modes accurately, increases dramatically the condition number of the 
matrices, and therefore the round-off errors).

\subsubsection{Regular modes}
The eigenvalue problem that we solve is also expected to have some regular or
quasi-regular solutions, as found for instance by \cite{DRV99} for solid-body
rotation.  Such modes are not focused towards any singular shear layer, and are
therefore significantly less damped than modes featuring an attractor.  Regular
modes have a kinetic energy distribution in the shell that is almost
independent of the dissipative parameters $E$ and $\mathcal{P}$. They may then exist at
vanishing viscosity and thermal diffusivity, and are expected to be only weakly
damped.  As our numerical method can compute the least-damped modes around a
given initial frequency guess (see section~\ref{sec:numerics}), regular modes
are expected to be easily found where they exist. Such kind of modes are 
only found when $m\ne0$, letting us think that regular or quasi-regular axisymmetric
modes do not exist in our setup. 

\begin{figure}
  \centering
   \includegraphics[width=\textwidth]{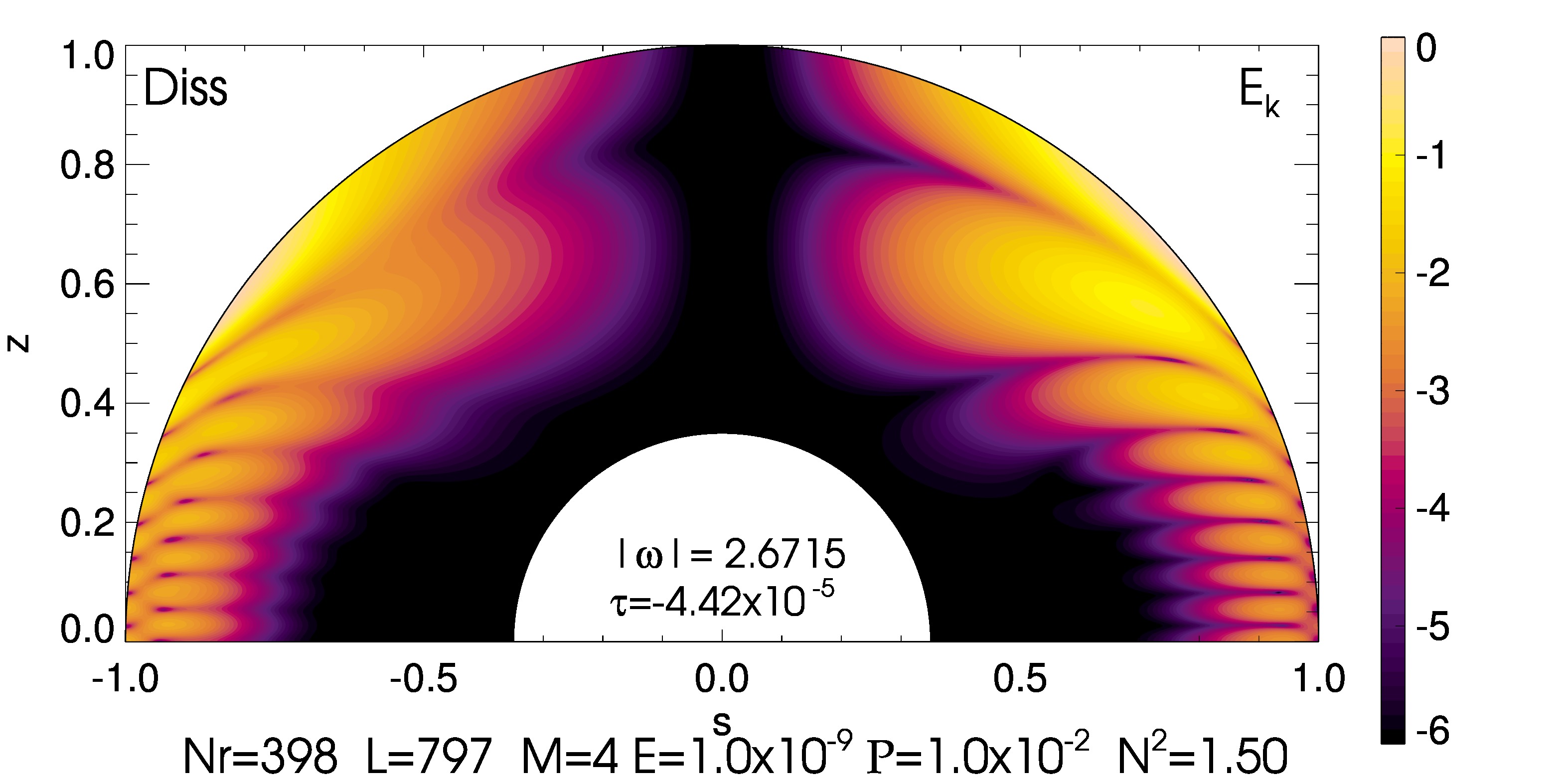} \\
   \includegraphics[width=\textwidth]{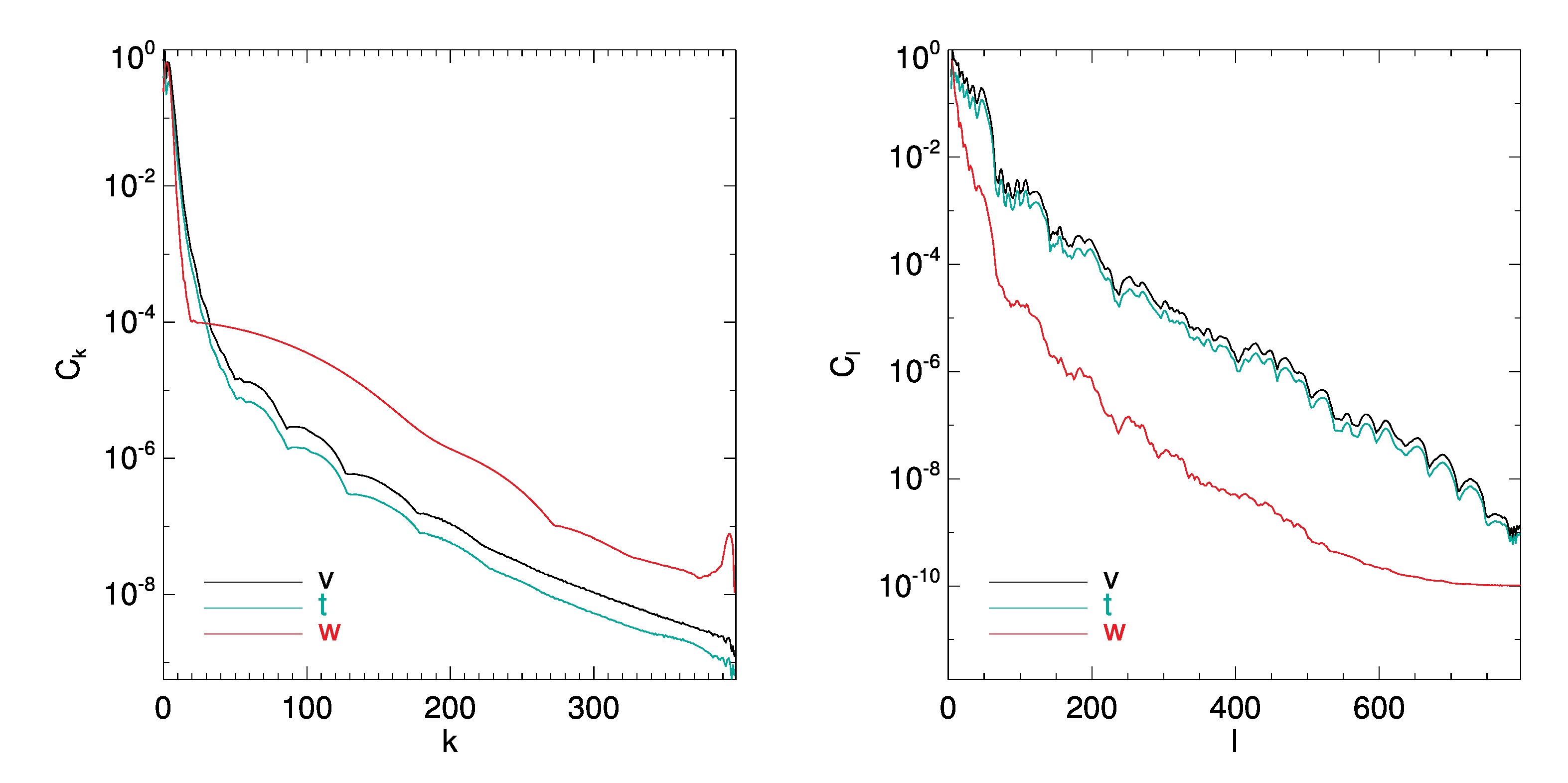} \\
   \includegraphics[angle=-90,origin=cc,width=0.48\textwidth]{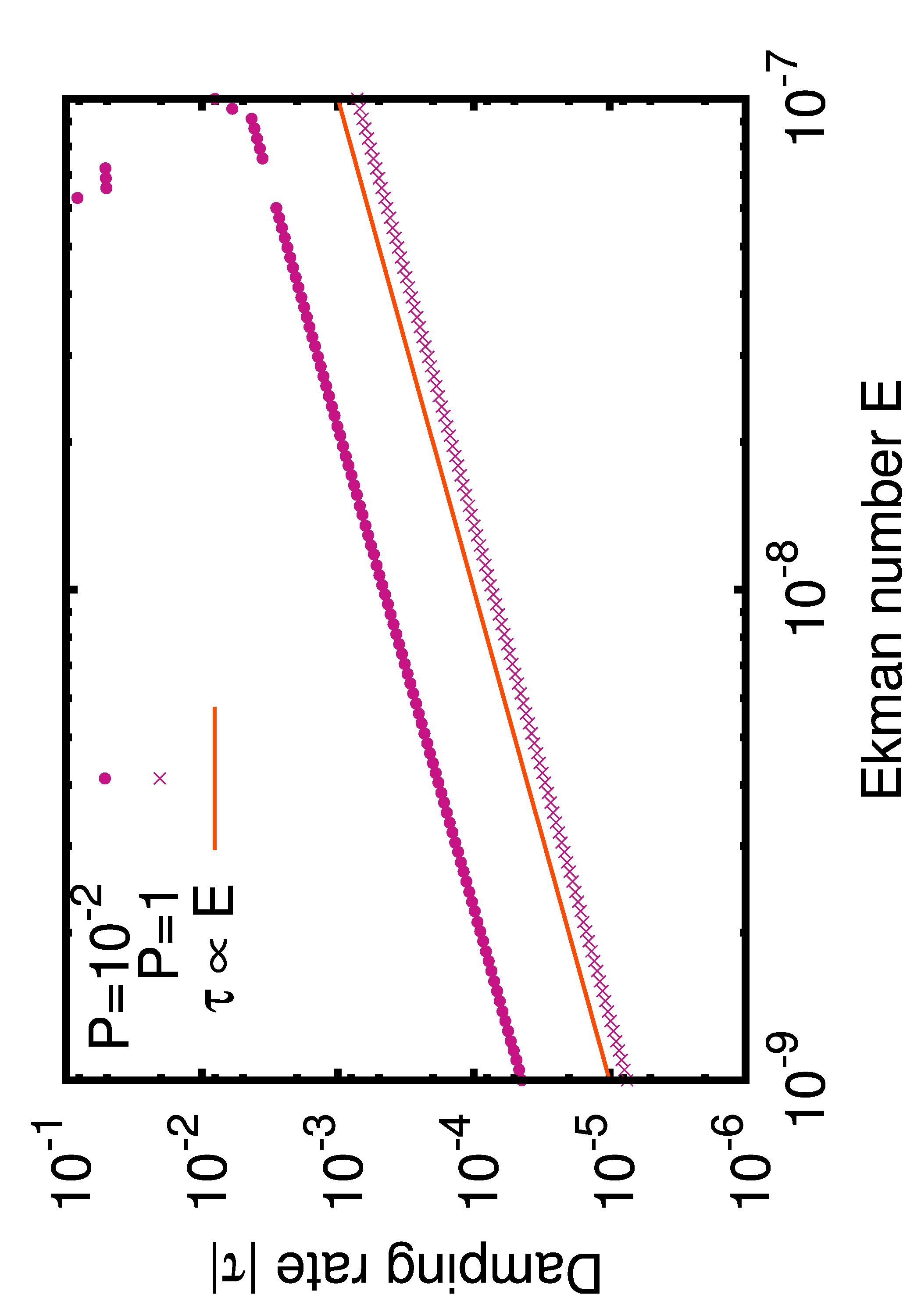}
   \includegraphics[angle=-90,origin=cc,width=0.48\textwidth]{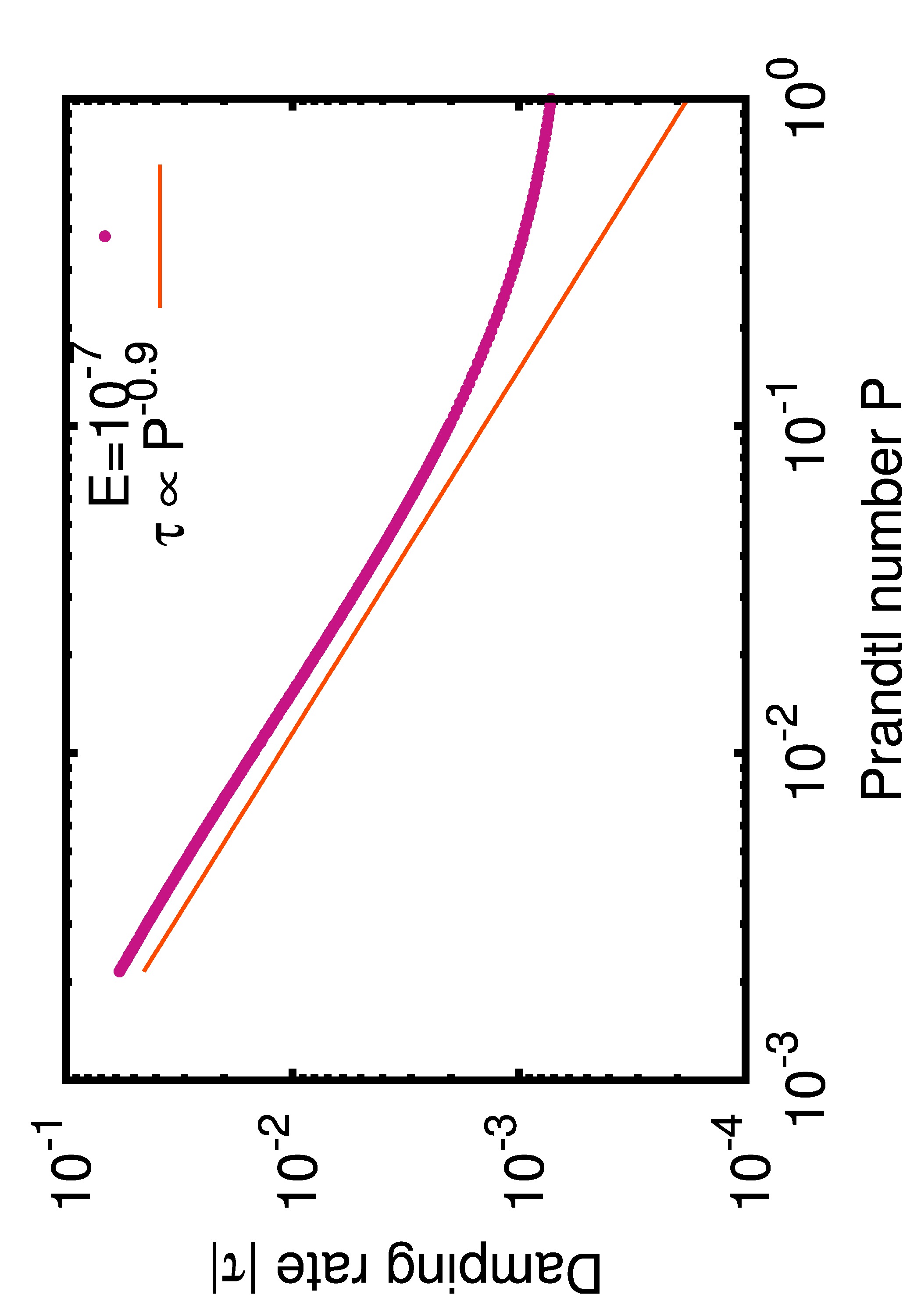} 
   \caption{
     Regular mode, at $\op= -2.67, m=4, \eta=0.35, E=10^{-9}, \mathcal{P}=10^{-2}$ and
     $N^2=1.5$.  Top: Meridional slice of the viscous dissipation (left
     quarter-panel) and the kinetic energy (right). 
     Both are normalised by their maximum value, and plotted in logarithmic scale.  Middle: Spectral
     decomposition on the Chebyshev (left) and the spherical harmonics (right)
     bases.  Bottom: Damping rate as a function of the Ekman number for
     $\mathcal{P}=10^{-2}$ and $1$ (left), and as a function of the Prandtl number for
     $E=10^{-7}$ (right).
     }
     \label{fig:regularmode}
\end{figure}

Figure~\ref{fig:regularmode} illustrates the properties of such a mode. Here,
the spectral convergence is easily reached compared to the previous modes.  As
the bottom plots show, we assess the scaling laws of the damping rate $|\tau|$
with the Ekman number at fixed Prandtl number (that is, while varying viscosity
and thermal diffusivity at the same rate) and with $\mathcal{P}$ at fixed $E$ (that is,
varying the thermal diffusivity while keeping the viscosity constant).  We find
$|\tau| \propto E$ at constant $\mathcal{P}$ and low $E$, and  $|\tau| \propto \mathcal{P}^{-0.9}$ at
constant $E$ and low $\mathcal{P}$.

\subsection{Interpretation of scaling laws}
\label{sec:scaling}
In the previous section we inspected numerically the behaviour of
gravito-inertial modes as the dissipation parameters are varied.  Clearly, the
damping rates often show a scaling law with either the Ekman number or the
Prandtl number, when these parameters tend to small values.  We now attempt to
interpret these behaviours, however without a detailed boundary layer analysis,
which is beyond the scope of this first exploration.  We rather resort to the
integral expression of the damping rate, which reads
\begin{equation}
  \label{eq:tau}
  \tau = -{\mathcal D}_v -{\mathcal D}_t +{\mathcal D}_r,
\end{equation}
that is, the sum of the viscous and thermal dissipations, and the differential
rotation driving. The aforementioned terms of equation~(\ref{eq:tau}) are
given by
\begin{eqnarray}
{\mathcal D}_v = \frac{1}{\mathcal{E}} \frac{E}{2} \displaystyle\int\limits_V  s_{ij}^2 \ dV,\quad 
{\mathcal D}_t &=& \frac{1}{\mathcal{E}} N^2\frac{E}{\mathcal{P}} \displaystyle\int\limits_V  |\del T|^2 \ dV,\quad
{\mathcal D}_r = \frac{1}{\mathcal{E}} \displaystyle\int\limits_V s {\mathcal R} \{v_r \prt_r\Omega v_\phi^\star\} \ dV,\qquad\label{eq:tau_terms}\\
{\rm where}\quad {\mathcal E} &=& \displaystyle\int\limits_V \vv^2 \ dV + N^2 \displaystyle\int\limits_V |T|^2 \ dV \nonumber
\end{eqnarray}
and $s_{ij}$ is the stress tensor, while ${\mathcal R}\{z\}$ and $z^\star$ 
designate the real part and the complex conjugate of $z$, respectively.
We follow the method detailed in Appendix B of \cite{DRV99} to derive equations
(\ref{eq:tau}) and (\ref{eq:tau_terms}).  It is clear that ${\mathcal D}_r$ is
the only term that can contribute positively to the growth rate, and explain
unstable gravito-inertial modes.

We compute the expected damping rates from the velocity and temperature fields
obtained in our simulations, through equation (\ref{eq:tau}).  Provided the
mode computation is accurately converged, the expected damping rate matches the
computed eigenvalue.  We present our results for some axisymmetric modes shown
earlier.

\begin{figure}
   \centering
   \includegraphics[angle=-90,origin=cc,width=0.48\textwidth]{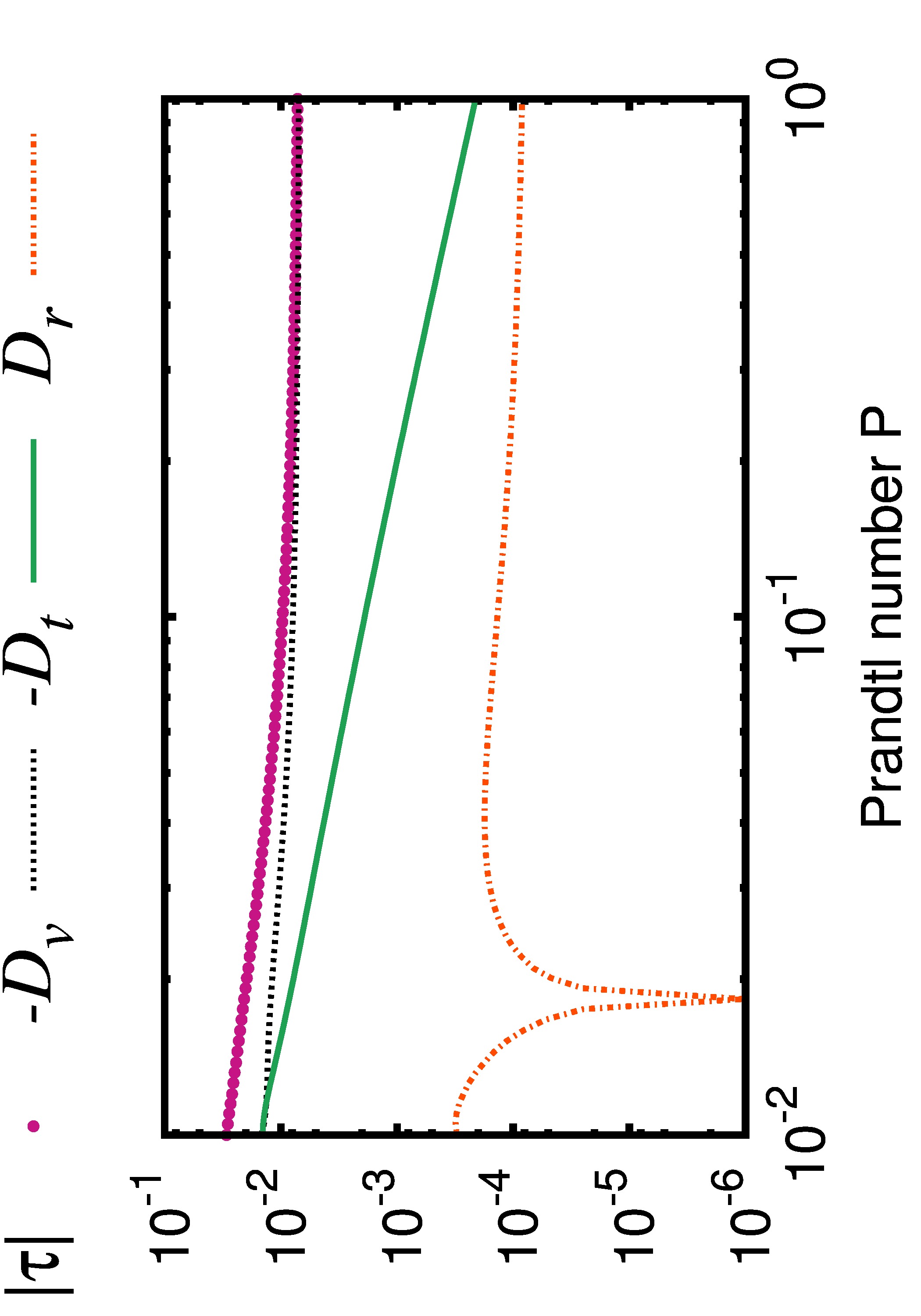} 
   \includegraphics[angle=-90,origin=cc,width=0.48\textwidth]{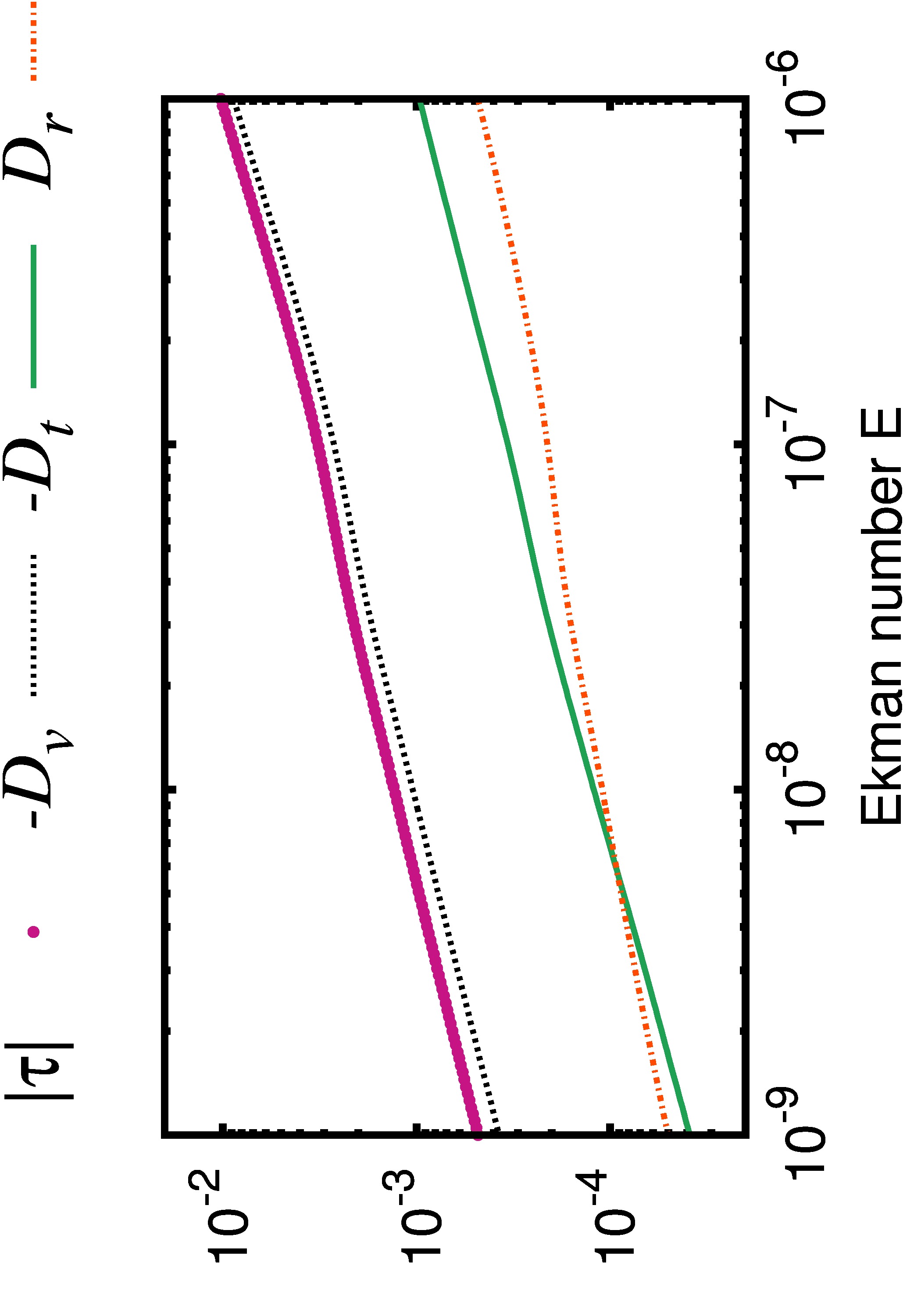}
   \caption{
     Left: Variation of the damping rate as a function of the Prandtl number for the H mode displayed 
     in the top right panel of figure~\ref{fig:ex_ax}. Right: Variation of the damping rate $|\tau|$ and its 
     components ${\mathcal D}_t, {\mathcal D}_v$ and ${\mathcal D}_r$, as a function of the Ekman number for 
     the HT mode displayed in figure~\ref{fig:DTmode}.
 }
     \label{fig:scalD}
\end{figure}

Figure \ref{fig:scalD} shows the damping rate and its three contributions for
two axisymmetric modes: the H mode in the top right panel of figure
\ref{fig:ex_ax}, and the HT mode of figure~\ref{fig:DTmode}.  We note that
the impact of differential rotation is marginal for both of these modes. 
Indeed, as shown in the left panel, the thermal dissipation ${\mathcal D}_t$
dominates at low Prandtl numbers, while the viscous dissipation ${\mathcal
D}_v$  dominates at higher values.  The thermal dissipation roughly scales as
the expected $\mathcal{P}^{-1}$, and both the viscous and differential rotation
dissipations do not scale with $\mathcal{P}$.  Interestingly, we see that the sign of the
term arising from differential rotation changes sign upon varying
$\mathcal{P}$.  However, since its magnitude is much smaller than the viscous and thermal
dissipation terms, it has a negligible impact on the net damping rate.

For the mode followed on the right panel of figure~\ref{fig:scalD}, we see that the viscous dissipation
dominates (as $\mathcal{P}=1$).  The thermal and viscous dissipation terms follow the
same scaling ${\mathcal D}_v, {\mathcal D}_t \sim E^{1/2}$. This is expected
from \cite{DRV99}: as the width of the shear layers around the attractor scales
with $E^{1/4}$, we expect the terms ${\mathcal D}_t$ and ${\mathcal D}_v$ to scale with $E^{1/2}$. 
Expanding the velocity and temperature perturbations in a series of powers of the square root of the Ekman
number, it appears that $v_\phi = i v_r$ at first order \cite[][]{RGV01}.  This
phase relation implies that the differential rotation integral ${\mathcal D}_r$
vanishes at first order.  Actually, the way ${\mathcal D}_r$ scales with $E$ or $\mathcal{P}$ is
different from the other terms and cannot be described without a precise
boundary layer analysis.

\begin{figure}
   \centering  
   \includegraphics[angle=-90,origin=cc,width=0.48\textwidth]{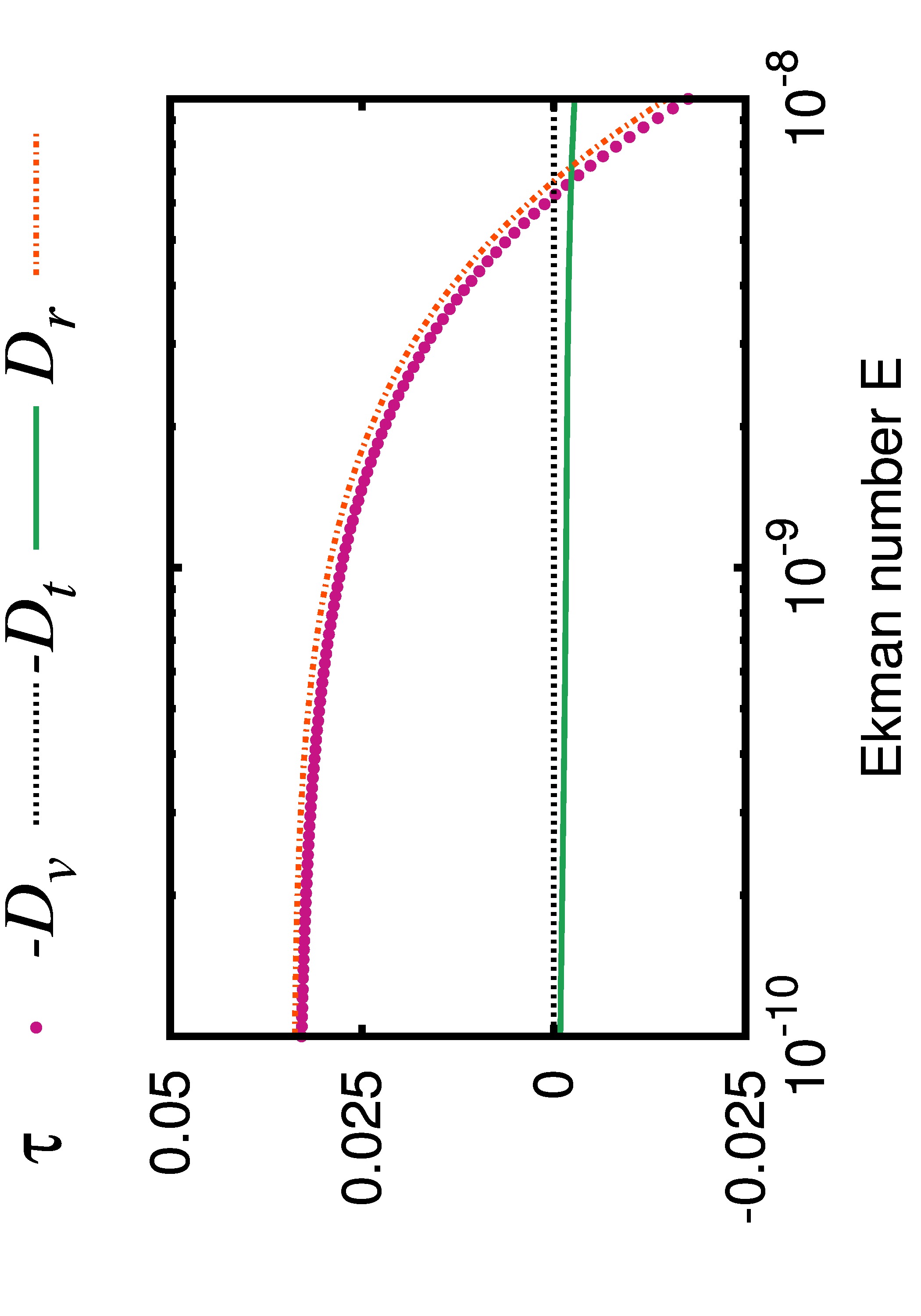} 
   \caption{
     Growth rate $\tau$ and dissipation terms as a function of the Ekman number, 
     for the unstable wedge-trapped mode displayed in figure~\ref{fig:wedgeinst}.
     }
     \label{fig:scalwedge}
\end{figure}
Figure \ref{fig:scalwedge} shows the contributions to the damping/growth rate of
the wedge-trapped mode shown in figure~\ref{fig:wedgeinst}. We find that the
contribution of the differential rotation term to the net damping/growth rate is 
by far the largest, and changes sign in this specific case.

For a regular mode, such as the one presented in
figure~\ref{fig:regularmode}~where the kinetic energy very slightly depends on the
Ekman and Prandtl numbers, equation (\ref{eq:tau}) predicts that the damping rate
should be a linear combination of the viscosity $\nu$ and the thermal
dissipation $\kappa$.  The scalings we find are in overall agreement with
this expectation and seem to confirm the regular nature of the modes in the
range of parameters that we have explored; it is likely that this property
extends to actual astrophysical values.
\section{Concluding remarks and astrophysical perspectives}
\label{sec:ccl}

In this paper, we have studied the properties of gravito-inertial modes in
a differentially rotating spherical shell. Our simplified model unveiled
a very rich dynamics that we investigated in two different ways: (i) by
studying the linearised oscillation problem in the non-dissipative limit,
and (ii) by solving numerically the fully-dissipative eigenproblem with
a spectral solver. This study emphasises the need to account for 
stratification, rotation and shear, as they all have a significant impact
on the modes' propagation properties and dissipations.

In the non-dissipative limit, the fluid equations can be recast as a
second-order partial differential equation of mixed type. From the paths
of characteristics associated with this equation, we found two kinds of
gravito-inertial modes : H modes that can propagate in the whole shell,
and HT modes that can propagate in a part of the shell bounded by turning
surfaces and the shell's boundaries.  Scanning the $(N^2,\op)$
parameter space ($N^2$ being the squared surface \BV frequency, and
$\op$ the wave's frequency in the inertial frame), we determined the
occurrence of both H and HT modes. We found that the frequency domain
reachable for HT modes with differential rotation is wider compared to
gravito-inertial modes with solid-body rotation, or inertial modes with
comparable differential rotation.  We have also described the various
geometries of HT modes. For both H and HT modes, we computed the paths of
characteristics, which often converge towards so-called attractors. We
have assessed the presence of attractors and determined the strength
of the associated focussing by estimating their Lyapunov exponent.
When turning surfaces form an acute angle with one of the boundaries of the
domain, we found that the attractors tend towards the singular wedge.
Another singularity we have seen is the presence of critical latitudes,
where characteristics are tangent to the inner or outer shell boundaries.
Non-axisymmetric HT modes also exhibit corotation resonances where
the Doppler-shifted frequency vanishes.  We determined the corotating
radius, which may be inside the propagation region of the shell or be
fully inside the elliptic region.  We have determined the associated range
of parameters in both cases. We have shown that the waves can cross the
corotation resonance when the latter is inside the propagation domain.
For both axisymmetric and non-axisymmetric modes, we also discussed the
baroclinic and shear instabilities expected to appear in our set-up and painted
out their probable occurrence.

High-resolution numerical solutions were carried out to explore
the waves' propagation properties in the fully dissipative case.
These simulations confirm many predictions of our analytic study of the
non-dissipative limit.  We computed the kinetic energy distribution
in the shell, confirming the expected propagation domain geometries.
Our calculations have highlighted a broad variety of gravito-inertial
modes: (i) modes with thin shear layers distributed over attractors of
characteristics, (ii) wedge-trapped modes potentially amplified by the
ABCD instability at low dissipation, (iii) modes that feature a strong
shear layer emitted at the critical latitude tangent to the inner core,
(iv) modes with corotation resonances inside the propagation domain which
are sometimes found to be linearly unstable, and (v) quasi-regular modes
whose  kinetic energy is almost independent of the dissipative parameters
in the studied range.

Studying the damping rates of the various categories of modes and
their dependence with the Ekman and Prandtl numbers, we often found
power laws, some of which could be related to the thickness of the
shear layers shaping the mode.  In the case of a positive growth rate,
we obtain clear evidence of the dominant role of
the differential rotation in the rise of the instability.

One of the interesting results of the foregoing study is that unstable modes
are not so common regarding the list of instabilities that are possibly
present in this set-up. Most probably, the criteria deciding of the
existence of such instabilities, which are based on a local analysis,
are too loose for the set of modes we have considered. Likely, at higher
$m$, with lower Ekman or Prandtl numbers unstable modes are more
frequent, but this needs to be confirmed by future work.

In the astrophysics perspective, the results of the present work are
interesting in two respects. First, the results show that even with a very
simplified set-up, the low-frequency spectrum of a rotating, stably
stratified spherical layer is quite complex, collecting many different
types of modes, shaped both by turning surfaces and various attractors
of characteristics. It leaves a daunting perspective when we have
to consider more realistic configurations including two-dimensional
differential rotation $\Omega(r,\theta)$, compressibility effects and
magnetic fields.

The good news however, is that some large-scale modes may be unstable
and oscillatory.  As we showed, the associated instability is directly
related to the differential rotation either for axisymmetric modes
or non-axisymmetric ones. The determination of the true nature of the
instability needs a specific study. 
Indeed, in such a setup, large-scale modes could destabilize and destroy
the background flow. This is however unlikely: the shear of baroclinic flow
is expected to feed small-scale turbulence \cite[e.g.][or our section \ref{sec:shear}]{zahn92}, resulting in a turbulent viscosity
that may prevent large-scale mode growth, or ease their saturation. If such unstable oscillations 
do not disappear into a turbulent flow, and saturate
in some way (by mode coupling for instance, e.g. \citealt{GD08}), 
the stellar oscillation spectrum will
be enriched with modes that are specifically dependent on the differential
rotation and thus open a window on the internal dynamics of the stars.

In another perspective, the properties of the oscillation spectrum of
a star also impact the response of the star or a planet to a tidal excitation. 
The modes shaped by attractors of characteristics are known to be effective
at dissipating energy \cite[][]{RV10}, but the existence of dynamically
unstable modes may change somehow the response of the star. However, as
for the free oscillations discussed above, this new kind of modes demands
fully non-linear calculations to determine the actual consequences of the
instability on the global dynamics of the star. We leave this study to
future work.

\begin{acknowledgements}
The authors thank the anonymous referees for their useful comments, 
Boris Dintrans, Vincent Prat and Fran\c{c}ois Ligni\`eres for useful
discussions.  The numerical calculations have been carried out on the CalMip
machine of the ``Centre Interuniversitaire de Calcul de Toulouse'' (CICT),
which is gratefully acknowledged.
\end{acknowledgements}

\appendix
\section{Derivation of the generalized Poincar\'e operator}
In this appendix, we give the full derivation of the generalized Poincar\'e operator defined in section~\ref{sec:caracs}.

The equation of motion, with no thermal diffusion and no viscosity, reads
\begin{equation}
\prt_t \uu + \Omega\prt_\phi\uu + 2 \bb{\Omega}\times\uu + s \left( \uu\bcdot\del\Omega \right)\bb{e}_\phi =  -\frac{1}{\rho_0}\del P + N^2 T \bb{r}.
\label{eq:poincare_dep}
\end{equation}

We replace time derivatives using $\prt_t = i\op$ and define $\opt = \op+m\Omega$. The heat equation then gives 
\begin{equation}
  T = -\frac{r u_r}{i\opt}.
\end{equation}

Projecting (\ref{eq:poincare_dep}) on the vector basis $(\bb{e_s}, \bb{e_\phi}, \bb{e_z})$, using 
$\bb{e}_r = \sin\theta \bb{e}_s + \cos\theta \bb{e}_z$, and $r\cos\theta = z , r\sin\theta = s$,
we get
\begin{eqnarray}
  \label{eq:appendix1}
  i\opt u_s - 2\Omega u_\phi &=& -\prt_s p - \frac{N^2}{i\opt} \left( s^2 u_s + sz u_z \right),\\
  \label{eq:appendix2}
  i\opt u_\phi + 2\Omega u_s + s\prt_s\Omega u_s + s\prt_z\Omega u_z &=& - \frac{imp}{s}, \\
  \label{eq:appendix3}
  i\opt u_z &=& - \prt_z p - \frac{N^2}{i\opt} \left( sz u_s + z^2 u_z \right).
\end{eqnarray}

Combining equations (\ref{eq:appendix1}) and (\ref{eq:appendix3}), we get
\begin{eqnarray}
  u_z &=& \frac{i\opt \prt_z p + N^2 sz u_s}{\opt^2 - N^2 z^2}, \\
  u_\phi &=&  \frac{1}{2\Omega} \left[\frac{i\opt (\opt^2 - N^2r^2)}{\opt^2-N^2z^2} u_s + \frac{N^2 sz}{\opt^2 - N^2z^2} \prt_z p + \prt_s p \right].
\end{eqnarray}

We replace $u_z$ and $u_\phi$ in equation (\ref{eq:appendix2}) to obtain
\begin{align}
  u_s = &- \frac{i\opt(\opt^2-N^2 z^2)}{A_s (\opt^2-N^2z^2)-\opt^2(\opt^2-N^2 r^2)+A_z N^2sz}  \nonumber\\
        & \qquad\qquad\qquad\qquad\qquad \times \left[ \frac{N^2sz+A_z}{\opt^2-N^2z^2}\prt_z p + \prt_s p - \frac{2im\Omega}{i\opt s} p \right],
\end{align}
with 
\begin{equation}
  A_s = \frac{2\Omega}{s} \frac{\prt}{\prt s} \left( s^2 \Omega \right) \quad {\rm and} \quad
  A_z = \frac{2\Omega}{s} \frac{\prt}{\prt z} \left( s^2 \Omega \right).
\end{equation}

Using these expressions in the mass conservation equation, keeping only the second-order terms, we finally get equation (\ref{eq:poincare}):
\begin{equation}
  (N^2z^2 - \opt^2) \frac{\prt^2 p}{\prt s^2} - (2N^2 sz + A_z) \frac{\prt^2 p}{\prt s \prt z} + (A_s + N^2s^2 - \opt^2) \frac{\prt^2 p}{\prt z^2} = 0.
\end{equation}

\bibliography{bibnew}

\begin{thebibliography}{53}
\expandafter\ifx\csname natexlab\endcsname\relax\def\natexlab#1{#1}\fi

\bibitem[{Barker} \& {Ogilvie}(2010)]{BO2010}
{\sc {Barker}, A.~J. \& {Ogilvie}, G.~I.} 2010 {On internal wave breaking and
  tidal dissipation near the centre of a solar-type star}. {\em MNRAS\/} {\bf
  404}, 1849--1868.

\bibitem[Baruteau \& Rieutord(2013)]{BR13}
{\sc Baruteau, C. \& Rieutord, M.} 2013 {Inertial waves in a differentially
  rotating spherical shell - I. Free modes of oscillation}. {\em J. Fluid
  Mech.\/} {\bf 719}, 47--81.

\bibitem[{Carr} {\em et~al.\/}(1998){Carr}, {Belton}, {Chapman}, {Davies},
  {Geissler}, {Greenberg}, {McEwen}, {Tufts}, {Greeley}, {Sullivan}, {Head},
  {Pappalardo}, {Klaasen}, {Johnson}, {Kaufman}, {Senske}, {Moore}, {Neukum},
  {Schubert}, {Burns}, {Thomas} \& {Veverka}]{Carr98}
{\sc {Carr}, M.~H., {Belton}, M.~J.~S., {Chapman}, C.~R., {Davies}, M.~E.,
  {Geissler}, P., {Greenberg}, R., {McEwen}, A.~S., {Tufts}, B.~R., {Greeley},
  R., {Sullivan}, R., {Head}, J.~W., {Pappalardo}, R.~T., {Klaasen}, K.~P.,
  {Johnson}, T.~V., {Kaufman}, J., {Senske}, D., {Moore}, J., {Neukum}, G.,
  {Schubert}, G., {Burns}, J.~A., {Thomas}, P. \& {Veverka}, J.} 1998 {Evidence
  for a subsurface ocean on Europa}. {\em Nature\/} {\bf 391}, 363.

\bibitem[Chandrasekhar(1961)]{chandra61}
{\sc Chandrasekhar, S.} 1961 {\em Hydrodynamic and hydromagnetic stability\/}.
  Clarendon Press, Oxford.

\bibitem[Dintrans {\em et~al.\/}(1999)Dintrans, Rieutord \& Valdettaro]{DRV99}
{\sc Dintrans, B., Rieutord, M. \& Valdettaro, L.} 1999 {Gravito-inertial waves
  in a rotating stratified sphere or spherical shell}. {\em J. Fluid Mech.\/}
  {\bf 398}, 271--297.

\bibitem[Drazin \& Reid(1981)]{DR81}
{\sc Drazin, P. \& Reid, W.} 1981 {\em Hydrodynamic stability\/}. Cambridge
  University Press.

\bibitem[{Dupret} {\em et~al.\/}(2004){Dupret}, {Thoul}, {Scuflaire},
  {Daszy{\'n}ska-Daszkiewicz}, {Aerts}, {Bourge}, {Waelkens} \&
  {Noels}]{dupret_etal04}
{\sc {Dupret}, M.-A., {Thoul}, A., {Scuflaire}, R.,
  {Daszy{\'n}ska-Daszkiewicz}, J., {Aerts}, C., {Bourge}, P.-O., {Waelkens}, C.
  \& {Noels}, A.} 2004 {Asteroseismology of the {$\beta$} Cep star HD 129929.
  II. Seismic constraints on core overshooting, internal rotation and stellar
  parameters}. {\em A\&A\/} {\bf 415}, 251--257.

\bibitem[{Espinosa Lara} \& {Rieutord}(2013)]{ELR13}
{\sc {Espinosa Lara}, F. \& {Rieutord}, M.} 2013 {Self-consistent 2D models of
  fast rotating early-type stars}. {\em A\&A\/} {\bf 552}, A35.

\bibitem[{Favier} {\em et~al.\/}(2014){Favier}, {Barker}, {Baruteau} \&
  {Ogilvie}]{FBBO14}
{\sc {Favier}, B., {Barker}, A.~J., {Baruteau}, C. \& {Ogilvie}, G.~I.} 2014
  {Non-linear evolution of tidally forced inertial waves in rotating fluid
  bodies}. {\em MNRAS\/} {\bf 439}, 845--860.

\bibitem[Fotheringham \& Hollerbach(1998)]{FH98}
{\sc Fotheringham, P. \& Hollerbach, R.} 1998 Inertial oscillations in a
  spherical shell. {\em Geophys. Astrophys. Fluid Dyn.\/} {\bf 89}, 23--43.

\bibitem[Friedlander(1982)]{Friedl82}
{\sc Friedlander, S.} 1982 Turning surface behaviour for internal waves subject
  to general gravitational fields. {\em Geophys. Astrophys. Fluid Dyn.\/} {\bf
  21}, 189--200.

\bibitem[Friedlander(1987)]{Friedl87}
{\sc Friedlander, S.} 1987 Internal waves in a rotating stratified spherical
  shell: asymptotic solutions. {\em Geophys. J. R. Astr. Soc.\/} {\bf 89},
  637--655.

\bibitem[{Friedlander}(1989)]{friedl89}
{\sc {Friedlander}, S.} 1989 {Hydromagnetic waves in a differentially rotating,
  stratified spherical shell}. {\em Geophysical and Astrophysical Fluid
  Dynamics\/} {\bf 48}, 53--67.

\bibitem[Friedlander \& Siegmann(1982{\natexlab{{\em a\/}}})]{FS82a}
{\sc Friedlander, S. \& Siegmann, W.} 1982{\natexlab{{\em a\/}}} Internal waves
  in a contained rotating stratified fluid. {\em J. Fluid Mech.\/} {\bf 114},
  123--156.

\bibitem[Friedlander \& Siegmann(1982{\natexlab{{\em b\/}}})]{FS82b}
{\sc Friedlander, S. \& Siegmann, W.} 1982{\natexlab{{\em b\/}}} Internal waves
  in a rotating stratified fluid in an arbitrary gravitational field. {\em
  Geophys. Astrophys. Fluid Dyn.\/} {\bf 19}, 267--291.

\bibitem[{Fuller}(2014)]{F14}
{\sc {Fuller}, J.} 2014 {Saturn ring seismology: Evidence for stable
  stratification in the deep interior of Saturn}. {\em Icarus\/} {\bf 242},
  283--296.

\bibitem[{Gastine} \& {Dintrans}(2008)]{GD08}
{\sc {Gastine}, T. \& {Dintrans}, B.} 2008 {Direct numerical simulations of the
  {$\kappa$}-mechanism. I. Radial modes in the purely radiative case}. {\em
  A\&A\/} {\bf 484}, 29--42.

\bibitem[{Gerkema} {\em et~al.\/}(2008){Gerkema}, {Zimmerman}, {Maas} \& {van
  Haren}]{Gerk08}
{\sc {Gerkema}, T., {Zimmerman}, J.~T.~F., {Maas}, L.~R.~M. \& {van Haren}, H.}
  2008 {Geophysical and astrophysical fluid dynamics beyond the traditional
  approximation}. {\em Reviews of Geophysics\/} {\bf 46}, 2004.

\bibitem[{Goldreich} \& {Schubert}(1967)]{GS67}
{\sc {Goldreich}, P. \& {Schubert}, G.} 1967 {Differential rotation in stars}.
  {\em ApJ\/} {\bf 150}, 571.

\bibitem[{Goodman} \& {Lackner}(2009)]{GL09}
{\sc {Goodman}, J. \& {Lackner}, C.} 2009 {Dynamical Tides in Rotating Planets
  and Stars}. {\em ApJ\/} {\bf 696}, 2054--2067.

\bibitem[{Hypolite} \& {Rieutord}(2014)]{HR14}
{\sc {Hypolite}, D. \& {Rieutord}, M.} 2014 {Dynamics of the envelope of a
  rapidly rotating star or giant planet in gravitational contraction}. {\em
  A\&A\/} {\bf 572}, A15.

\bibitem[{Knobloch} \& {Spruit}(1983)]{KS83}
{\sc {Knobloch}, E. \& {Spruit}, H.~C.} 1983 {The molecular weight barrier and
  angular momentum transport in radiative stellar interiors}. {\em A\&A\/} {\bf
  125}, 59--68.

\bibitem[{Lainey} {\em et~al.\/}(2015){Lainey}, {Jacobson}, {Tajeddine},
  {Cooper}, {Murray}, {Robert}, {Tobie}, {Guillot}, {Mathis}, {Remus},
  {Desmars}, {Arlot}, {De Cuyper}, {Dehant}, {Pascu}, {Thuillot}, {Le
  Poncin-Lafitte} \& {Zahn}]{lainey_cassini15}
{\sc {Lainey}, V., {Jacobson}, R.~A., {Tajeddine}, R., {Cooper}, N.~J.,
  {Murray}, C., {Robert}, V., {Tobie}, G., {Guillot}, T., {Mathis}, S.,
  {Remus}, F., {Desmars}, J., {Arlot}, J.-E., {De Cuyper}, J.-P., {Dehant}, V.,
  {Pascu}, D., {Thuillot}, W., {Le Poncin-Lafitte}, C. \& {Zahn}, J.-P.} 2015
  {New constraints on Saturn's interior from Cassini astrometric data}. {\em
  ArXiv e-prints\/} .

\bibitem[{Ligni\`eres} {\em et~al.\/}(1999){Ligni\`eres}, {Califano} \&
  {Mangeney}]{LCM99}
{\sc {Ligni\`eres}, F., {Califano}, F. \& {Mangeney}, A.} 1999 Shear layer
  instability in a highly diffusive stably stratified atmosphere. {\em A\&A\/}
  {\bf 349}, 1027--1036.

\bibitem[Maas \& Lam(1995)]{ML95}
{\sc Maas, L. \& Lam, F.-P.} 1995 Geometric focusing of internal waves. {\em J.
  Fluid Mech.\/} {\bf 300}, 1--41.

\bibitem[Maeder(2009)]{maeder09}
{\sc Maeder, A.} 2009 {\em Physics, Formation and Evolution of Rotating
  stars\/}. Springer.

\bibitem[{Marcus} {\em et~al.\/}(2015){Marcus}, {Pei}, {Jiang}, {Barranco},
  {Hassanzadeh} \& {Lecoanet}]{marcus15}
{\sc {Marcus}, P.~S., {Pei}, S., {Jiang}, C.-H., {Barranco}, J.~A.,
  {Hassanzadeh}, P. \& {Lecoanet}, D.} 2015 {Zombie Vortex Instability. I. A
  Purely Hydrodynamic Instability to Resurrect the Dead Zones of Protoplanetary
  Disks}. {\em ApJ\/} {\bf 808}, 87.

\bibitem[{Marcus} {\em et~al.\/}(2013){Marcus}, {Pei}, {Jiang} \&
  {Hassanzadeh}]{marcus13}
{\sc {Marcus}, P.~S., {Pei}, S., {Jiang}, C.-H. \& {Hassanzadeh}, P.} 2013
  {Three-Dimensional Vortices Generated by Self-Replication in Stably
  Stratified Rotating Shear Flows}. {\em Physical Review Letters\/} {\bf
  111}~(8), 084501.

\bibitem[{Maslowe}(1986)]{Maslowe86}
{\sc {Maslowe}, S.~A.} 1986 {Critical layers in shear flows}. {\em Annual
  Review of Fluid Mechanics\/} {\bf 18}, 405--432.

\bibitem[{Mathis} {\em et~al.\/}(2014){Mathis}, {Neiner} \& {Tran
  Minh}]{mathis_etal14}
{\sc {Mathis}, S., {Neiner}, C. \& {Tran Minh}, N.} 2014 {Impact of rotation on
  stochastic excitation of gravity and gravito-inertial waves in stars}. {\em
  A\&A\/} {\bf 565}, A47.

\bibitem[Morel(1997)]{Morel97}
{\sc Morel, P.} 1997 {CESAM: a code for stellar evolution calculations}. {\em A
  \& A Suppl. Ser.\/} {\bf 124}, 597--614.

\bibitem[Ogilvie(2009)]{O09}
{\sc Ogilvie, G.} 2009 {Tidal dissipation in rotating fluid bodies: a
  simplified model}. {\em MNRAS\/} {\bf 396}, 794--806.

\bibitem[{Ogilvie}(2014)]{O14}
{\sc {Ogilvie}, G.~I.} 2014 {Tidal Dissipation in Stars and Giant Planets}.
  {\em Ann. Rev. Astron. Astrophys.\/} {\bf 52}, 171--210.

\bibitem[{Paxton} {\em et~al.\/}(2011){Paxton}, {Bildsten}, {Dotter}, {Herwig},
  {Lesaffre} \& {Timmes}]{paxton_etal11}
{\sc {Paxton}, B., {Bildsten}, L., {Dotter}, A., {Herwig}, F., {Lesaffre}, P.
  \& {Timmes}, F.} 2011 {Modules for Experiments in Stellar Astrophysics
  (MESA)}. {\em Astrophys. J. Supp. Ser.\/} {\bf 192}, 3.

\bibitem[Rieutord(1987)]{R87}
{\sc Rieutord, M.} 1987 {Linear theory of rotating fluids using spherical
  harmonics. I. Steady flows}. {\em Geophys. Astrophys. Fluid Dyn.\/} {\bf 39},
  163.

\bibitem[Rieutord(2006)]{R06}
{\sc Rieutord, M.} 2006 The dynamics of the radiative envelope of rapidly
  rotating stars. i. a spherical boussinesq model. {\em A\&A\/} {\bf 451},
  1025--1036.

\bibitem[{Rieutord}(2008)]{R08}
{\sc {Rieutord}, M.} 2008 {The solar dynamo}. {\em C. R. Physique\/} {\bf 9},
  757--765.

\bibitem[Rieutord \& Beth(2014)]{RB14}
{\sc Rieutord, M. \& Beth, A.} 2014 {Dynamics of the envelope of rapidly
  rotating stars. I Effects of spin-down of the outer layers}. {\em to appear
  in A\&A\/} {\bf 1}, 1.

\bibitem[{Rieutord} \& {Espinosa Lara}(2013)]{REL13}
{\sc {Rieutord}, M. \& {Espinosa Lara}, F.} 2013 {Ab Initio Modelling of Steady
  Rotating Stars}. In {\em {SeIsmology for studies of stellar Rotation and
  Convection}\/} (ed. M.~{Goupil}, K.~{Belkacem}, C.~{Neiner},
  F.~{Ligni{\`e}res} \& J.~J. {Green}), {\em Lecture Notes in Physics, Berlin
  Springer Verlag\/}, vol. 865, pp. 49--73, astro--ph/1208.4926.

\bibitem[Rieutord {\em et~al.\/}(2000)Rieutord, Georgeot \& Valdettaro]{RGV00}
{\sc Rieutord, M., Georgeot, B. \& Valdettaro, L.} 2000 Waves attractors in
  rotating fluids: a paradigm for ill-posed cauchy problems. {\em Phys. Rev.
  Lett.\/} {\bf 85}, 4277--4280.

\bibitem[Rieutord {\em et~al.\/}(2001)Rieutord, Georgeot \& Valdettaro]{RGV01}
{\sc Rieutord, M., Georgeot, B. \& Valdettaro, L.} 2001 Inertial waves in a
  rotating spherical shell: attractors and asymptotic spectrum. {\em J. Fluid
  Mech.\/} {\bf 435}, 103--144.

\bibitem[{Rieutord} {\em et~al.\/}(2012){Rieutord}, {Triana}, {Zimmerman} \&
  {Lathrop}]{RTZL12}
{\sc {Rieutord}, M., {Triana}, S.~A., {Zimmerman}, D.~S. \& {Lathrop}, D.~P.}
  2012 {Excitation of inertial modes in an experimental spherical Couette
  flow}. {\em Phys. Rev. E\/} {\bf 86}~(2), 026304.

\bibitem[Rieutord \& Valdettaro(1997)]{RV97}
{\sc Rieutord, M. \& Valdettaro, L.} 1997 {Inertial waves in a rotating
  spherical shell}. {\em J. Fluid Mech.\/} {\bf 341}, 77--99.

\bibitem[Rieutord \& Valdettaro(2010)]{RV10}
{\sc Rieutord, M. \& Valdettaro, L.} 2010 Viscous dissipation by tidally forced
  inertial modes in a rotating spherical shell. {\em J. Fluid Mech.\/} {\bf
  643}, 363--394.

\bibitem[{Spruit} \& {Knobloch}(1984)]{SK84}
{\sc {Spruit}, H.~C. \& {Knobloch}, E.} 1984 {Baroclinic instability in stars}.
  {\em A\&A\/} {\bf 132}, 89--96.

\bibitem[Swart {\em et~al.\/}(2010)Swart, Manders, Harlander \& Maas]{Swart10}
{\sc Swart, A., Manders, A., Harlander, U. \& Maas, L.R.M.} 2010 {Experimental
  observation of strong mixing due to internal wave focusing over sloping
  terrain}. {\em Dyn. of Atmos. and Oceans\/} {\bf 50}~(1), 16 -- 34.

\bibitem[{Unno} {\em et~al.\/}(1989){Unno}, {Osaki}, {Ando}, {Saio} \&
  {Shibahashi}]{unno_etal89}
{\sc {Unno}, W., {Osaki}, Y., {Ando}, H., {Saio}, H. \& {Shibahashi}, H.} 1989
  {\em {Nonradial oscillations of stars}\/}. University of Tokyo Press.

\bibitem[Valdettaro {\em et~al.\/}(2007)Valdettaro, Rieutord, Braconnier \&
  Fraysse]{VRBF07}
{\sc Valdettaro, L., Rieutord, M., Braconnier, T. \& Fraysse, V.} 2007
  Convergence and round-off errors in a two-dimensional eigenvalue problem
  using spectral methods and arnoldi-chebyshev algorithm. {\em J. Comput. and
  Applied Math.\/} {\bf 205}, 382--393.

\bibitem[{Witte} \& {Savonije}(1999)]{WS99a}
{\sc {Witte}, M.~G. \& {Savonije}, G.~J.} 1999 {The dynamical tide in a
  rotating 10M$_\odot$ main sequence star. A study of g- and r-mode
  resonances}. {\em A\&A\/} {\bf 341}, 842--852.

\bibitem[Zahn(1974)]{zahn74}
{\sc Zahn, J.-P.} 1974 {Rotational instabilities and stellar evolution}. In
  {\em IAU Symp. 59: Stellar Instability and Evolution\/}, pp. 185--194.

\bibitem[Zahn(1992)]{zahn92}
{\sc Zahn, J.-P.} 1992 Circulation and turbulence in rotating stars. {\em
  A\&A\/} {\bf 265}, 115.

\bibitem[{Zahn}(1993)]{zahn93houches}
{\sc {Zahn}, J.-P.} 1993 {Instabilities and turbulence in rotating stars.} In
  {\em Astrophysical Fluid Dynamics - Les Houches 1987\/} (ed. J.-P. {Zahn} \&
  J.~{Zinn-Justin}), pp. 561--615.

\bibitem[{Zhevakin}(1963)]{Z63}
{\sc {Zhevakin}, S.~A.} 1963 {Physical Basis of the Pulsation Theory of
  Variable Stars}. {\em Ann. Rev. Astron. Astrophys.\/} {\bf 1}, 367.

\end{thebibliography}
\bibliographystyle{jfm}

\end{document}